\documentclass[11pt,a4paper]{article}
\usepackage{jheppub}
\usepackage{tabularx}
\usepackage[flushleft]{threeparttable}
\usepackage{multirow}
\usepackage{xspace,xfrac}
\makeatletter
\usepackage{subfig}
\usepackage{tikz-feynman} 
\newcommand\footnoteref[1]{\protected@xdef\@thefnmark{\ref{#1}}\@footnotemark}
\makeatother
\usepackage{ifthen} 
\usepackage{feynmp-auto}
\newboolean{articletitles}
\setboolean{articletitles}{true} 
\usepackage{fix-cm}
\usepackage{anyfontsize}

\newcommand{\GeV}{\ensuremath{~\text{GeV}}\xspace}
\newcommand{\TeV}{\ensuremath{~\text{TeV}}\xspace}

\newcommand{\ETmiss}{\ensuremath{E_{\mathrm{T}}^{\text{miss}}}\xspace}
\newcommand{\dPhi}{\ensuremath{\Delta\phi}\xspace}
\newcommand{\ZPrime}{\ensuremath{Z^{\prime}}\xspace}
\newcommand{\mZPrime}{\ensuremath{m_{\ZPrime}}\xspace}
\newcommand{\Pt}{\ensuremath{p_{\mathrm{T}}}\xspace}
\newcommand{\pT}{\ensuremath{p_{\mathrm{T}}}\xspace}
\newcommand{\PtISR}{\ensuremath{\Pt^{\text{ISR}}}\xspace}
\newcommand{\kt}{\ensuremath{k_{\mathrm{T}}}\xspace}
\newcommand{\mT}{\ensuremath{m_{\mathrm{T}}}\xspace}
\newcommand{\Eta}{\ensuremath{|\eta|}\xspace}
\newcommand{\drap}{\ensuremath{y*}\xspace}
\newcommand{\dR}{\ensuremath{\Delta R}\xspace}
\newcommand{\rinv}{\ensuremath{r_{\text{inv}}}\xspace}

\newcommand{\rT}{\ensuremath{r_{\text{T}}}\xspace}
\newcommand{\rinvave}{\ensuremath{\overline{r}_{\text{inv}}}\xspace}
\newcommand{\maos}{\ensuremath{m_{\text{MAOS}}}\xspace}
\newcommand{\rinvavemaos}{\ensuremath{\overline{r}_{\text{inv}}^{\prime}}\xspace}

\newcommand{\MADGRAPH}{\textsc{MadGraph}\xspace}
\newcommand*{\MGMCatNLOV}[1]{\textsc{MadGraph5}\_aMC@NLO~#1\xspace}
\newcommand*{\PYTHIAV}[1]{\textsc{Pythia}~#1\xspace}
\newcommand*{\DELPHES}[1]{\textsc{Delphes}~#1\xspace}
\newcommand*{\PYHF}[1]{\textsc{pyhf}~#1\xspace}

\newcommand{\Ncdark}{\ensuremath{N_{c}^{\mathrm{dark}}}\xspace}
\newcommand{\Nfdark}{\ensuremath{N_{f}^{\mathrm{dark}}}\xspace}
\newcommand{\MTtwo}{\ensuremath{m_{\mathrm{T2}}}\xspace}
\newcommand{\ptvecmiss}{\ensuremath{\vec{p}_{\mathrm{T}}^{\,\mathrm{miss}}}\xspace}
\newcommand{\ptvecmissOne}{\ensuremath{\vec{p}_{\mathrm{T}}^{\,\mathrm{miss}_1}}\xspace}
\newcommand{\ptvecmissTwo}{\ensuremath{\vec{p}_{\mathrm{T}}^{\,\mathrm{miss}_2}}\xspace}
\newcommand{\ptmissOne}{\ensuremath{p_{\mathrm{T}}^{\mathrm{miss}_1}}\xspace}

\newcommand{\pzmissOne}{\ensuremath{p_{z}^{\mathrm{miss}_1}}\xspace}

\newcommand{\pxmissi}{\ensuremath{p_{x}^{\mathrm{miss}_i}}\xspace}

\newcommand{\pymissi}{\ensuremath{p_{y}^{\mathrm{miss}_i}}\xspace}
\newcommand{\mmissOne}{\ensuremath{m^{\mathrm{miss}_1}}\xspace}

\newcommand{\ptvecj}{\ensuremath{\vec{p}_{\mathrm{T}}^{\,j}}\xspace}
\newcommand{\ptvecjOne}{\ensuremath{\vec{p}_{\mathrm{T}}^{\,j_1}}\xspace}
\newcommand{\ptvecjTwo}{\ensuremath{\vec{p}_{\mathrm{T}}^{\,j_2}}\xspace}
\newcommand{\ptjOne}{\ensuremath{p_{\mathrm{T}}^{j_1}}\xspace}

\newcommand{\pzjOne}{\ensuremath{p_{z}^{j_1}}\xspace}

\newcommand{\pxji}{\ensuremath{p_{x}^{j_i}}\xspace}

\newcommand{\pyji}{\ensuremath{p_{y}^{j_i}}\xspace}
\newcommand{\aOne}{\ensuremath{\alpha_{1}}\xspace}
\newcommand{\aTwo}{\ensuremath{\alpha_{2}}\xspace}
\newcommand{\mdecomp}{\ensuremath{m_{\text{decomp}}}\xspace}

\begin{document}

\title{Semi-visible jets + X: Illuminating Dark Showers with Radiation}

\author[a]{Bingxuan Liu}
\author[b]{Kevin Pedro}
\affiliation[a]{School of Science, Sun Yat-sen University,Shenzhen Campus, 66 Gongchang Road, Shenzhen, Guangdong 518107, PRC}
\affiliation[b]{Fermi National Accelerator Laboratory, Batavia, IL 60510, USA}
\emailAdd{liubx28@mail.sysu.edu.cn}
\emailAdd{pedrok@fnal.gov}

\abstract{
We investigate the potential to search for semi-visible jets (SVJs) at the Large Hadron Collider (LHC) using initial-state radiation (ISR). Both photon ISR and jet ISR channels are considered, using a benchmark signal model with the decay of a leptophobic \ZPrime mediator forming two SVJs. We compare and extend several techniques to decompose the missing transverse momentum into per-jet contributions, in order to reconstruct the mediator mass and to define a new observable measuring the fraction of invisible dark hadrons. The presence of ISR facilitates the identification of the SVJs, and the resulting boost improves the resolution of the observables, especially for models with high invisible fractions. We combine the two observables to propose a complete search strategy and discuss an extension of the strategy to probe the whole model parameter space.
}

\preprint{FERMILAB-PUB-24-0563-CSAID-PPD}
\maketitle
\clearpage
\section{Introduction}
\label{sec:intro} 

Although the standard model (SM) of particle physics has been proven to be a
very successful theory whose theoretical predictions agree well with most
experimental results, it does not explain all phenomena, such as the origin of
dark matter (DM)~\cite{dm1,dm2,dm3,dm4,dm5}. Searching for DM has been one of
the most critical tasks in particle physics. The Large Hadron Collider (LHC) is
the only particle collider in the world that can reach a center-of-mass energy
of 13.6 TeV, which offers a unique opportunity to search for a wide range of
potentially accessible DM candidates. The LHC experiments, such as the CMS and
ATLAS collaborations, have diverse DM search programs. Until very recently,
they have been focused on models with weakly interacting massive particles
(WIMPs)~\cite{wimps}, where the DM candidates do not leave any trace in the
detector, resulting in an imbalanced distribution of transverse momentum.
Therefore, these classic DM searches consider events with large missing
transverse momentum (\ETmiss, the magnitude of the two-vector \ptvecmiss),
and are categorized by the types of visible
particles that recoil against the DM candidates. A large variety of scenarios
have been explored, such as DM candidates recoiling against a
jet~\cite{cmsmonojet,atlasmonojet}, a top
quark~\cite{cmsmonotop,atlasmonotop,atlasmonotoppair}, a vector
boson~\cite{cmsmonoz,atlasmonov}, a photon~\cite{cmsmonophoton}, or a Higgs
boson~\cite{cmsmonohiggs,atlasmonohiggstautau,atlasmonohiggsbb}. This type of
search assumes the DM candidates are clearly separated from the objects they
recoil against, so they explicitly require \ptvecmiss not to overlap with those
objects. As a consequence, the traditional search strategies have limited
sensitivity to models without this feature.     

Dark sectors containing a new confining force $SU(N)$, an analog to quantum
chromodynamics (QCD) in the SM, are well motivated from both the theoretical
and experimental perspectives. If the scale of the new confining force,
referred to as the dark QCD~\cite{darkqcd}, is related to SM QCD, the models
are consistent with the DM relic density~\cite{hiddenvalleydm}. Unlike the WIMP
models where the DM candidates are produced directly, here the DM candidates
are generated during the dark showering. The dark sector contains several
flavors of dark quarks ($\chi_i$), which form bound states called dark hadrons
that may be either stable or unstable. The stable dark hadrons are DM
candidates that will escape detection, while the unstable dark hadrons will
decay promptly to SM particles. The mixture of visible and invisible decay
products creates a striking feature called a ``semi-visible'' jet
(SVJ)~\cite{svjtheory,svjlhc}. The previous DM search program at the LHC is not
optimal for such signatures, because of the required separation of \ptvecmiss from visible objects
mentioned above, as discussed in Refs.~\cite{cmssvj,atlassvj}.

SVJs are attracting more attention in the LHC physics program.  CMS performed
the first search for SVJs~\cite{cmssvj}, which considers a heavy mediator,
\ZPrime, produced via a $s$-channel resonant production and decaying to dark
quarks. A recent ATLAS analysis probes non-resonant production of
SVJs~\cite{atlassvj}. The community has also identified a set of common model
parameters in order to be able to compare results across
experiments~\cite{snowmasssvj}. One model parameter, \rinv, which is the
fraction of dark hadrons that are stable and invisible
($N_{\mathrm{stable}}/(N_{\mathrm{stable}} + N_{\mathrm{unstable}})$), governs
the experimental signatures. When \rinv approaches zero, the signal jets are
fully visible rather than semi-visible. But they are still different from those
jets expected in SM, as the complicated dark showering leads to changes in jet substructure.
ATLAS has published a result searching for such dark jets~\cite{atlasdarkjets}. The typical dijet
searches~\cite{atlasdijet} have also explored a large phase space relevant to
this scenario. If \rinv goes to one, the jets become fully invisible, so
traditional DM searches relying on \ETmiss become sensitive. CMS has recently
demonstrated the complementary sensitivity of these
strategies~\cite{cmsdarksector}. So far, the dedicated SVJ search strategies
only cover a portion of the overall phase space for intermediate \rinv values.

The ATLAS and CMS searches for SVJs require either \ETmiss or transverse
momentum sum triggers. For possible signal processes, the former would exploit the missing energy from the
invisible particles from the dark shower, while the latter would rely on the high
momentum of the visible decay products.
Consequently, it is challenging to optimize the search strategy for the entire
range of \rinv values. In addition, the trigger thresholds are too high to
include lower mediator masses.  One common strategy adopted in dijet resonance
searches to overcome this is to trigger on the initial state radiation (ISR).
Though processes with substantial ISR have significantly reduced cross
sections, this technique has been demonstrated to be sensitive to mediator
masses from 700 down to 200\GeV~\cite{cmsdijetisr,atlasdijetisr}. The ISR
production channel has not yet been explored in the context of dark QCD
searches. In this work, we show that it is a promising avenue. The unique
features of SVJs also allow us to construct more powerful discriminating
variables when the SVJ system is boosted against the energetic ISR. Depending
on the mediator mass and the ISR momentum, the two SVJs can either be merged
into one large-radius jet, or reconstructed as two separate small-radius jets.
In this paper, we consider the latter case.     

\begin{figure}[ht]
  \begin{center}
   \includegraphics[width=0.8\columnwidth]{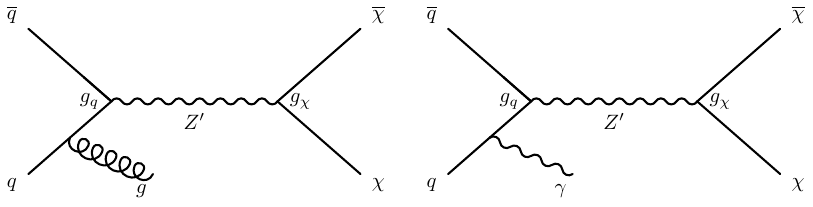}
   \caption{Feynman diagrams of the two \ZPrime production channels considered in this work: \ZPrime production in association with an ISR gluon (left) and \ZPrime production in association with an ISR photon (right).}
  \label{fig:diagrams}
  \end{center}
\end{figure}

The benchmark signal process is the $s$-channel production of a new heavy
particle (\ZPrime) that subsequently decays to two dark
quarks~\cite{svjtheory,svjlhc}. Figure~\ref{fig:diagrams} summarizes the
leading-order production diagrams. We consider the scenarios where the ISR
object is a jet or a photon, leaving other possibilities, such as a vector
boson, to future work.

The article is organized as follows: Section~\ref{sec:samples} details the
simulation procedure; Section~\ref{sec:kine} discusses the kinematic
properties of the signal and background processes; Section~\ref{sec:analysis}
lays out a simple search strategy, followed by a summary of the sensitivity in
Section~\ref{sec:results}; and finally, Section~\ref{sec:summary} gives an outlook
and offers some further thoughts.

\section{Simulated Event Samples}
\label{sec:samples}

All samples used in this work are produced using
\MGMCatNLOV{3.5.1}~\cite{madgraph} for generation,
\PYTHIAV{8.306}~\cite{pythia} for showering, and
\DELPHES{3.5.0}~\cite{delphes} for reconstruction.
The CMS detector geometry and performance are used for reconstruction. 

\subsection{Signal Process}

The signal grid is defined by the \ZPrime mass and \rinv. Two mass points, 0.5
TeV and 1.0 TeV, are included to cover the most relevant kinematic region.
Three \rinv values, 0.1, 0.5, and 0.9, span the whole \rinv range. Two
additional mass points, 0.75 TeV and 1.25 TeV, with \rinv set to 0.9, are also
generated to evaluate the sensitivity in Section~\ref{sec:results}. Since the
only change for different \rinv values occurs during the showering step, the
same generated events are reused. The CKKW-L scheme~\cite{Lonnblad:2011xx} is used to perform matrix
element merging in \PYTHIAV{8.306}, with a merging scale of 45. The details of
the signal model follow Ref.~\cite{cmssvj}, using the \MADGRAPH
implementation from Ref.~\cite{svjlhc}. Important parameters include $g_{q}$,
the coupling between \ZPrime and SM quarks, which is set to 0.25 for all quark
flavors, and $g_{\chi}$, the coupling between \ZPrime and the dark quarks,
which is set to $1/\sqrt{\smash[b]{\Ncdark\Nfdark}} = 0.5$. The latter takes
into account \Ncdark, the number of dark colors, and \Nfdark, the number of
dark flavors, which are both set to 2 in this model. The resulting branching
fraction for $\ZPrime\to \chi \overline{\chi}$, $\mathcal{B}(\ZPrime\to \chi
\overline{\chi})$, is therefore 47\%, in agreement with the benchmark for
simplified DM models~\cite{Albert:2017onk}. The dark hadron masses are set to
20 GeV. 

Handling the signal processes with an ISR photon is straightforward,
as we do not expect energetic photons from \ZPrime decays. The signal samples
in this channel are generated with an additional photon. Processes with up to
one additional parton besides the photon are included as well. The minimum
\Pt of the additional photon is set to 150 GeV in order to have appreciable acceptance
from typical single photon triggers. Each signal
sample contains 100,000 events.

The ISR jet channel is more complicated as there are also jets from \ZPrime
decays. Furthermore, the \Pt of the jets from \ZPrime decays depend on \rinv,
which makes it necessary to generate the samples inclusively. Given
that a typical single jet momentum trigger threshold approaches 500 GeV, the acceptance
of the inclusive production mode is low. Therefore, each signal sample contains
500,000 events in the ISR jet channel. Up to two additional partons are
included. Table~\ref{tab:generator} summarizes the processes generated in \MGMCatNLOV{3.5.1}.

\begin{table}[htbp]
  \begin{center}
  \caption{The \MADGRAPH processes used in the signal generation.
  The symbols xd and xd$\sim$ stand for dark quarks and dark antiquarks in this implementation,
  while j stands for an SM parton (quark or gluon).}
\makebox[0pt]{
\renewcommand{\arraystretch}{1.2}
\begin{tabular}{|c|c|c|}
\hline
Channel & Process & Generator Selection \\
\hline
\multirow{2}{*}{ISR Photon} & generate p p $>$ xd xd$\sim$ $\gamma$ & \multirow{2}{*}{$p_{\mathrm{T}}^{\gamma} > 150\GeV$} \\
           & add process p p $>$ xd xd$\sim$ $\gamma$ j &                     \\
\hline
\multirow{3}{*}{ISR Jet} & generate p p $>$ xd xd$\sim$ & \multirow{3}{*}{-} \\
        & add process p p $>$ xd xd$\sim$ j &  \\
        & add process p p $>$ xd xd$\sim$ j j &  \\
\hline
\end{tabular}
}
  \label{tab:generator}
  \end{center}
\end{table}

\subsection{Background Process}

The main background for this search comes from SM QCD multijet production. Two sets
of samples are created. For the photon ISR channel, the sample includes multijet processes
with one photon and up to three jets, where the photon is required to
have \Pt larger than 150 GeV. For the jet ISR channel, all multijet processes
with up to four jets are included, where at least one jet has to pass $\Pt >
500\GeV$. Each sample has 500,000 events produced.
While applying selections at the generation level can introduce biases,
the choices made here are sufficient for the scope of this study.

\section{Kinematic Properties}
\label{sec:kine}

This section discusses some interesting characteristics of the event kinematic
properties for both the signal and background processes, in order to motivate the
analysis strategy established in Section~\ref{sec:analysis}. 

\subsection {Basic Kinematics}

Very loose fiducial selections are applied so that only events containing at
least two jets with $\Pt > 25\GeV$ and $\Eta < 4.5$ are considered. This is
because of the proposed search strategy that relies on observables constructed
using the two jets from the \ZPrime and \ETmiss, as described in
Section~\ref{sec:analysis}. The \mT is constructed using the two leading non-ISR jets
and \ETmiss.   

\begin{figure}[ht]
\begin{center}
 \includegraphics[width=0.22\columnwidth]{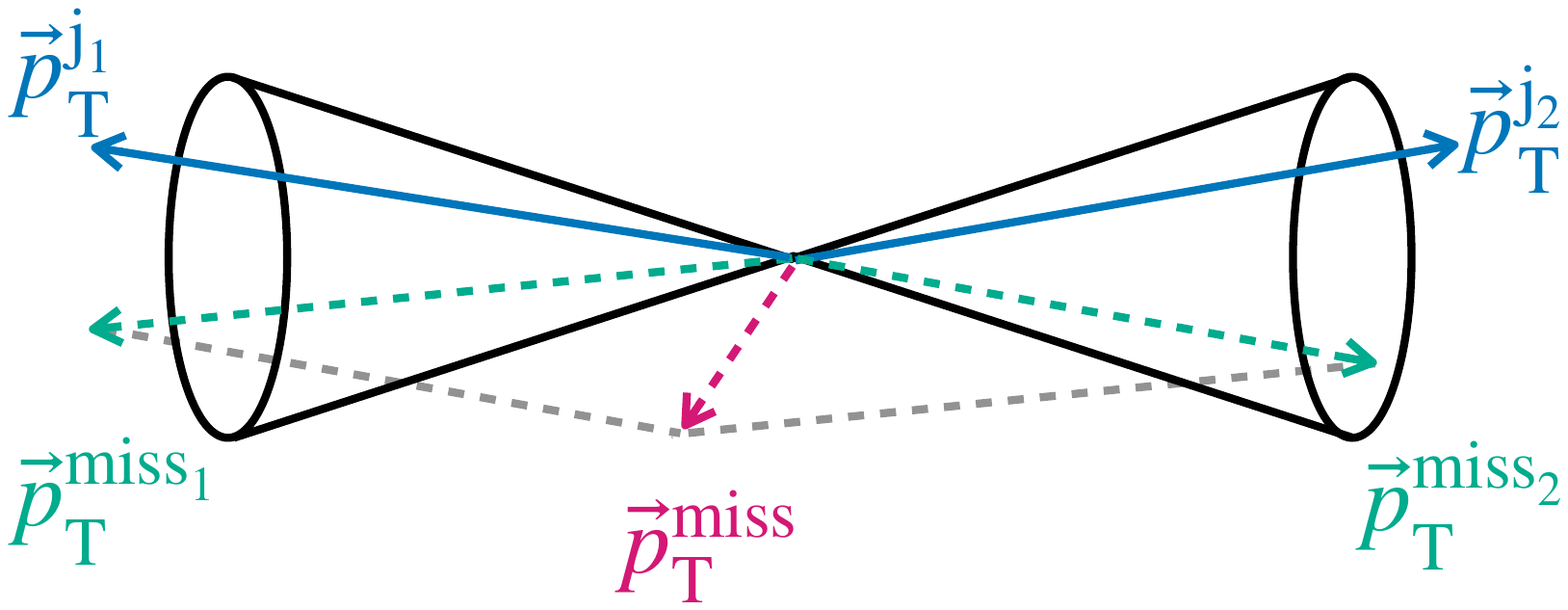}
 \includegraphics[width=0.22\columnwidth]{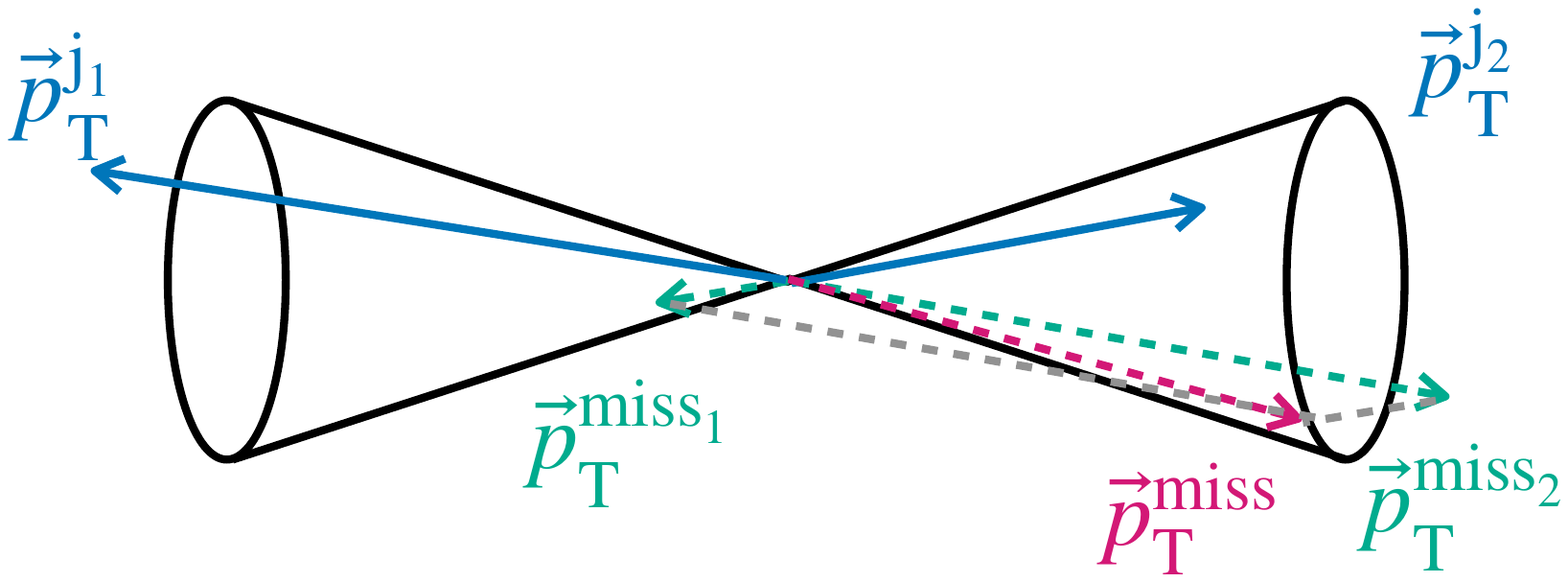}
 \includegraphics[width=0.22\columnwidth]{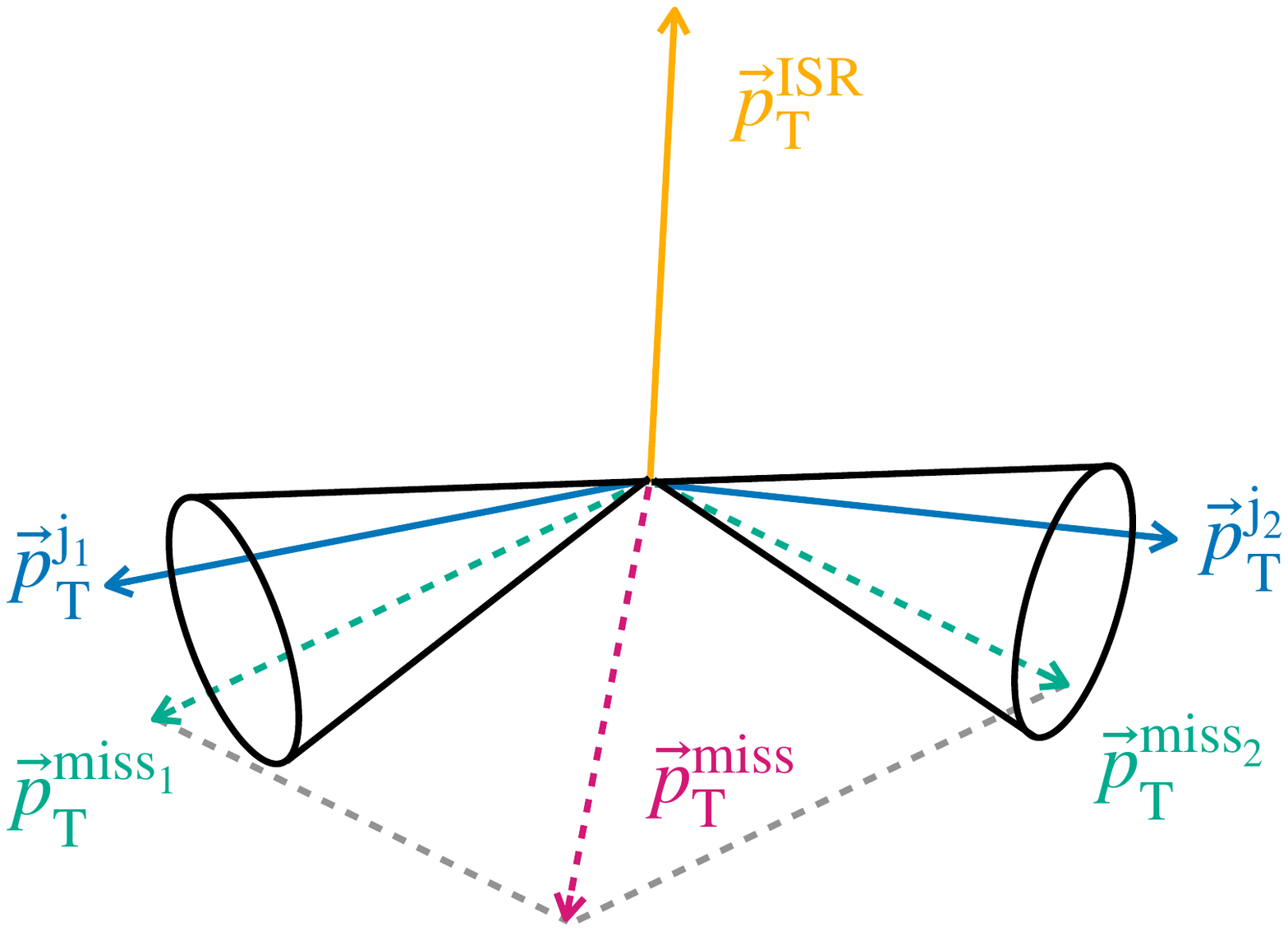}
 \includegraphics[width=0.22\columnwidth]{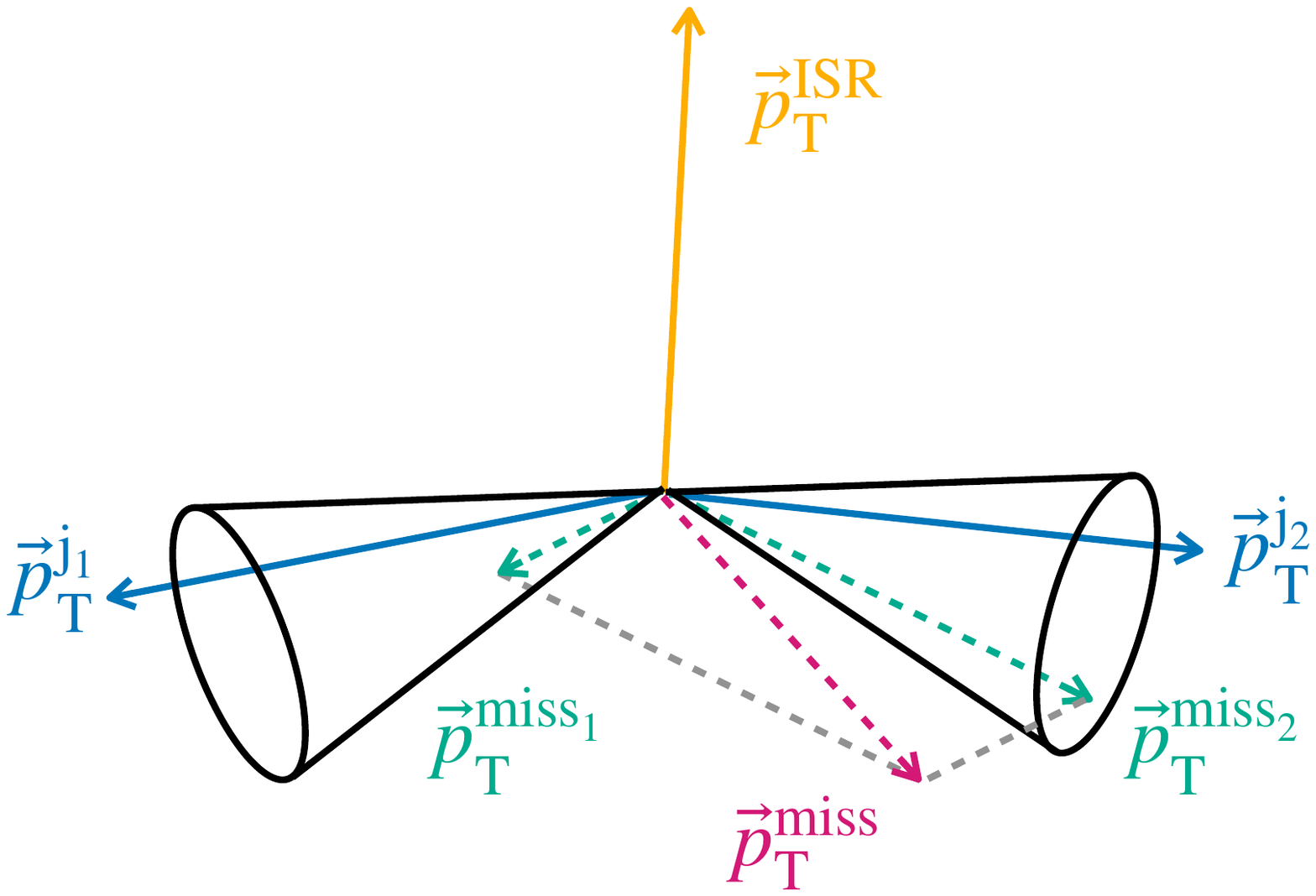}
\caption{Diagrams illustrating the back-to-back topology and ISR topology, with symmetric and asymmetric invisible components in the SVJs.}
\label{fig:topo_cartoon}
\end{center}
\end{figure}

Some signal kinematic variables are strongly correlated with the \Pt of the ISR
object. Each signal jet originates from a dark quark that undergoes a showering
process, producing a stochastic mixture of stable and unstable dark hadrons. If
both jets have similar numbers of stable dark hadrons, the transverse component
of the missing momentum will be symmetric between the jets and will mostly
cancel, even for large \rinv.  However, when there is an energetic ISR object
recoiling against the \ZPrime system, the resulting SVJs become more
collimated. Events with either symmetric or asymmetric invisible components in
the SVJs become more similar topologically, as illustrated in
Figure~\ref{fig:topo_cartoon}.

\begin{figure}[htbp]
\begin{center}
 \includegraphics[width=0.49\columnwidth]{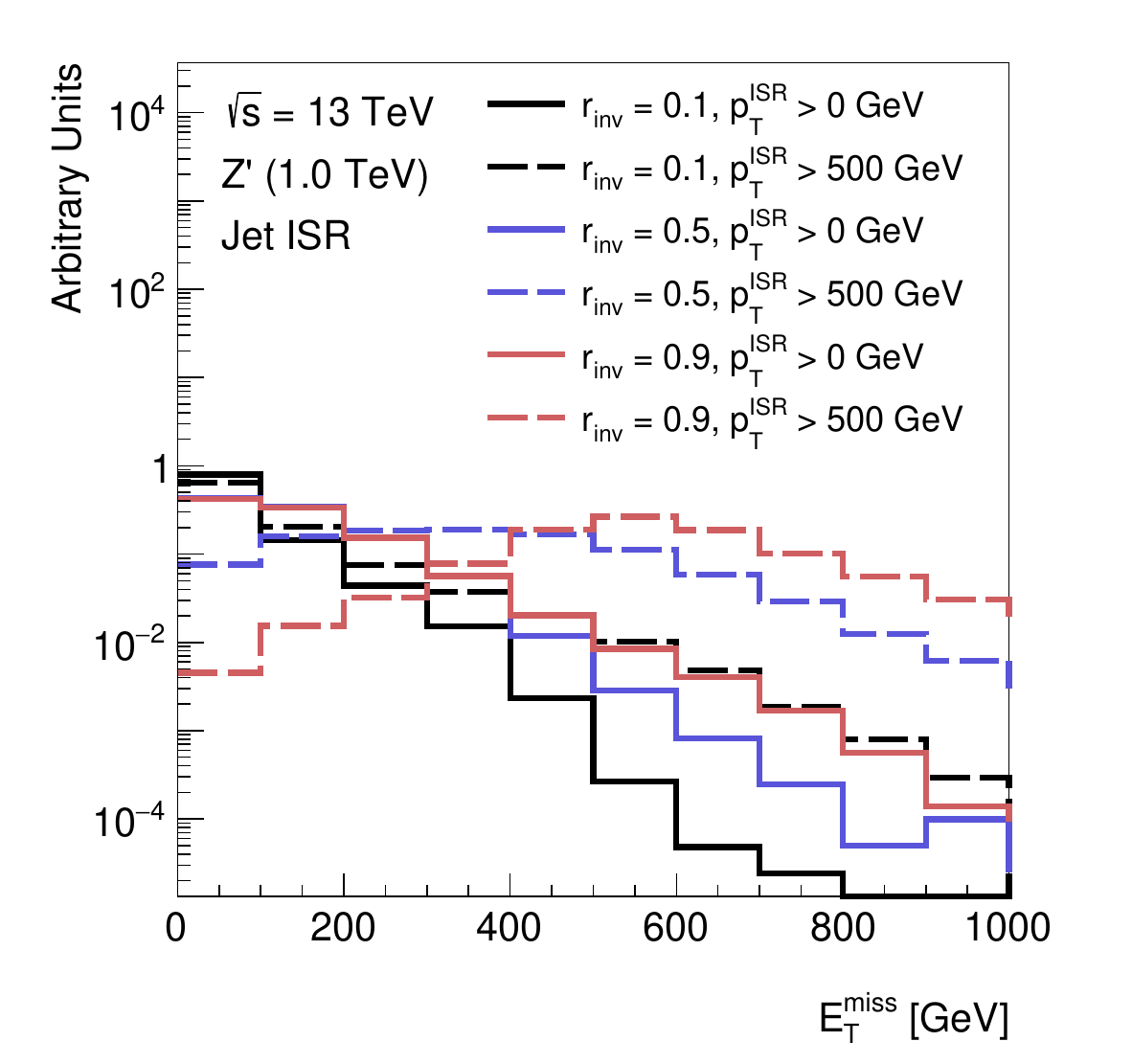}
 \includegraphics[width=0.49\columnwidth]{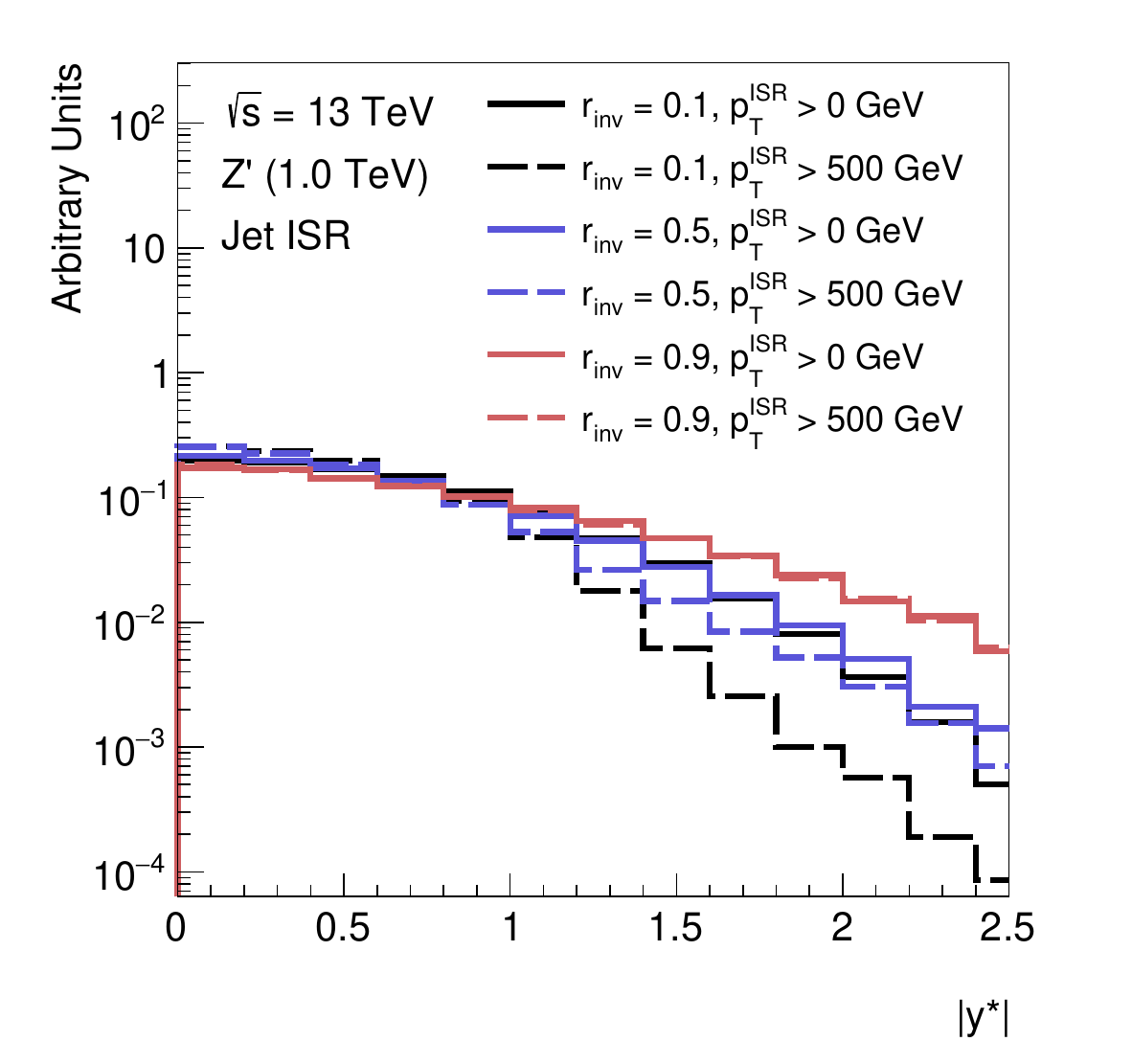}
\caption{Left: \ETmiss distributions of the \ZPrime samples for $\mZPrime = 1.0\TeV$ (right), with various \rinv and $\PtISR$. Right: $|\drap|$ distributions of the \ZPrime samples for $\mZPrime = 1.0\TeV$ (right), with various \rinv and $\PtISR$.}
\label{fig:met_com_rinv}
\end{center}
\end{figure}

Figure~\ref{fig:met_com_rinv} shows that \ETmiss increases with \rinv when the
\Pt of the ISR object is high, while without an energetic ISR object, the
\ETmiss depends less on \rinv, because of the cancellation of the symmetric
invisible components. This enhancement of \ETmiss has remarkable impacts on the
observables that will be discussed later. Another variable often used in heavy
particle search is the rapidity difference \drap, defined as $(Y_{1} -
Y_{2})/2$, where $Y_{i}$ is the rapidity of a given jet~\cite{atlasdijet}. In
heavy particle decays, \drap for the two resulting jets has a smaller absolute
value compared to the multijet background. This quantity is less affected than
\ETmiss by the ISR object.

\begin{figure}[ht]
\begin{center}
 \includegraphics[width=0.49\columnwidth]{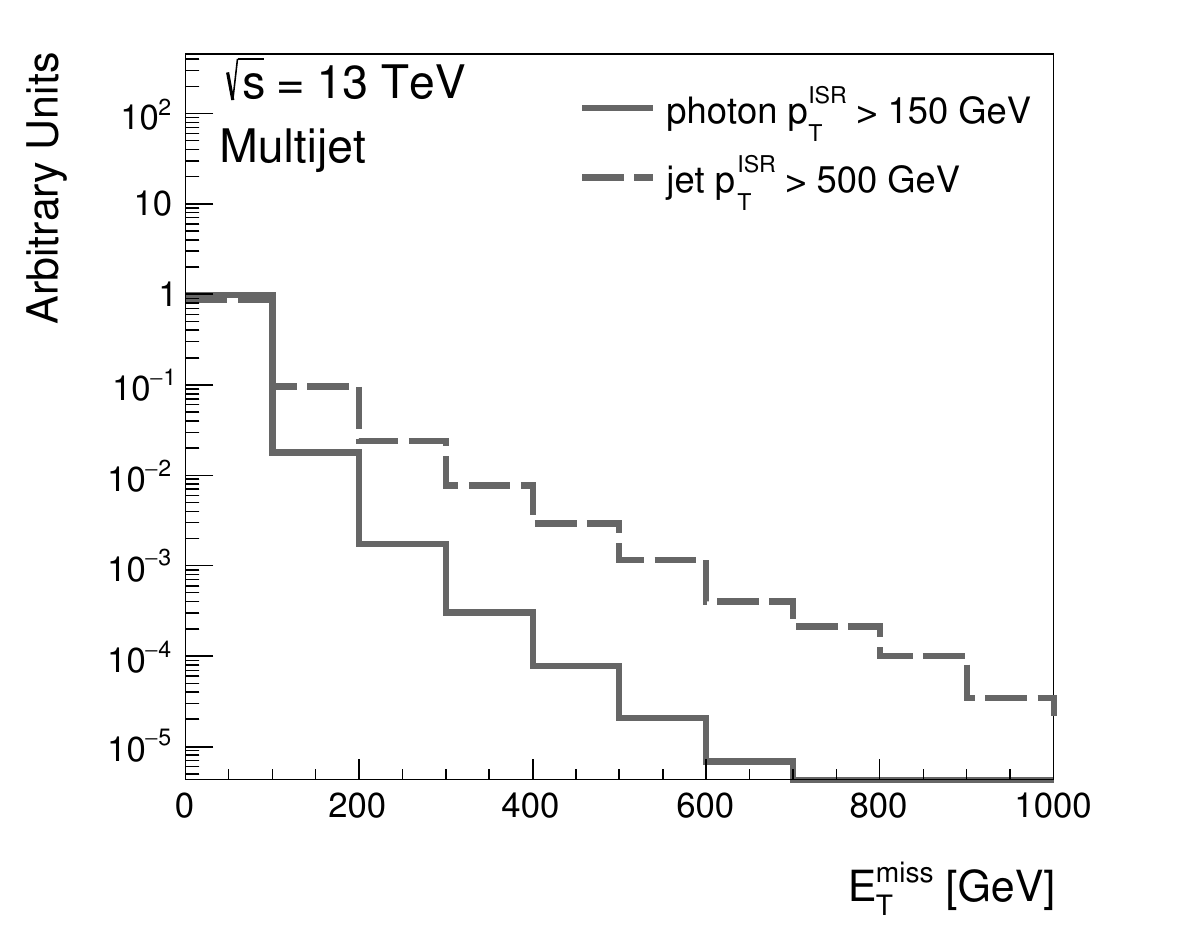}
 \includegraphics[width=0.49\columnwidth]{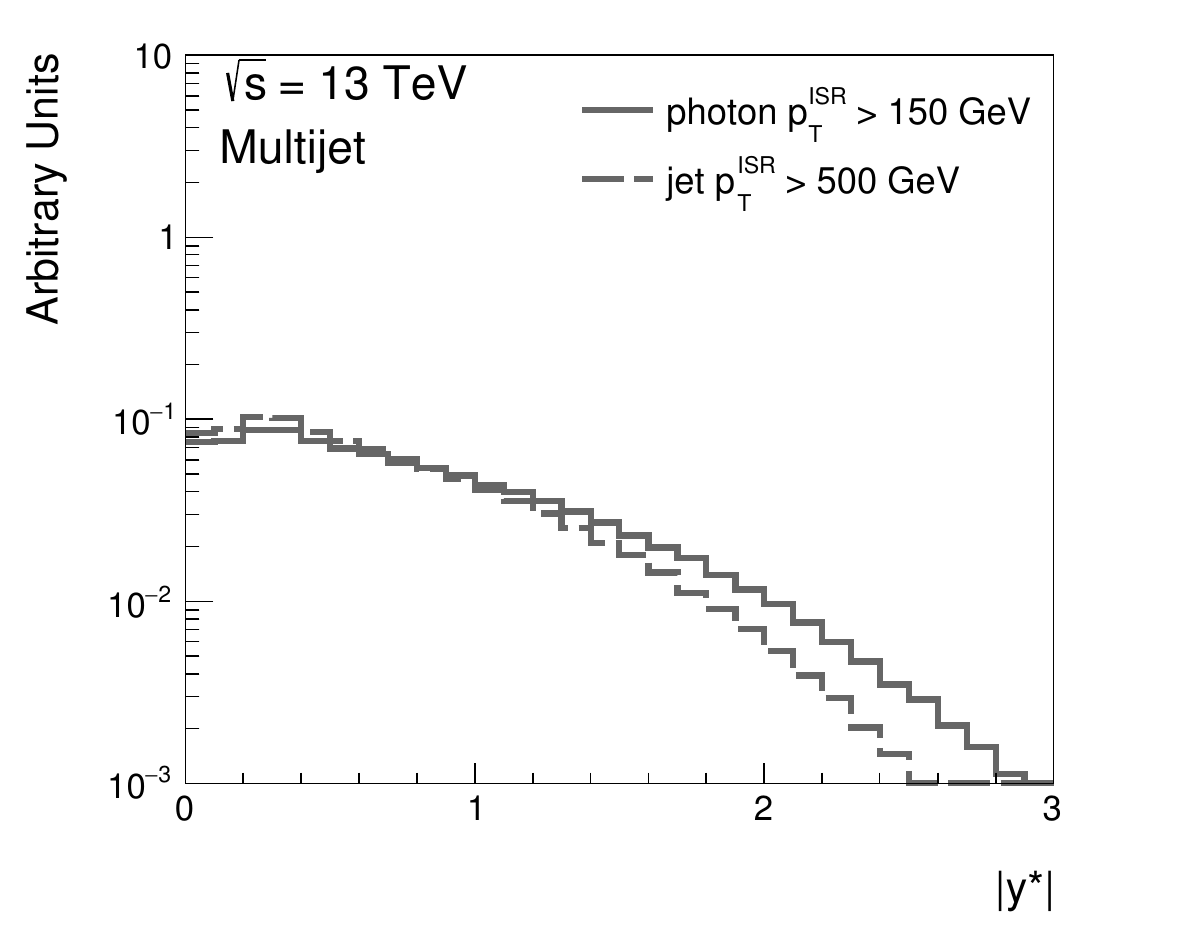}
\caption{\ETmiss (left) and $|\drap|$ (right) distributions of multijet events for $\PtISR = 150\GeV$ (solid line) and $\PtISR = 500\GeV$ (dashed line).}
\label{fig:multijet_com_isr}
\end{center}
\end{figure}

The kinematic distributions of the multijet background process are also
affected by the ISR object, as seen in Figure~\ref{fig:multijet_com_isr}. The
\ETmiss increases moderately as the \Pt of the ISR object goes from 0 to 500
\GeV, because there is no genuine source of sizable \ETmiss in the multijet
process. The main source of \ETmiss in multijet events comes from jet
mis-reconstruction caused by detector inefficiency and noise, such as inactive
calorimeter cells and beam remnants. By construction, \ETmiss is balanced
against the other visible objects, so when there exists an energetic ISR
object, \ETmiss increases correspondingly. The increase is smooth, so in the
next section, we will see it does not introduce discontinuities in the main
observables. \drap changes similarly for the multijet background and the
signal.

\subsection{\texorpdfstring{\ptvecmiss}{Missing Transverse Momentum Vector} Decomposition}
\label{subsec:metdecomp}

The challenge in this search is reconstructing the multiple missing transverse momentum contributions that arise from the production of stable bound states during the dark quark hadronization.
We compare multiple methods of decomposing the measured \ptvecmiss two-vector into components matching the two semi-visible jets in each event.

The first method is the calculation of the \MTtwo mass variable~\cite{Lester:1999tx} (computed using Ref.~\cite{Lester:2014yga}).
\MTtwo is defined as:
\begin{equation}
\MTtwo \equiv \min \left[ \max \left\{ \mT(j_1, \ptvecmissOne), \mT(j_2, \ptvecmissTwo) \right\} \right],
\end{equation}
where the minimization considers all possible values satisfying:
\begin{equation}
\ptvecmiss = \ptvecmissOne + \ptvecmissTwo,
\label{eq:met_constraint}
\end{equation}
and the transverse mass \mT is defined as:
\begin{equation}
\mT^{2}(j,\ptvecmiss) = (E_{\mathrm{T}}^{j} + \ETmiss)^2 - (\ptvecj + \ptvecmiss)^{2} .
\end{equation}
This method was developed for any final-state system with two invisible particles.
It naturally produces two missing transverse momentum components, each associated with one of the two jets included in the calculation.

A newer, analytic decomposition was developed specifically for semivisible jet final states~\cite{svjdecomp}.
This technique assumes that the visible and invisible transverse momentum of each semivisible jet are aligned. The ratio between the visible and invisible transverse momentum is strongly correlated with $\rinv/(1 - \rinv)$.
This leads to the following system of equations that can be solved to find the two coefficients \aOne and \aTwo, using Eq.~\eqref{eq:met_constraint} as a constraint:
\begin{align}
\ptvecmissOne &= \aOne \cdot \ptvecjOne, \\ 
\ptvecmissTwo &= \aTwo \cdot \ptvecjTwo.
\end{align}

\subsection{Mass Reconstruction}

The transverse mass \mT, with the visible component as the massive dijet system, was proposed in Ref.~\cite{svjtheory} and used in the CMS $s$-channel search~\cite{cmssvj} to reconstruct the \ZPrime mediator mass.
By default, the multijet background distribution for this type of mass variable has a smoothly falling spectrum, facilitating background estimation.
Simply requiring large \ETmiss sculpts the background \mT distribution
significantly, motivating the introduction of \rT, the ratio between \ETmiss
and \mT, which was shown to bring a clear sensitivity gain with minimal
background sculpting.

Another, more sophisticated method to recover the \ZPrime mass uses the \MTtwo construction: the
\MTtwo-assisted on shell (MAOS) technique~\cite{maos}.
MAOS produces the variable \maos by promoting the decomposed two-vectors \ptvecmissOne and \ptvecmissTwo to four-vectors, assigning $\pzmissOne = \pzjOne (\ptmissOne/\ptjOne)$ and $\mmissOne = 0$, and similarly for the second component.
The invariant mass of the two visible four-vectors and the two invisible four-vectors is then an estimate of the mediator mass.
This variable was recently shown to be nearly optimal in Ref.~\cite{svjpedro} for SVJ searches without any ISR.

\begin{figure}[ht]
\begin{center}
 \includegraphics[width=0.32\columnwidth]{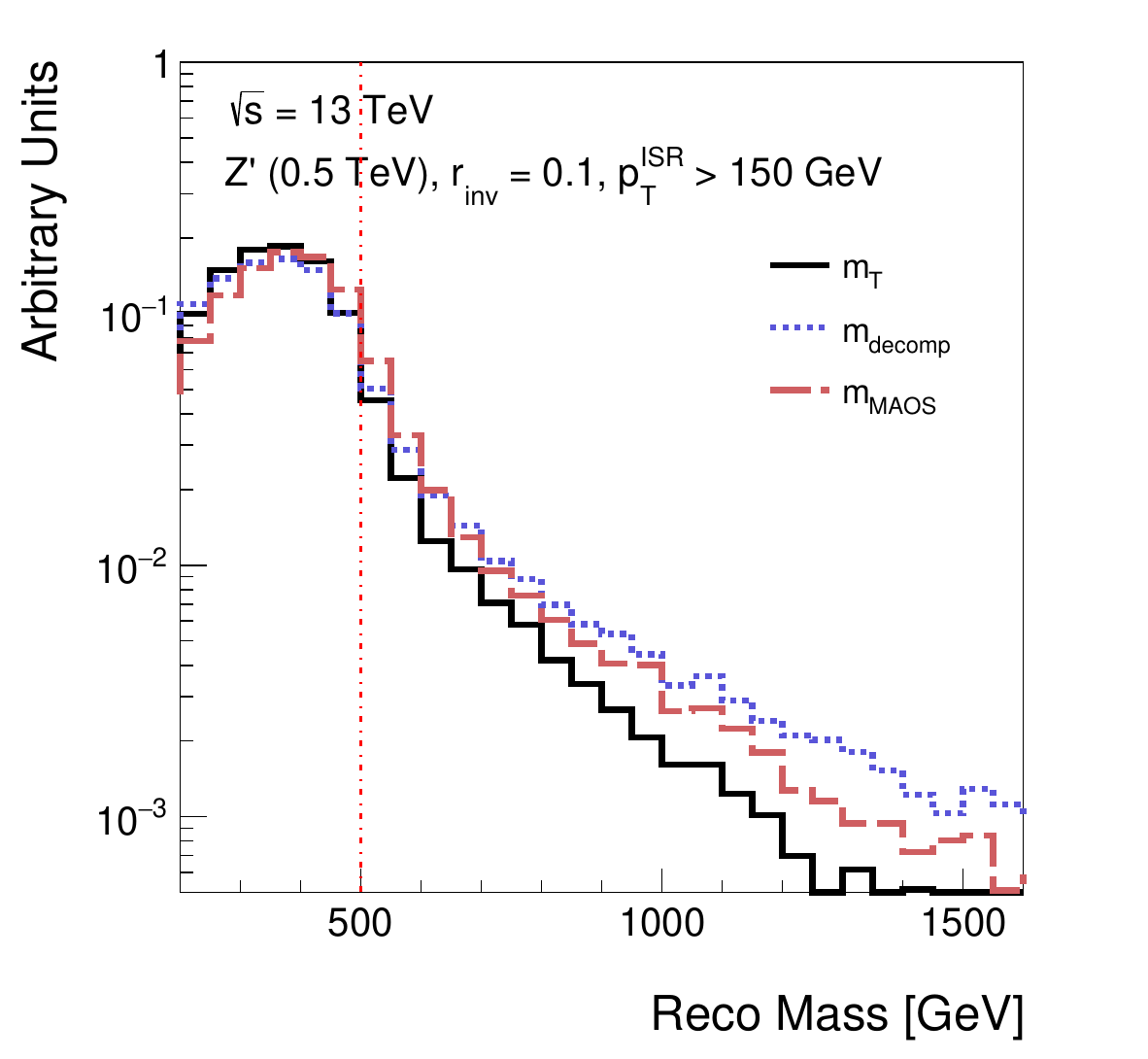}
 \includegraphics[width=0.32\columnwidth]{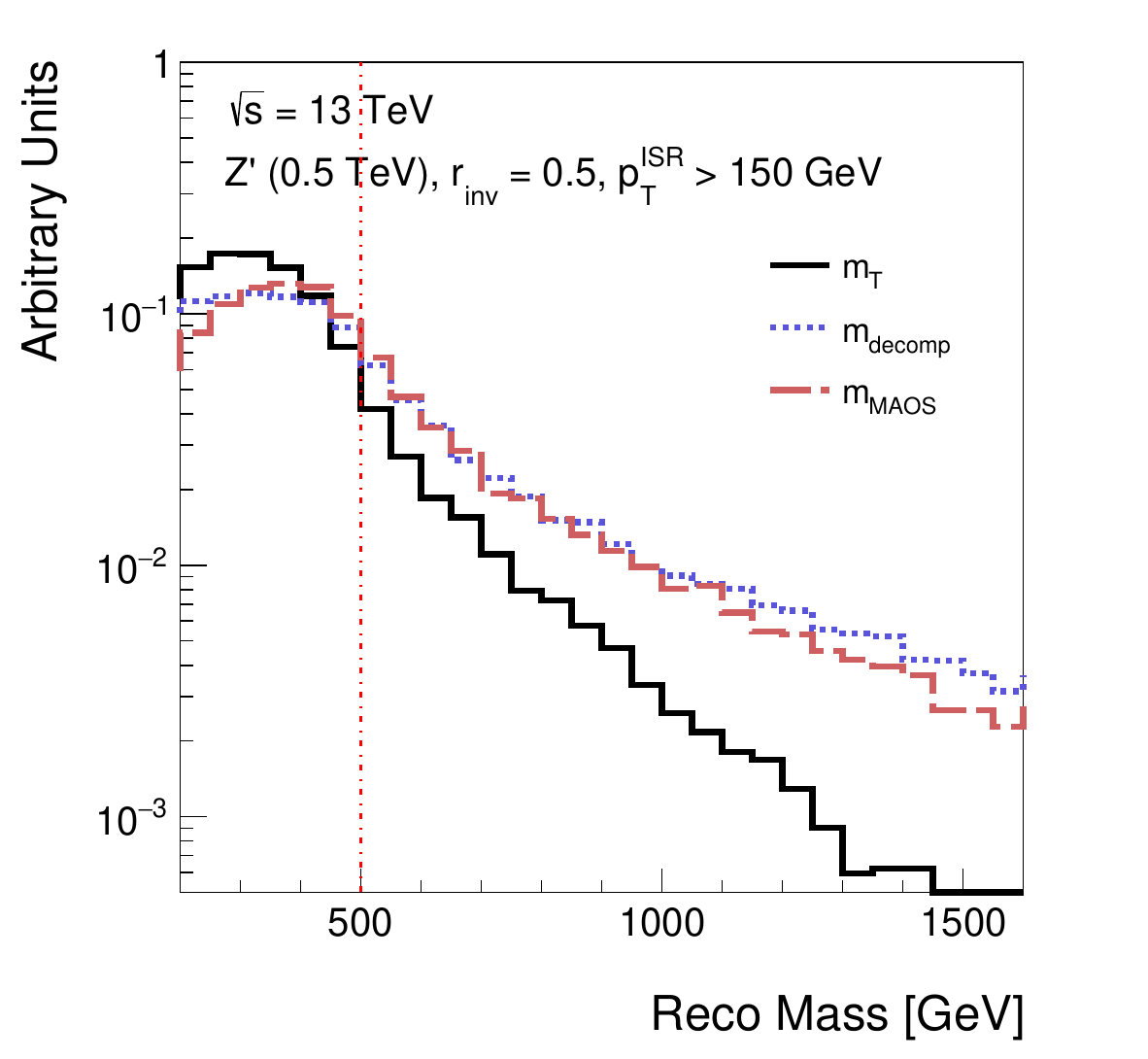}
 \includegraphics[width=0.32\columnwidth]{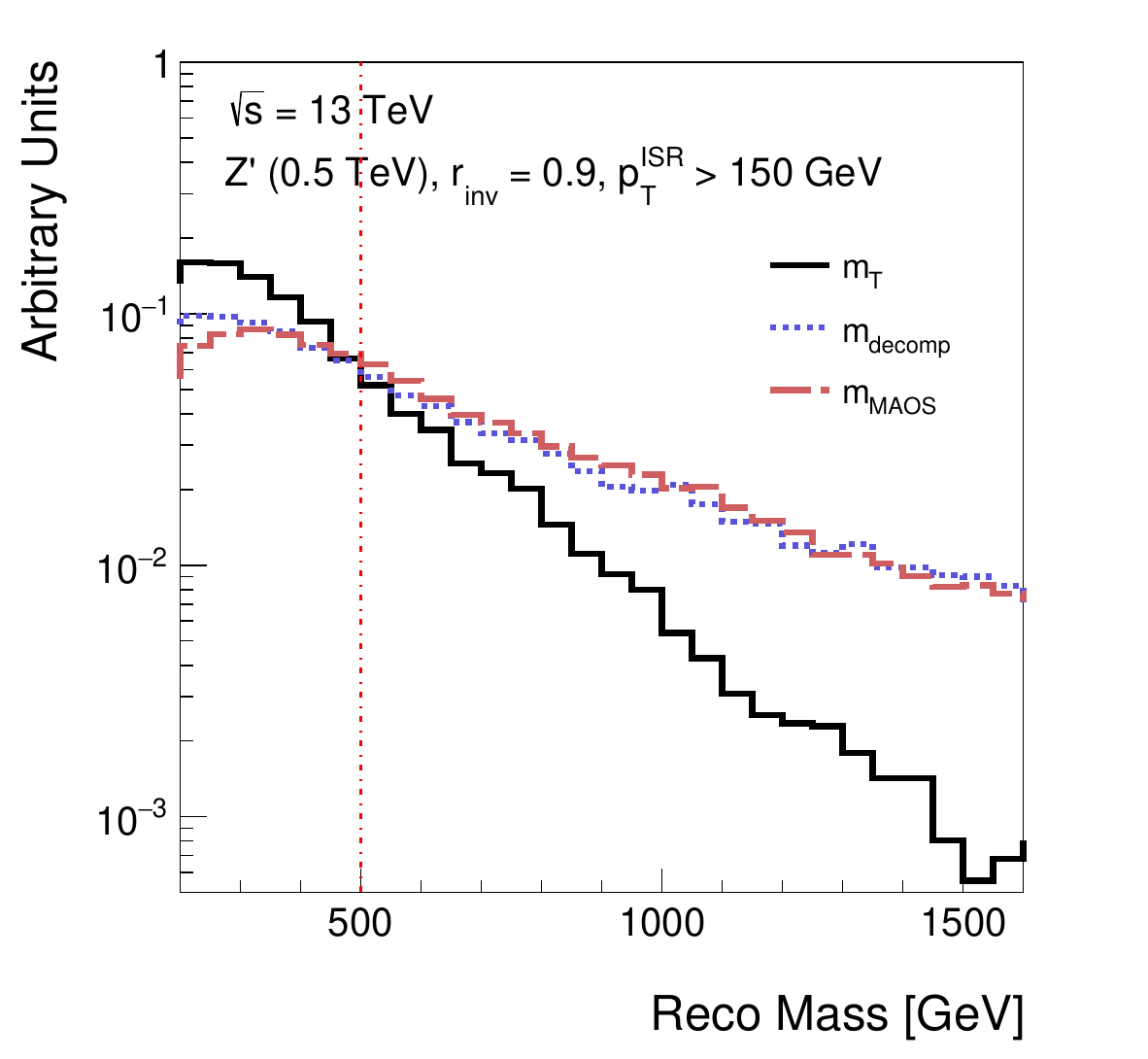}
\caption{Distributions of different reconstructed mass variables for $\mZPrime = 0.5\TeV$ with $\PtISR = 150\GeV$ and $\rinv = 0.1$ (left), $\rinv = 0.5$ (center), and $\rinv = 0.9$ (right). The red dashed vertical line shows the theoretical \mZPrime value.}
\label{fig:mass1}
\end{center}
\end{figure}

\begin{figure}[ht]
\begin{center}
 \includegraphics[width=0.32\columnwidth]{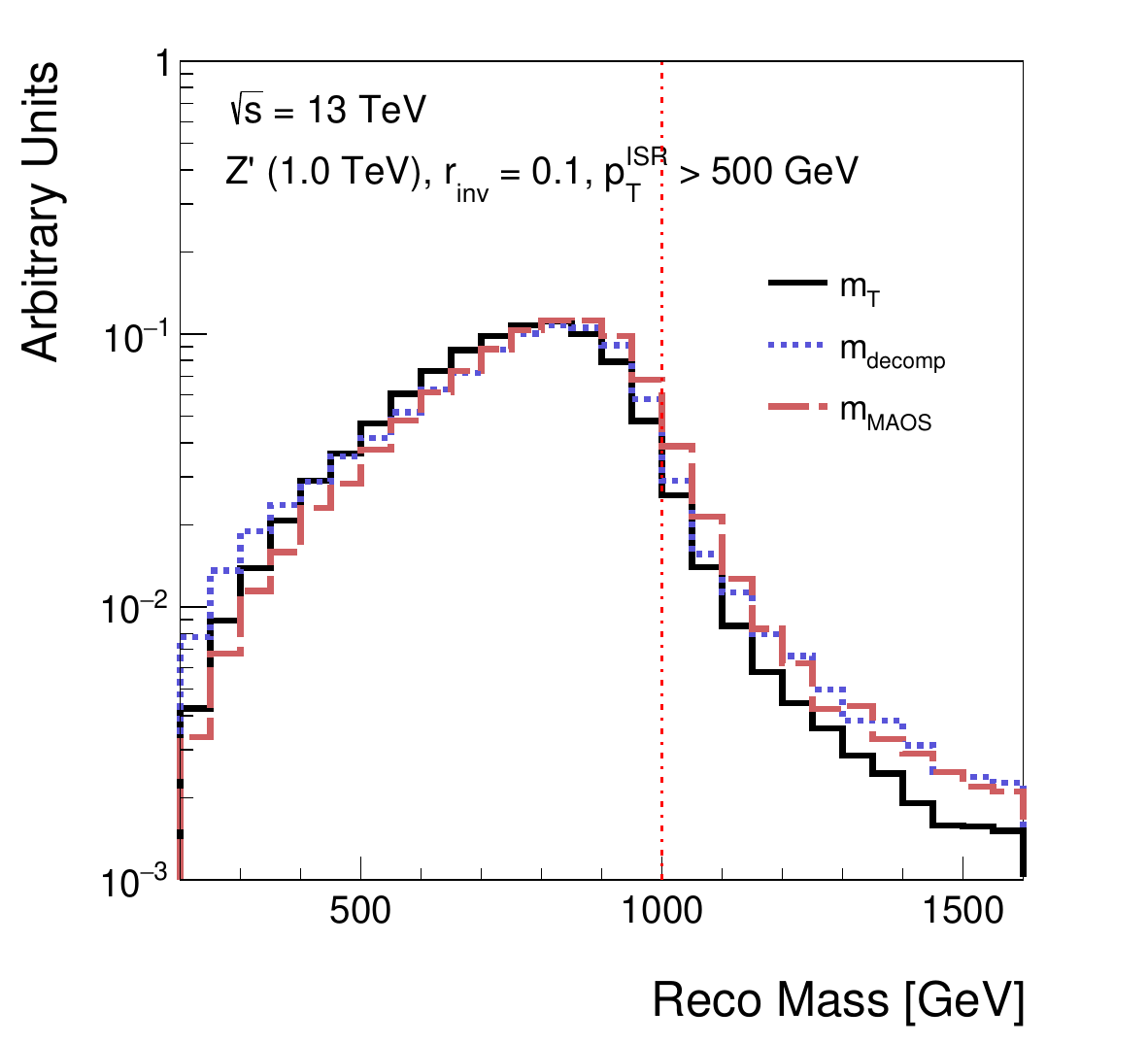}
 \includegraphics[width=0.32\columnwidth]{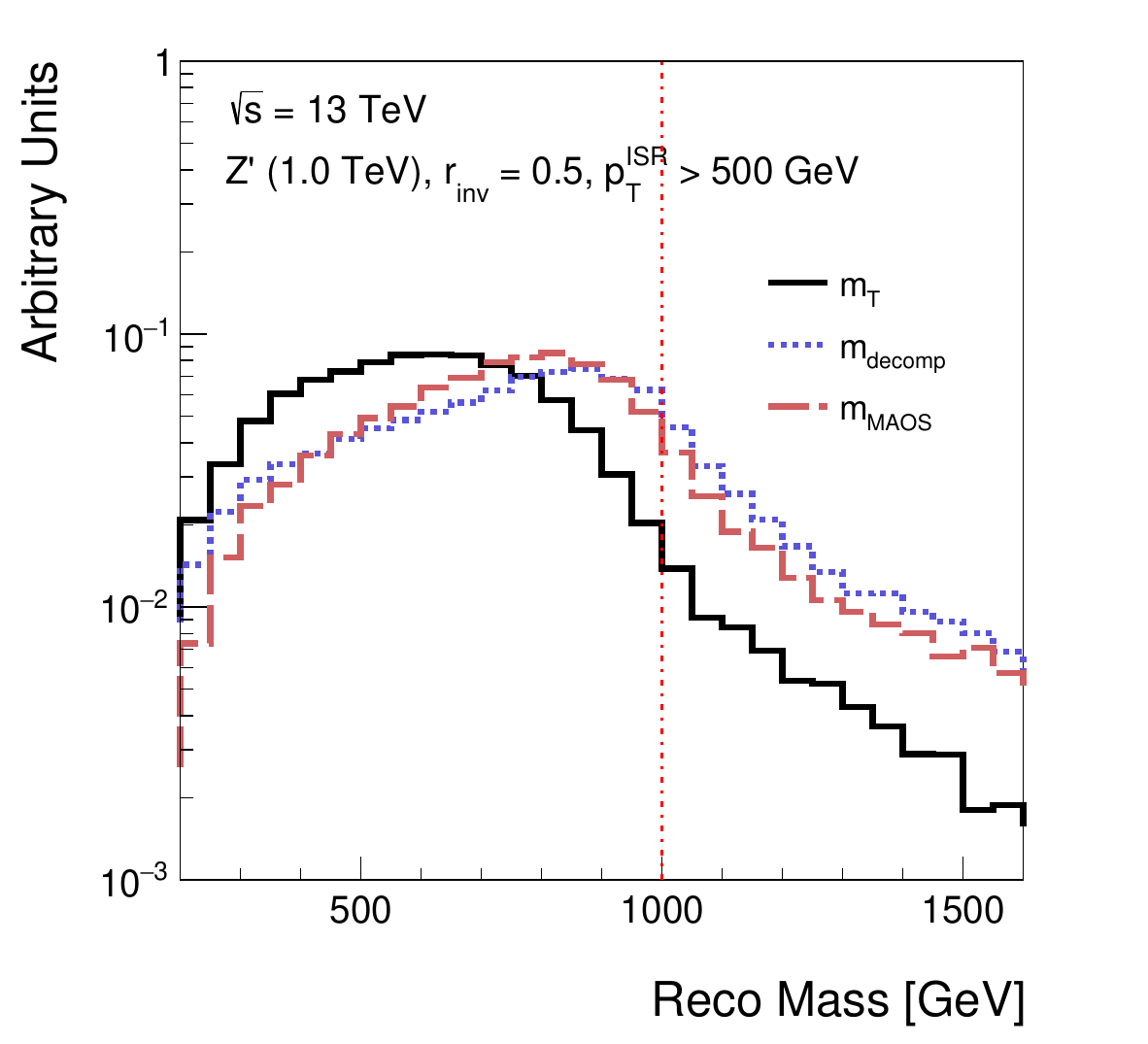}
 \includegraphics[width=0.32\columnwidth]{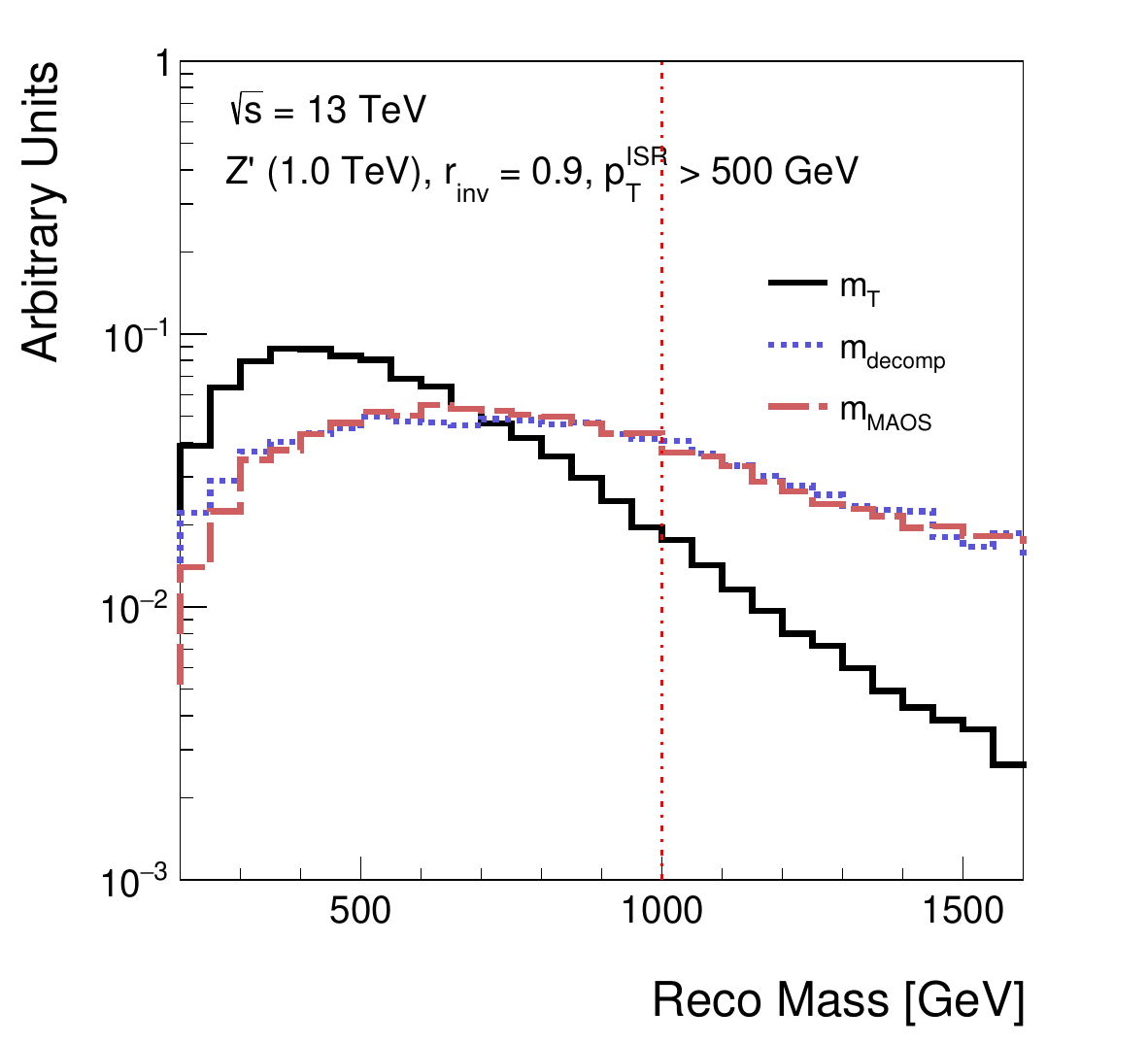}
\caption{Distributions of different reconstructed mass variables for $\mZPrime = 1.0\TeV$ with $\PtISR = 500\GeV$ and $\rinv = 0.1$ (left), $\rinv = 0.5$ (center), and $\rinv = 0.9$ (right). The red dashed vertical line shows the theoretical \mZPrime value.}
\label{fig:mass2}
\end{center}
\end{figure}

We apply the same procedure to the invisible components from the analytic decomposition, resulting in the new variable \mdecomp.
This combination has not been studied before.
We compare all three mass variables for $\mZPrime = 0.5\TeV$ with $\PtISR = 150\GeV$ in Figure~\ref{fig:mass1} and for $\mZPrime = 1.0\TeV$ with $\PtISR = 500\GeV$ in Figure~\ref{fig:mass2}.
For this comparison, we assume that the semivisible jets in each event are always correctly identified (further discussed in Section~\ref{sec:jetmatch}).
Both \maos and \mdecomp retain better performance than \mT as \rinv increases.
\maos is found to have the best resolution.
Importantly, all three variables have similarly falling spectra in multijet background events. 

\subsection{\texorpdfstring{\rinv}{Invisible Fraction} Reconstruction}

We further extend the use of \ptvecmiss decomposition techniques to measure \rinv on a per-event basis.
This is most natural for the analytic decomposition, which is explicitly based on \rinv.
In this case, we can write $\rinv^{i} = \alpha_i/(1 + \alpha_i)$ for jet $i$,
and we observe that the signal
events tend to have larger $\rinv^{i}$ for both jets. As a consequence, \rinvave, the
average of $\rinv^{i}$, can be used to distinguish signal events from
the background. The \rinvave is smaller than the corresponding \rinv for
various reasons. Events are already required to have two jets with $\Pt >
25\GeV$, which acts as a requirement on the visible decay fraction in each jet.
In addition, the jet axes are only approximations of the invisible components.

To apply the same principle to the numerical \MTtwo decomposition, we must account for the lack of an explicit $\alpha$ coefficient relating the $x$ and $y$ momenta of each invisible component.
We therefore take the average of the corresponding formula applied separately to the $x$ and $y$ momenta:
\begin{equation}
\rinv^{i} = \frac{1}{2} \left( \frac{\pxmissi}{\pxji + \pxmissi} + \frac{\pymissi}{\pyji + \pymissi} \right).
\end{equation}
We then define the average \rinvavemaos in the same way as \rinvave above.

\begin{figure}[ht]
\begin{center}
 \includegraphics[width=0.32\columnwidth]{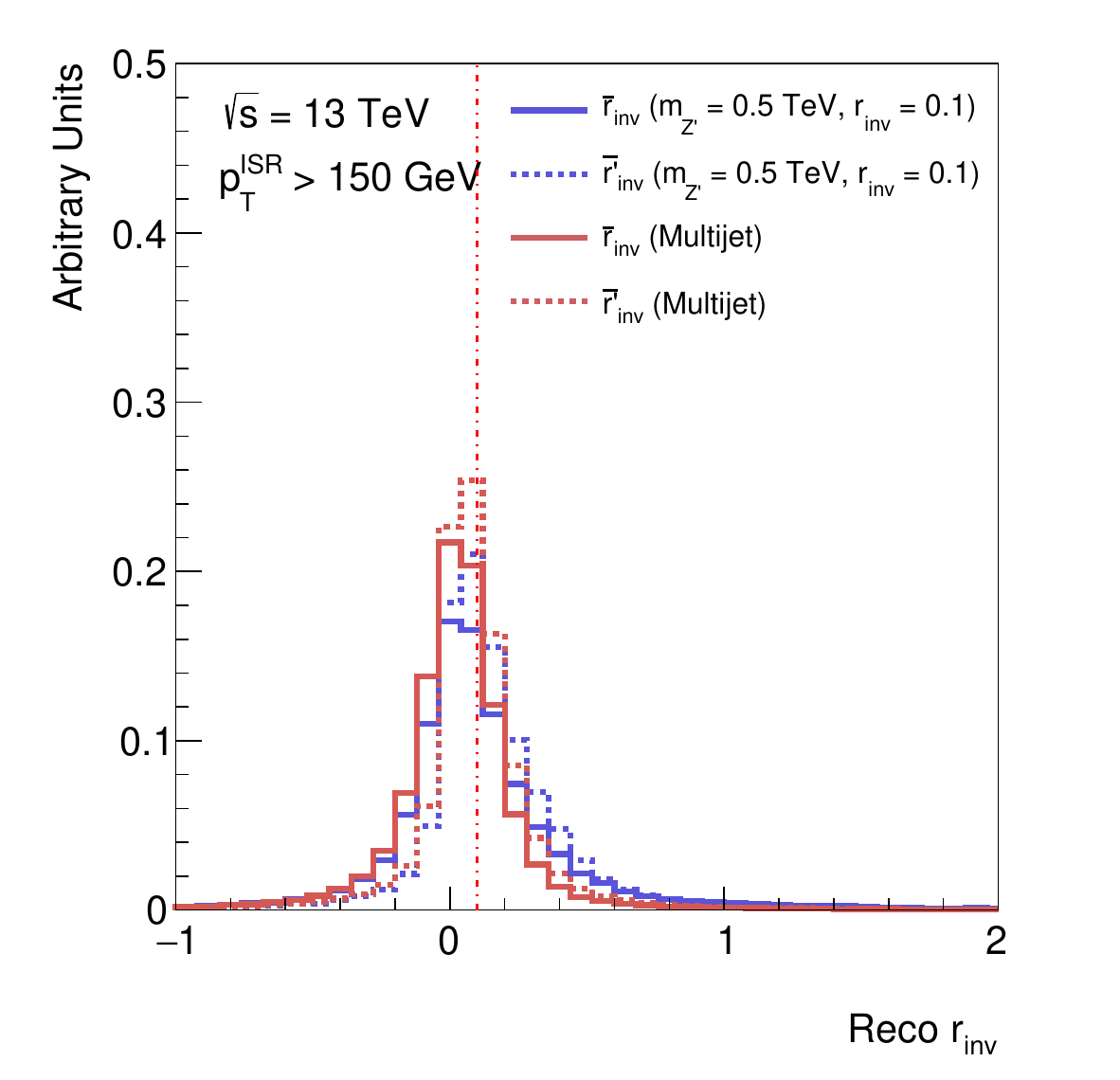}
 \includegraphics[width=0.32\columnwidth]{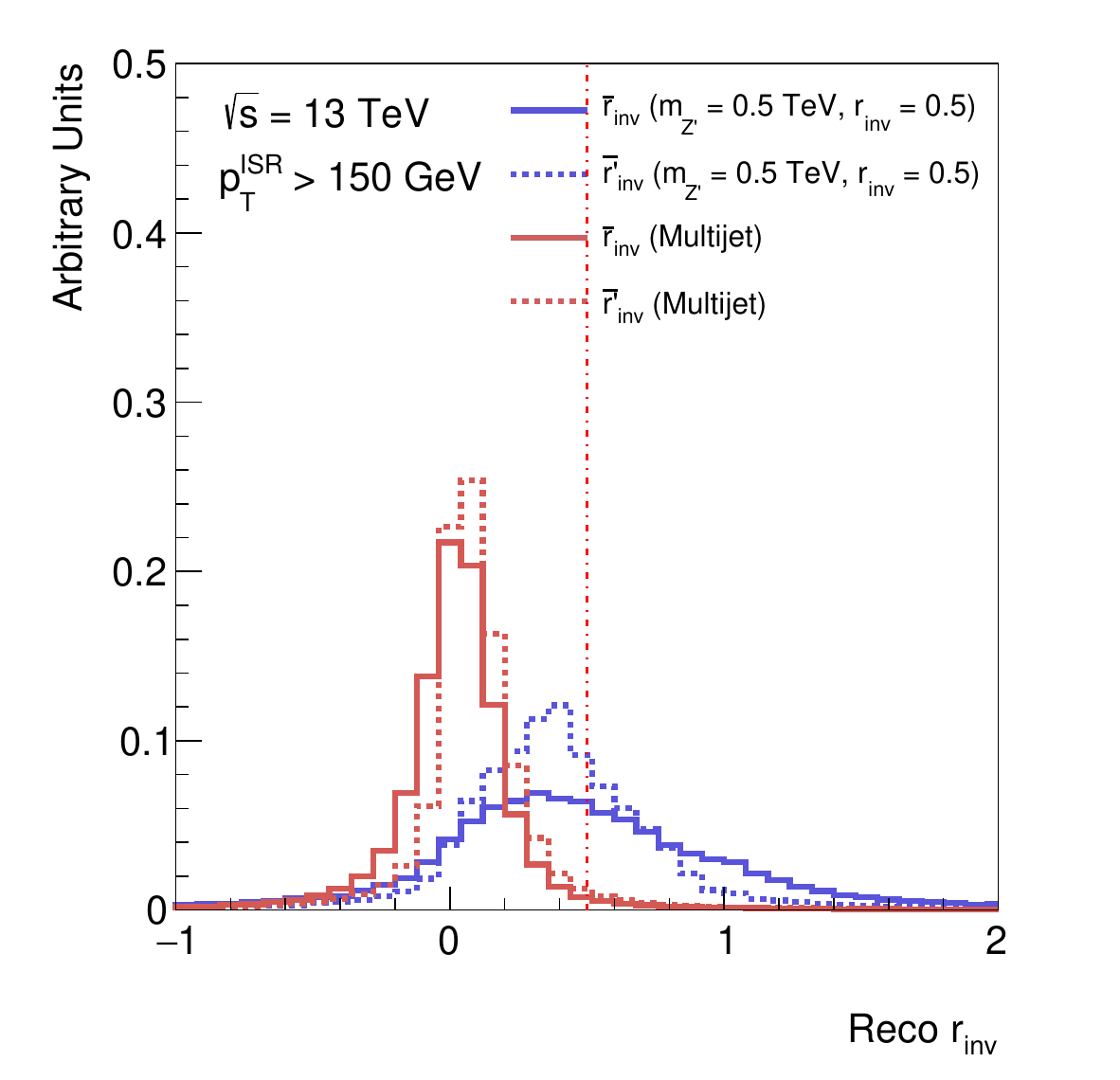}
 \includegraphics[width=0.32\columnwidth]{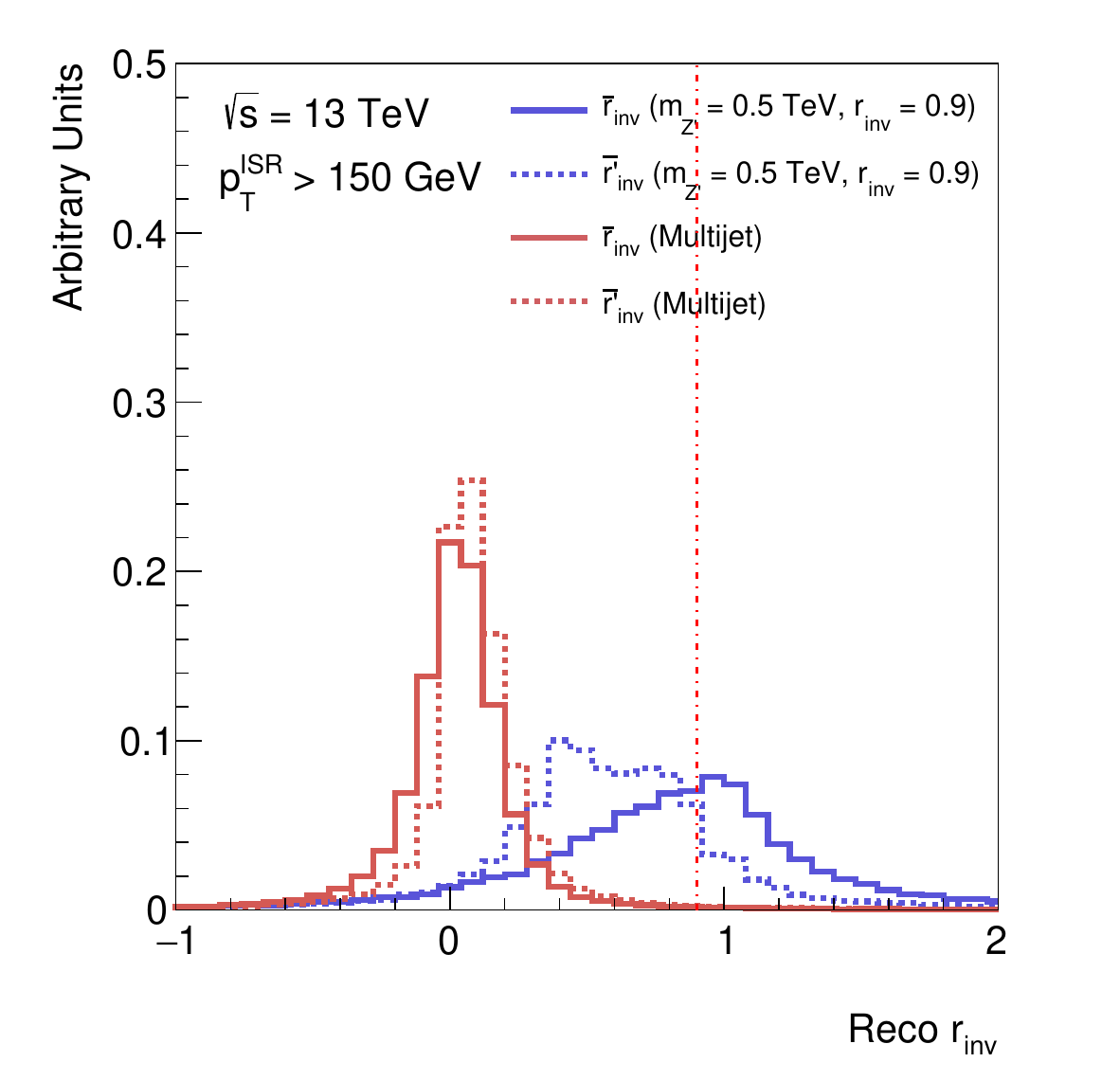}
\caption{Distributions of \rinv estimators for the multijet background and signals with $\mZPrime = 0.5\TeV$, $\PtISR = 150\GeV$, and $\rinv = 0.1$ (left), $\rinv = 0.5$ (center), and $\rinv = 0.9$ (right). The red dashed vertical line shows the theoretical \rinv value.}
\label{fig:rinv1}
\end{center}
\end{figure}

\begin{figure}[ht]
\begin{center}
 \includegraphics[width=0.32\columnwidth]{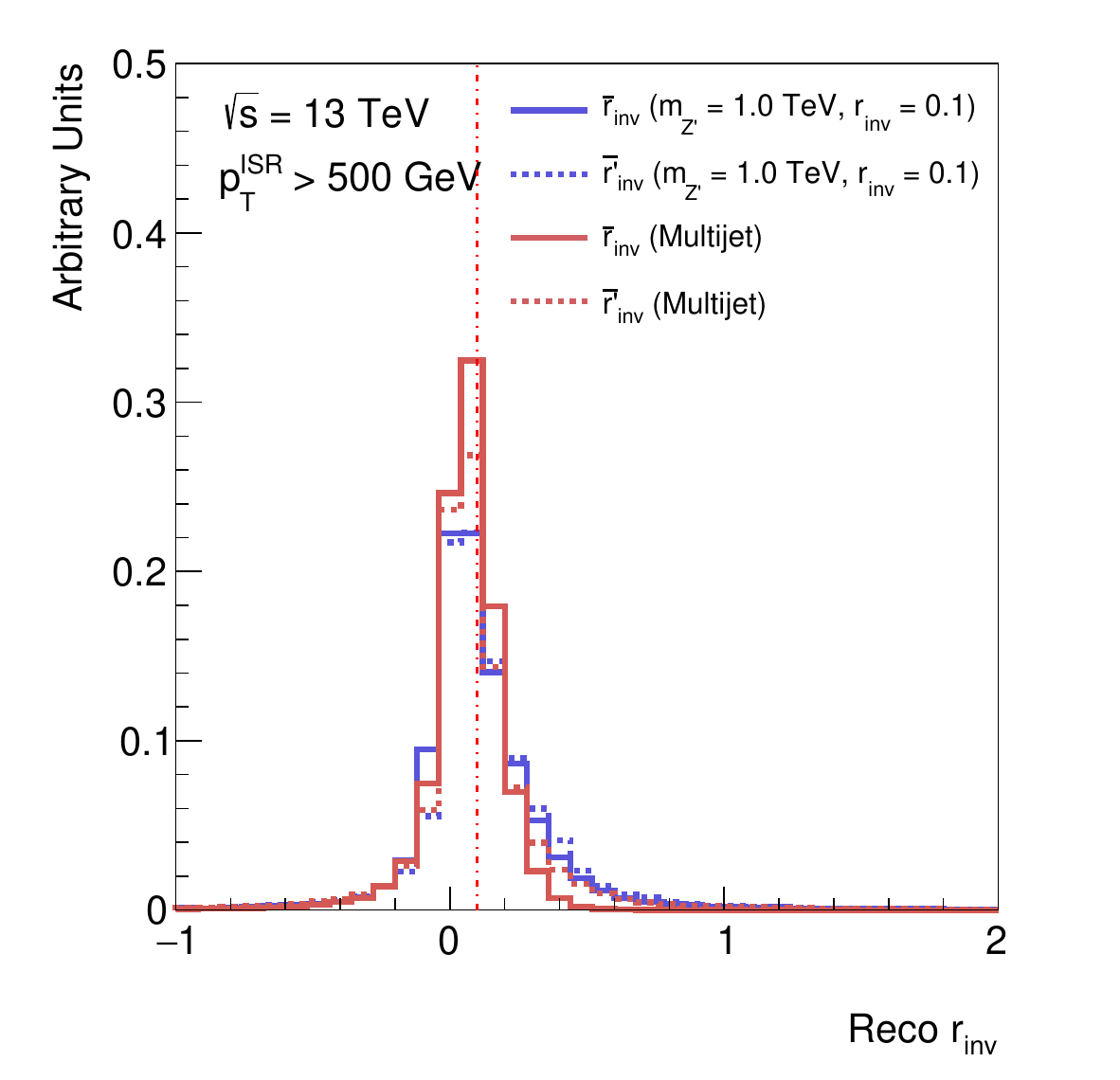}
 \includegraphics[width=0.32\columnwidth]{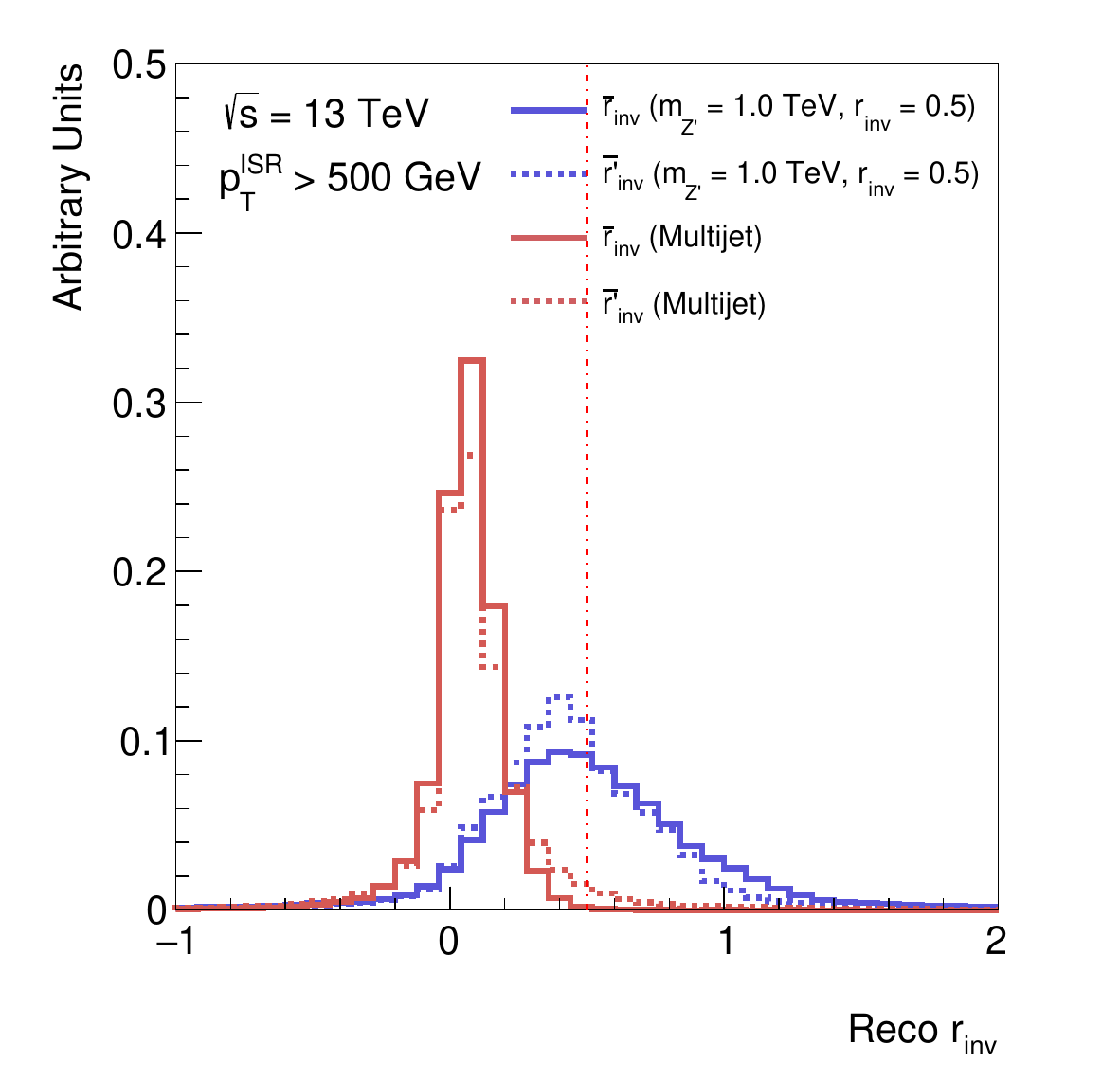}
 \includegraphics[width=0.32\columnwidth]{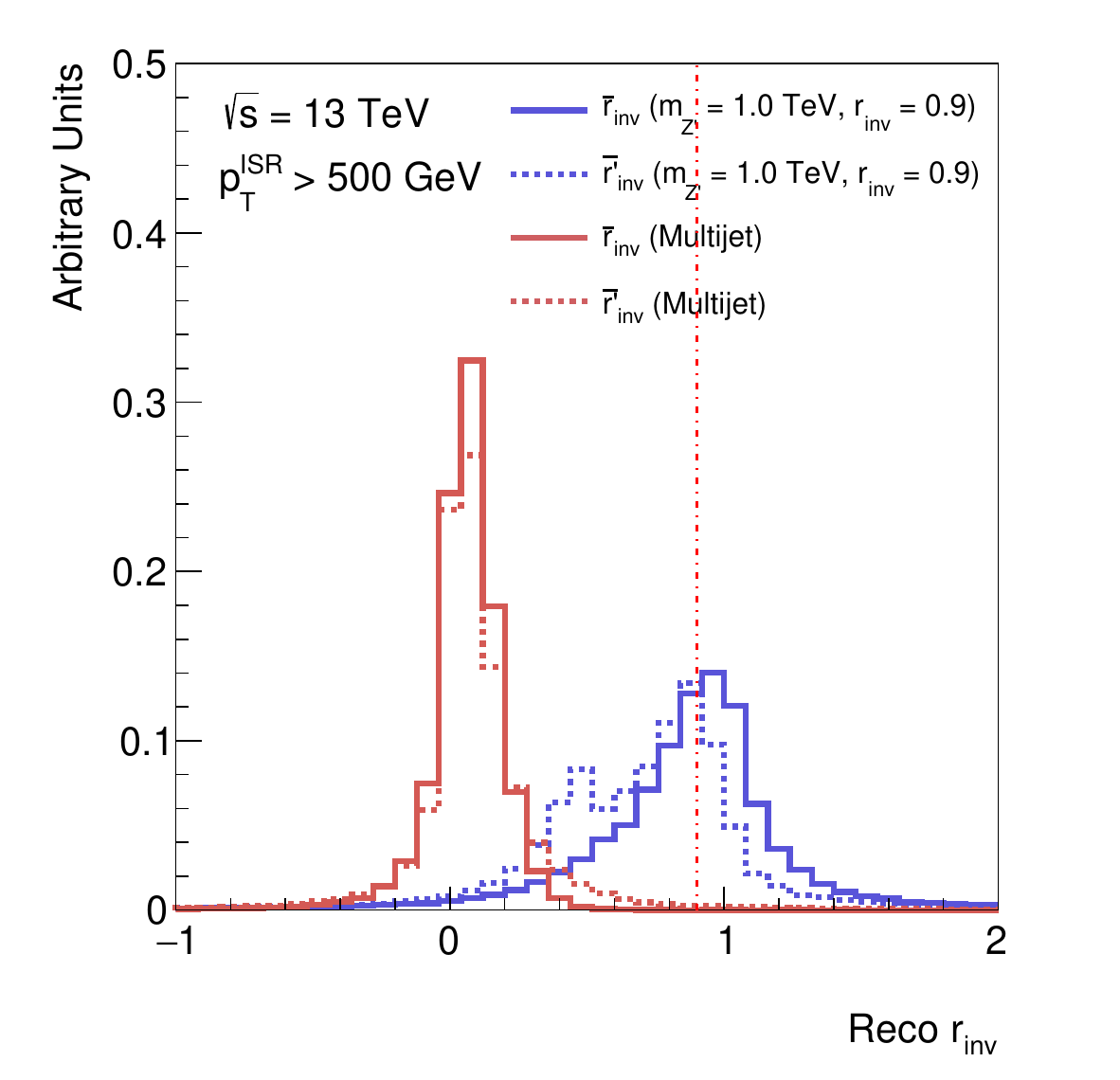}
\caption{Distributions of \rinv estimators for the multijet background and signals with $\mZPrime = 1.0\TeV$, $\PtISR = 500\GeV$, and $\rinv = 0.1$ (left), $\rinv = 0.5$ (center), and $\rinv = 0.9$ (right). The red dashed vertical line shows the theoretical \rinv value.}
\label{fig:rinv2}
\end{center}
\end{figure}

\begin{figure}[ht]
\begin{center}
 \includegraphics[width=0.49\columnwidth]{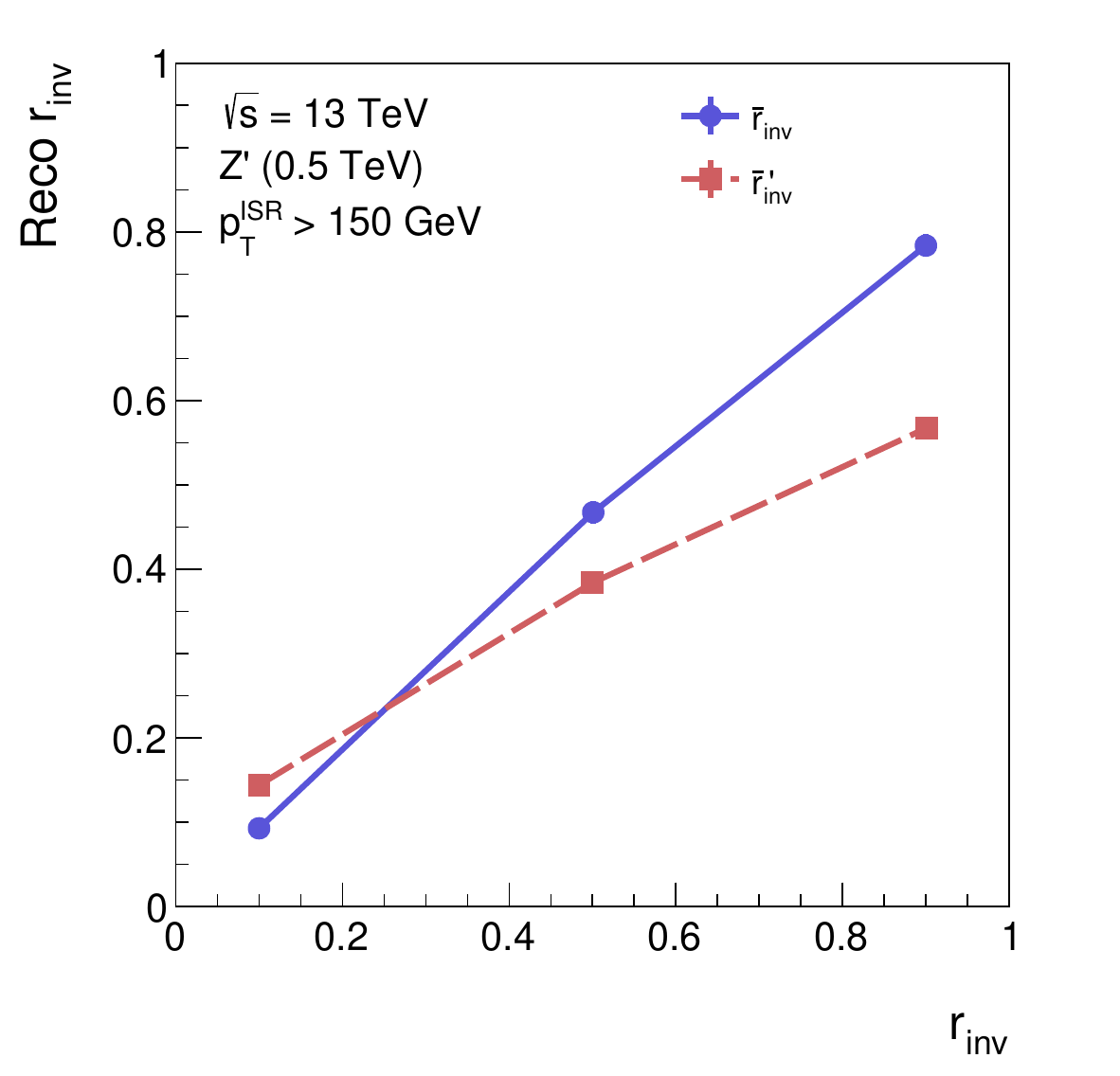}
 \includegraphics[width=0.49\columnwidth]{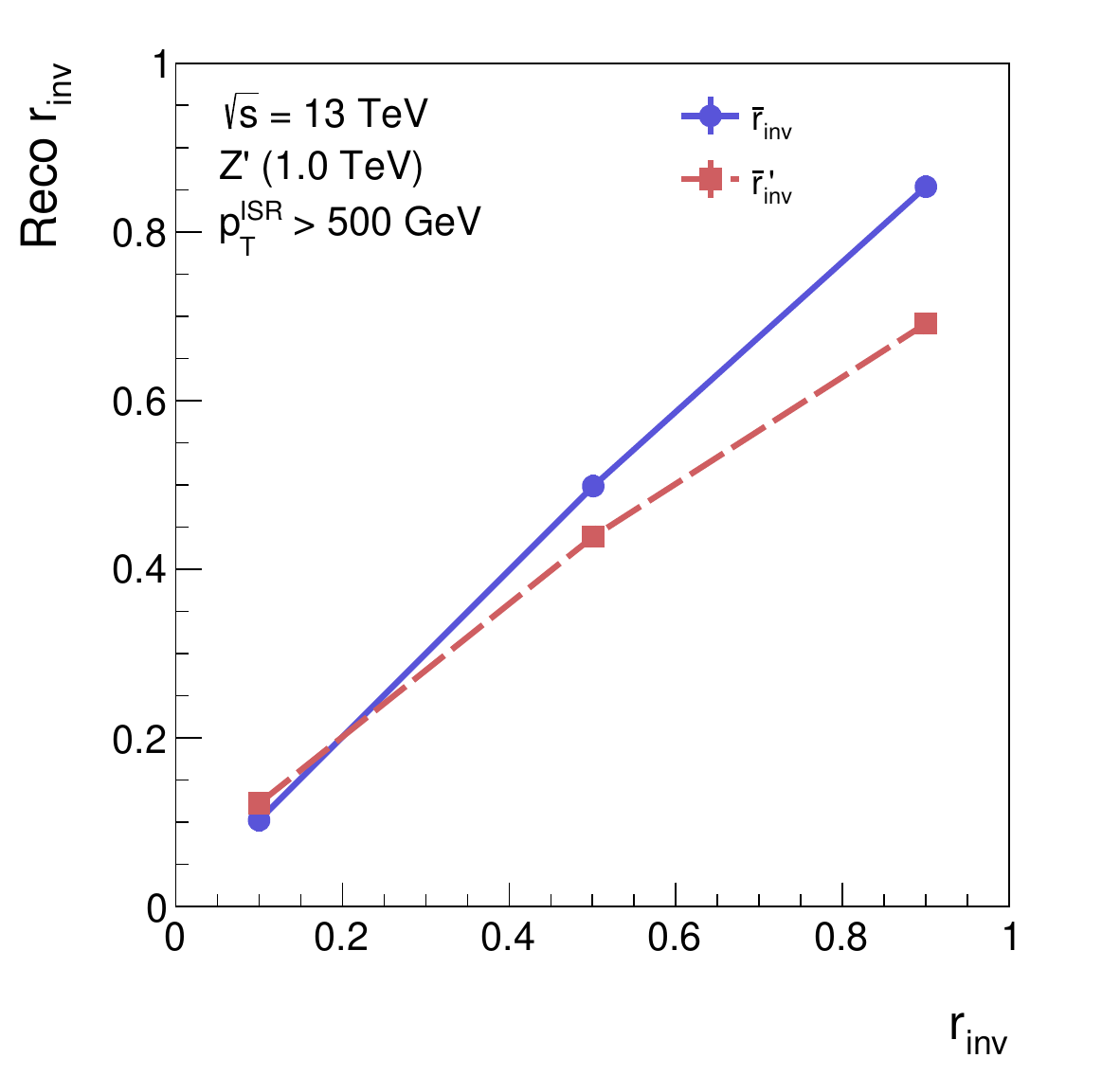}
\caption{Comparisons of the average values of the two \rinv estimators with the theoretical \rinv value, for signals with $\mZPrime = 0.5\TeV$, $\PtISR = 150\GeV$ (right) and $\mZPrime = 1.0\TeV$, $\PtISR = 500\GeV$ (left).}
\label{fig:rinv3}
\end{center}
\end{figure}

Comparisons between these two \rinv estimators are shown in Figures~\ref{fig:rinv1} and~\ref{fig:rinv2}, and their average values are compared to the theoretical \rinv value (specified during signal simulation) in Fig.~\ref{fig:rinv3}.
As \rinv increases, both variables become more powerful.
However, as studied later in
Section~\ref{sec:SRs}, \rinvavemaos is not as powerful as \rinvave.
There is a clear relationship with the theoretical \rinv value, which could be used for calibration when measuring this quantity.
Both \rinvave and \rT can be used to enhance
sensitivity, as will be investigated in Section~\ref{sec:analysis}.

\subsection{Combining Observables}
\label{subsec:observable}

\begin{figure}[ht]
\begin{center}
 \includegraphics[width=0.8\columnwidth]{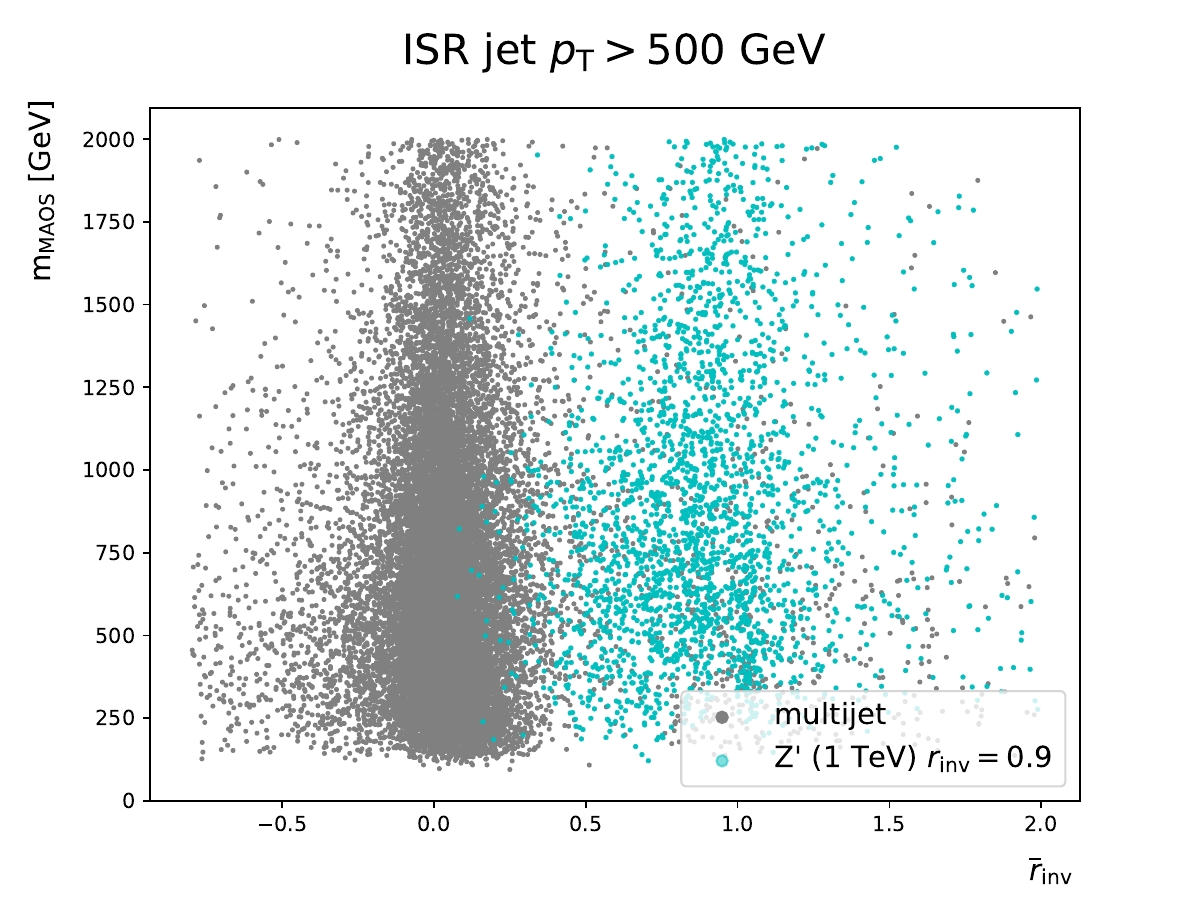}
\caption{Scatterplot of \maos and \rinvave for multijet (grey, 20000 events) and \ZPrime (cyan, 10000 events). The \ZPrime sample has a mass of 1 TeV and an \rinv parameter of 0.9. The ISR jet is required to have $\Pt > 500\GeV$. }
\label{fig:maosjj_rinv}
\end{center}
\end{figure}

As both \rinvave and \maos have excellent discriminating power, combining the
two may further enhance the search power.  Figure~\ref{fig:maosjj_rinv} shows
the \maos-\rinvave 2D distribution for both the multijet background and the
signal. The signal events are clustered in the region determined by the \ZPrime
mass and \rinv. This opens up new ways to perform this search, as discussed in
Section~\ref{sec:results}. 

Similar to \mT, \maos and \rinvave benefit from the energetic ISR object. The
performance of \rinvave drops dramatically as the \Pt of the ISR object
decreases. This is because the decomposition is influenced by the angular
separation. This can be clearly appreciated from an extreme example where
\ptvecmiss is perfectly aligned with one of the jets. In this situation, the
decomposition is ambiguous unless the system is boosted.

This event-level topology can be generalized as a two-body decay with non-interacting final state particles in each decay leg.
This behavior is expected in various scenarios such as leptonic
$H\rightarrow\tau^{+}\tau^{-}$ and $H\rightarrow W^{+}W^{-}$ decays. Methods
proposed to improve reconstruction of the Higgs mass in semivisible decays can be applied to SVJs as well; in fact, this was the original genesis of the MAOS method.
The performance improvement in collinear events has also been discussed in the context of
$H\rightarrow\tau^{+}\tau^{-}$~\cite{htautau_colinear}. SVJ searches will likely
benefit further from more interactions between various communities.  

\section{Analysis Strategy}
\label{sec:analysis}

In this section, we demonstrate the feasibility of constructing a search for a
resonant \ZPrime decaying to dark quarks in the ISR production channels, where
the two jets originating from the dark quarks are reconstructed as two separated
small-radius ($R = 0.5$) jets. When the \ZPrime mass is very small, using large-radius
jets containing the entire boosted \ZPrime system is a better choice, but this
is a different strategy not pursued here.

\subsection{Jet Matching Method}\label{sec:jetmatch}

\begin{figure}[ht]
\begin{center}
 \includegraphics[width=0.4\columnwidth]{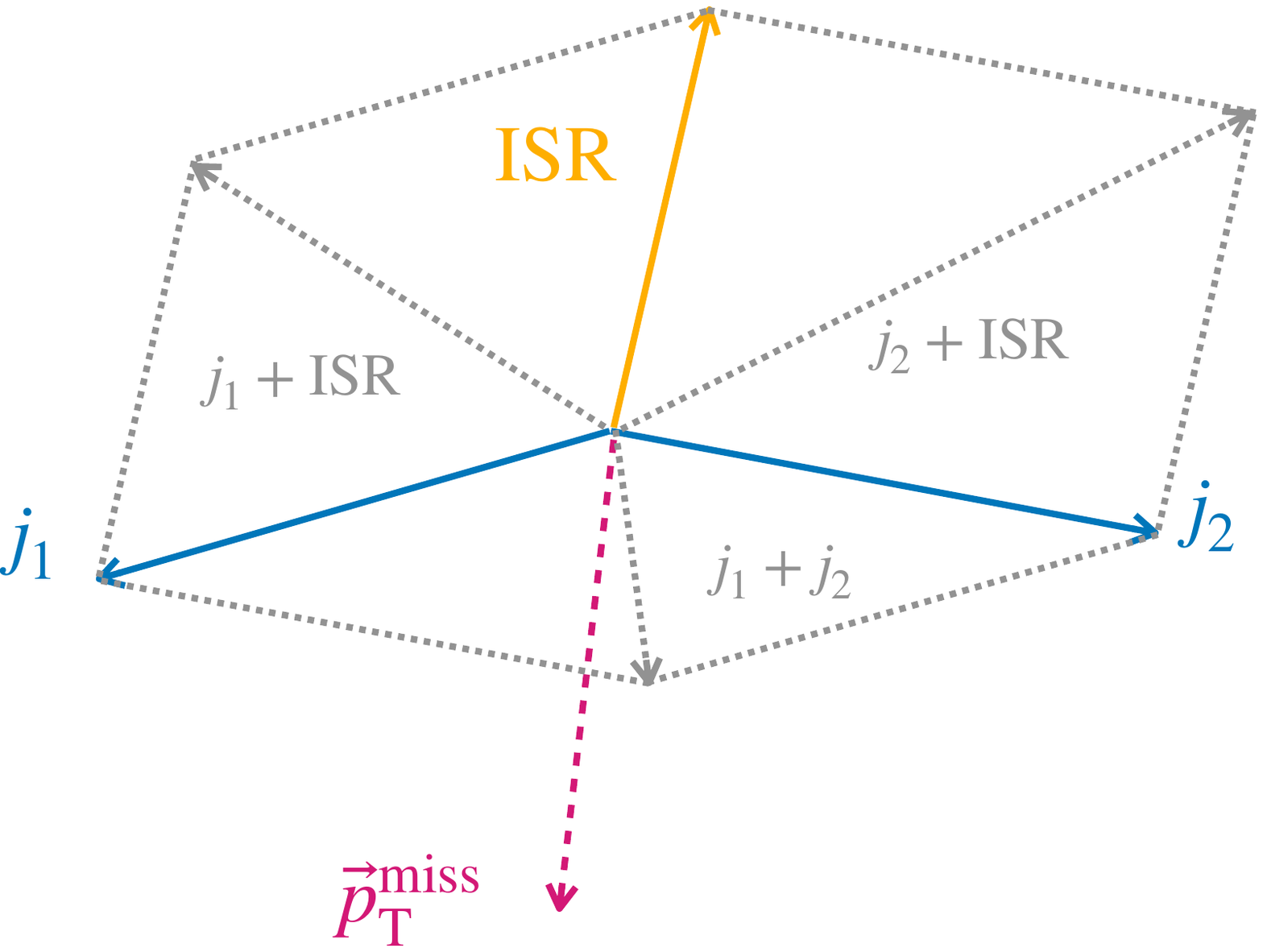}
\caption{Transverse view of the two jets, ISR object and \ptvecmiss. }
\label{fig:jet_match}
\end{center}
\end{figure}

Experimentally, the two jets from \ZPrime have to be distinguished from the ISR
object. In the photon ISR case, simply selecting the leading two jets is
sufficient. However, as reported in other resonance searches using an ISR jet,
the jet matching is a challenging task. In particular, once the ISR jet and the
two jets from the heavy particles have similar \Pt, the three jets are well
balanced on the transverse plane, forming a triangular topology. It is
extremely difficult to disentangle these three jets using basic kinematic
properties only. In SVJ models, thanks to the appreciable \ETmiss, we are able
to identify the two signal jets using \dPhi between the jet pair and
\ptvecmiss, as illustrated in Figure~\ref{fig:jet_match}. Since the two target
jets and \ETmiss come from the \ZPrime, when the \ZPrime system is boosted,
these three objects are collimated, resulting in smaller separations in \dPhi.
While identifying the correct jets in signal events is critical, ensuring
minimal background sculpting is equally important. It is found that minimizing
the variable $(\Delta\phi_{j_{i},j_{k}} + \Delta\phi_{j_{i}j_{k},
\ptvecmiss})/2$ gives a smooth background and acceptable matching efficiency.
$\Delta\phi_{j_{i},j_{k}}$ is the \dPhi between two jets under consideration,
and $\Delta\phi_{j_{i}j_{k}, \ptvecmiss}$ is the \dPhi between the
corresponding dijet system and \ptvecmiss. Before the minimization, the three
jets are required to be well-separated by requiring \dPhi between each pair to
be larger than 0.8. The two jets from the \ZPrime, $j_{1}$ and $j_{2}$, are
identified by matching the dark quarks to reconstructed jets by requiring $\Delta R <
0.4$ in order to evaluate the efficiency.

\begin{figure}[ht]
\begin{center}
 \includegraphics[width=0.4\columnwidth]{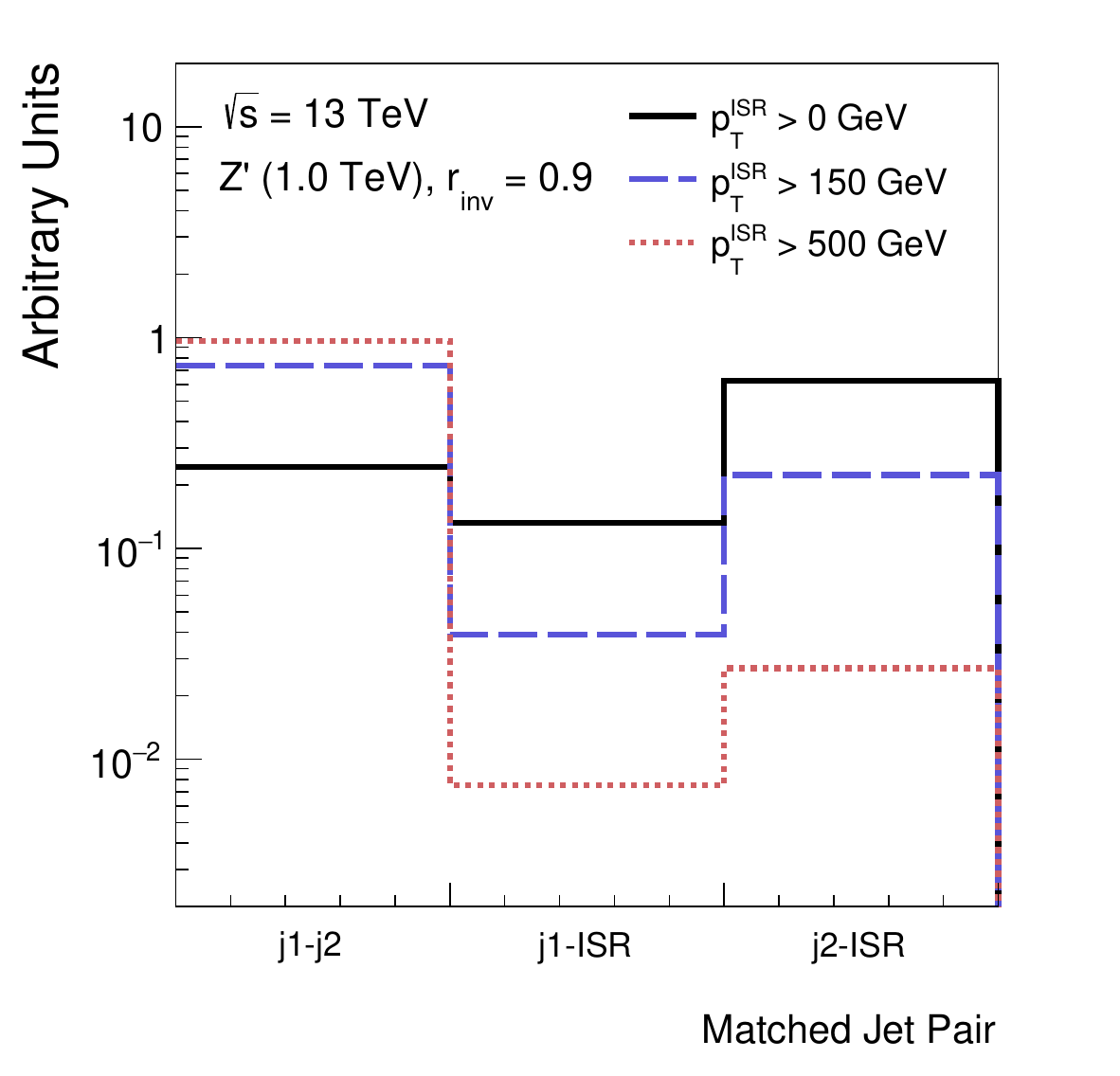}
 \includegraphics[width=0.4\columnwidth]{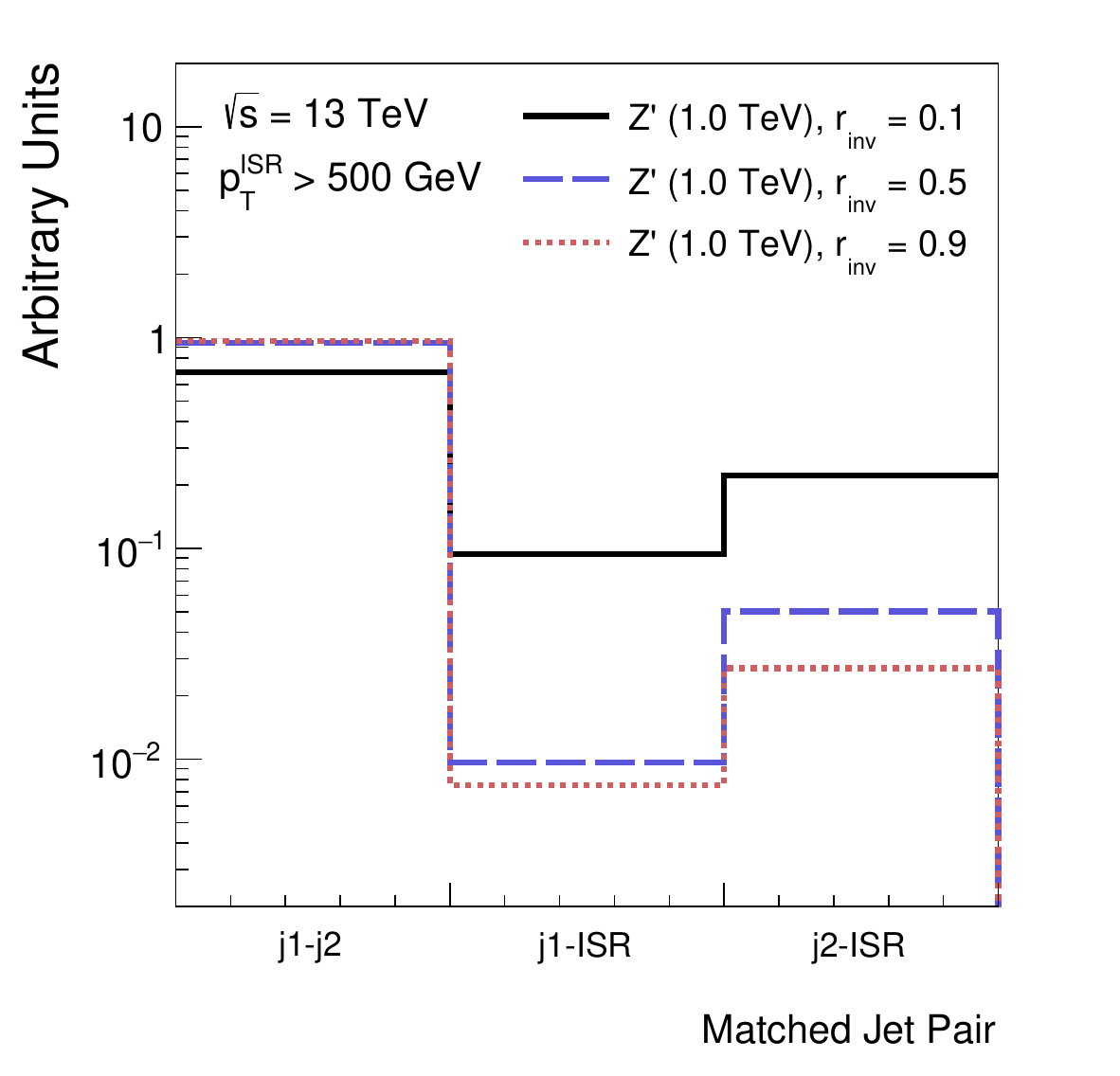}
\caption{Fractions of various matching configurations for one \ZPrime sample with different ISR \Pt selections (left), and for three \ZPrime samples with different \rinv parameters when the same ISR \Pt selection is applied (right). The first bin on the $x$-axis corresponds to the case where the matching procedure correctly identifies the two jets from the \ZPrime.}
\label{fig:jetmatching_isr}
\end{center}
\end{figure}

Figure~\ref{fig:jetmatching_isr} shows the probabilities of different jet pairs
to be selected via the above procedure. As the \Pt of the ISR jet increases,
the matching efficiency of this method improves. For 500 GeV, the method is
over 90\% efficient for large \rinv values.            

\subsection{Basic Selections}

The typical threshold for a single isolated photon trigger is
about 140 \GeV, while the corresponding threshold for a single jet trigger
is about 420 GeV~\cite{atlasdijetisr}. Therefore, the events are categorized
into two channels: one requires an energetic photon with $\Pt > 150\GeV$, and
the other requires a jet with \Pt larger than 500 \GeV. The thresholds are
slightly increased so that the analysis considers events above the trigger turn-on
region. 

Furthermore, the events are required to contain at least two (three) $R=0.5$ jets clustered with the anti-\kt algorithm~\cite{Cacciari:2008gp,Cacciari:2011ma},
passing $\Pt > 25\GeV$, with $\Eta < 4.5$, for the photon (jet) ISR channel.
(The choice of jet clustering radius is discussed further in Appendix~\ref{app:radius}.)
In the photon channel, the photon must not overlap with the two leading jets ($\dR
> 0.4$). The two signal jets are identified as the leading two jets in the photon
ISR channel, and the two signal jets are identified via the method mentioned
above in the jet ISR channel. The $|\drap|$ variable has been used in numerous
dijet resonance searches in the past, so a similar strategy is adopted here.
The threshold is chosen to be 0.8, based on a recently published ATLAS
result~\cite{atlasdijetisr} that considers similar kinematic regions. 

\subsection{Signal Region Categorization}
\label{sec:SRs}

\begin{figure}[ht]
\begin{center}
 \includegraphics[width=0.49\columnwidth]{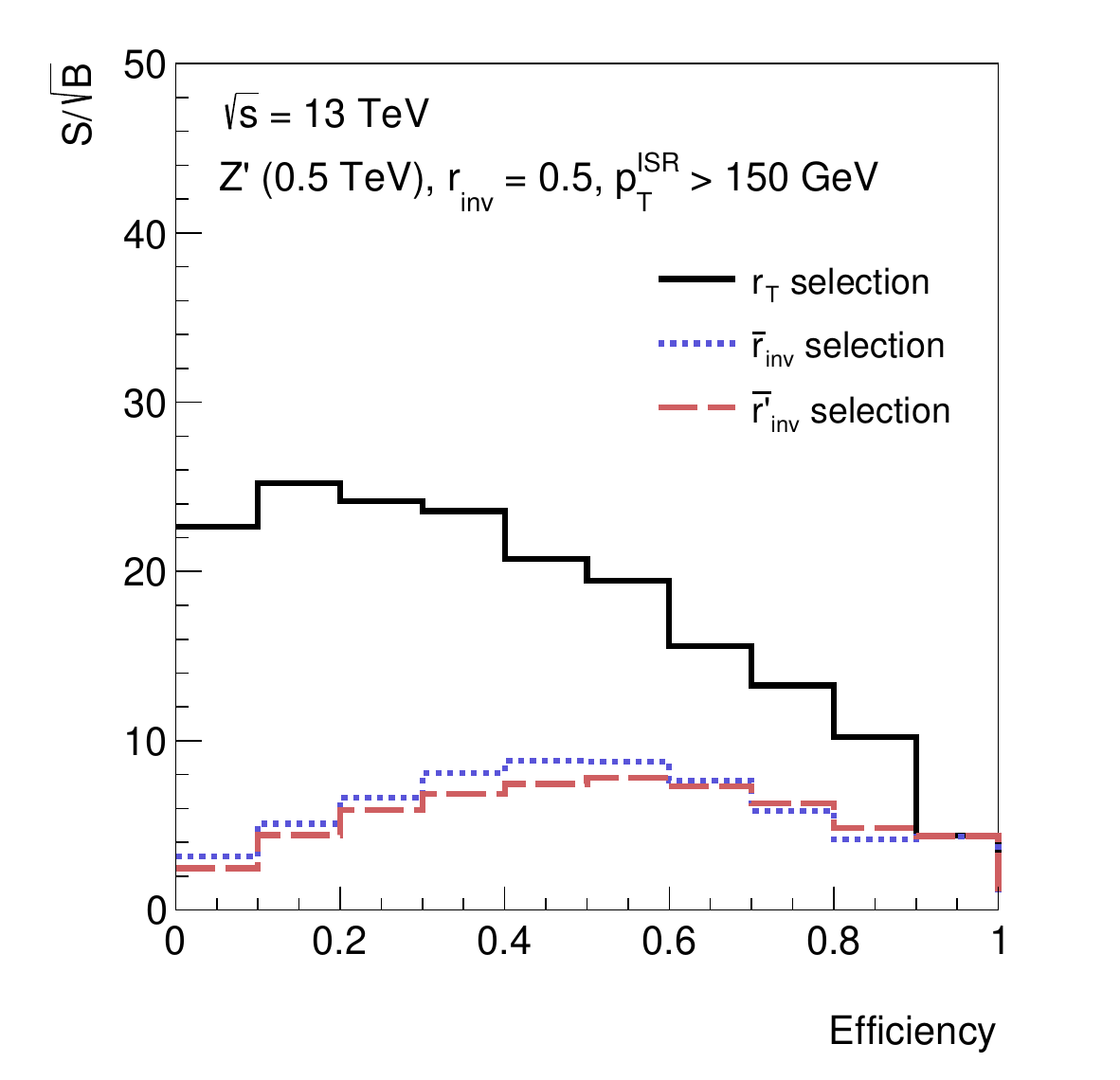}
 \includegraphics[width=0.49\columnwidth]{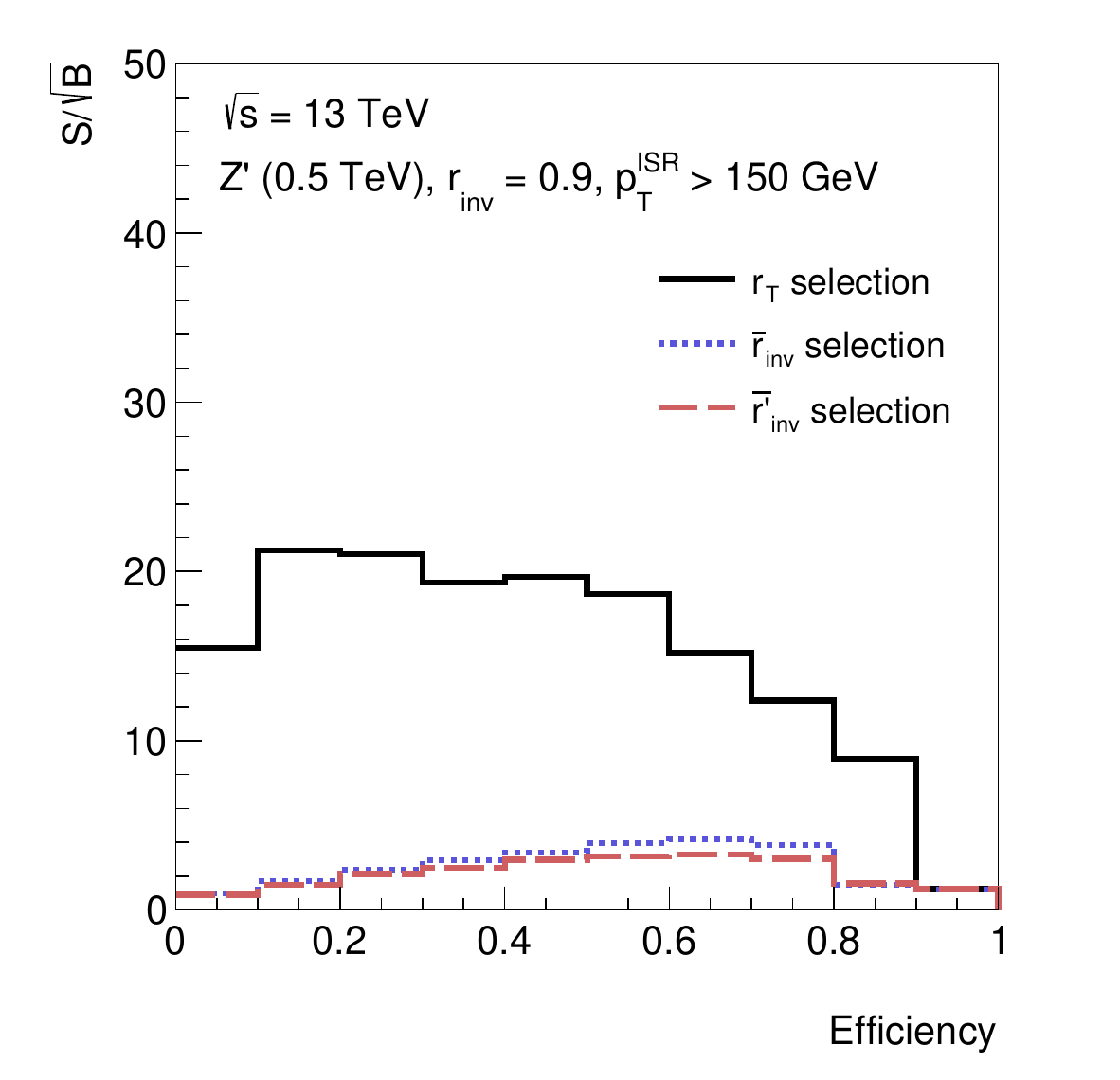}
 \includegraphics[width=0.49\columnwidth]{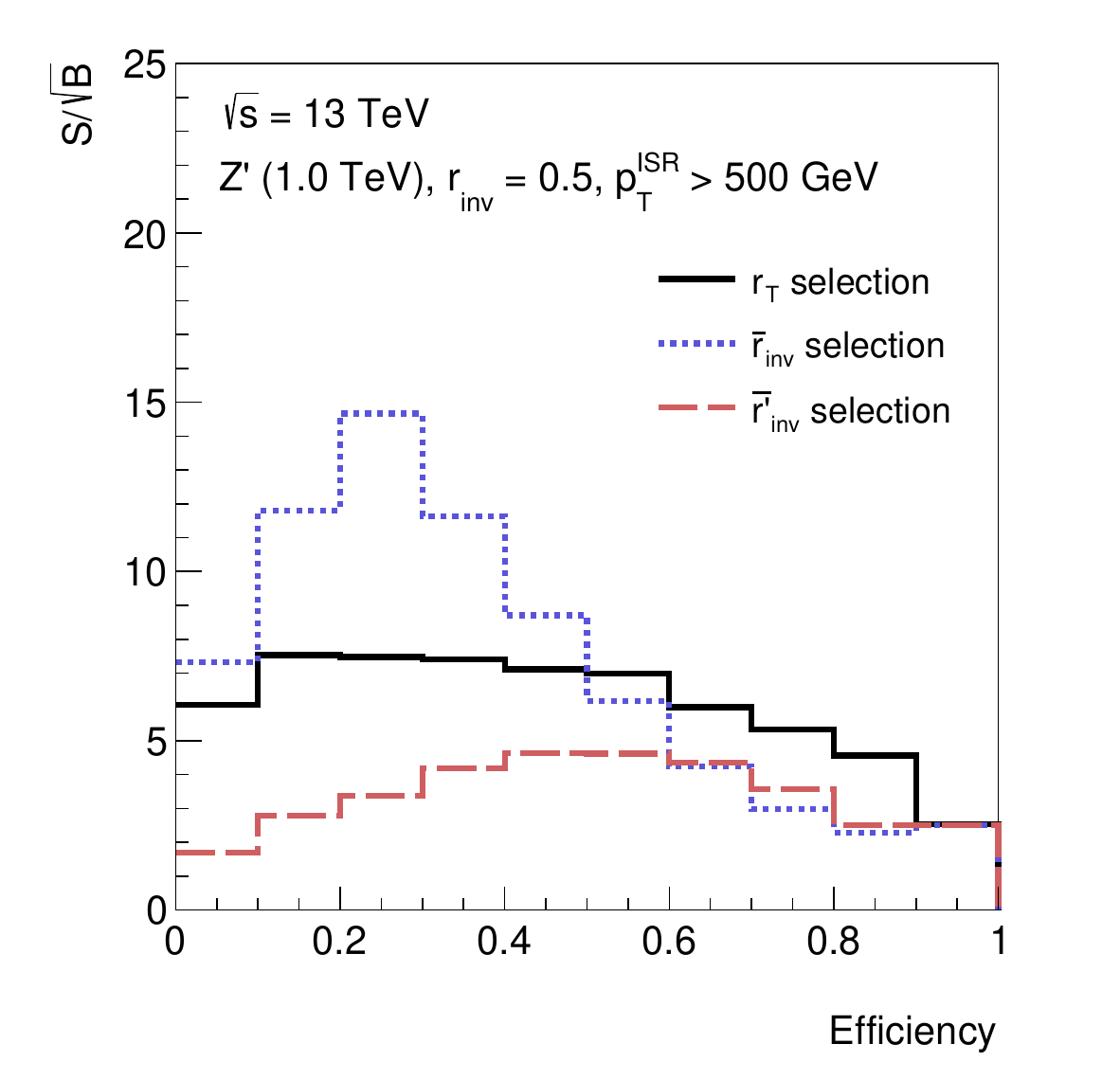}
 \includegraphics[width=0.49\columnwidth]{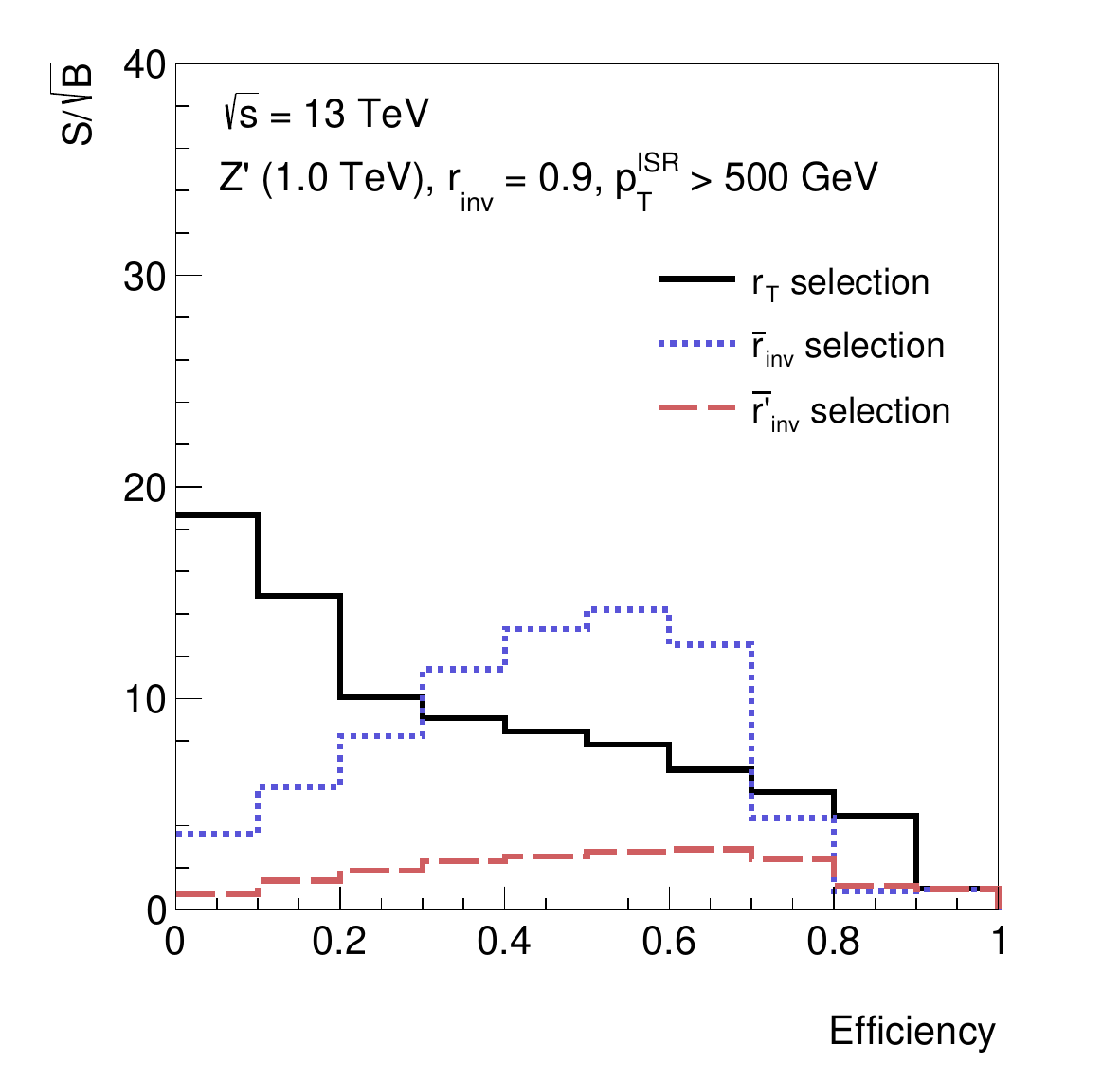}
\caption{Top: Significance in the photon ISR channel as a function of the signal efficiency for selections on \rT, \rinvave, or \rinvavemaos. The 0.5 TeV \ZPrime samples with an \rinv parameter of 0.5 (left) and 0.9 (right) are considered. Bottom: Significance in the jet ISR channel as a function of the signal efficiency for selections on \rT, \rinvave, or \rinvavemaos. The 1.0 TeV \ZPrime samples with an \rinv parameter of 0.5 (left) and 0.9 (right) are considered.}
\label{fig:sig_scan}
\end{center}
\end{figure}

As discussed in Section~\ref{sec:kine}, both \rinvave and \rinvavemaos have
excellent discriminating power for signal with large \rinv. Here, we also compare these variables to \rT, since they all involve the \ETmiss, though in different ways.
The approximate significance $S/\sqrt{B}$, based on the signal and background
yields $S$ and $B$, is evaluated for a range of values in each variable.
The signal yields $S$ and $B$ are evaluated for $\sfrac{1}{2}\mZPrime < \maos < \sfrac{3}{2}\mZPrime$, in order to consider the most pertinent background events in the region near the signal mass peak, which provides the most sensitivity.
Figure~\ref{fig:sig_scan}
shows the results for the 0.5 (1.0) TeV \ZPrime in the photon (jet) ISR
channels. In the photon ISR channel where the boost is modest, \rT brings
better sensitivity, while in the jet ISR channel, \rinvave performs the best
except in the very high signal efficiency region. This is expected as \rinvave
is more impacted by the boost. It is
also clear that \rinvavemaos is not as efficient as \rinvave in the jet ISR channel. 

\begin{figure}[ht]
\begin{center}
 \includegraphics[width=0.45\columnwidth]{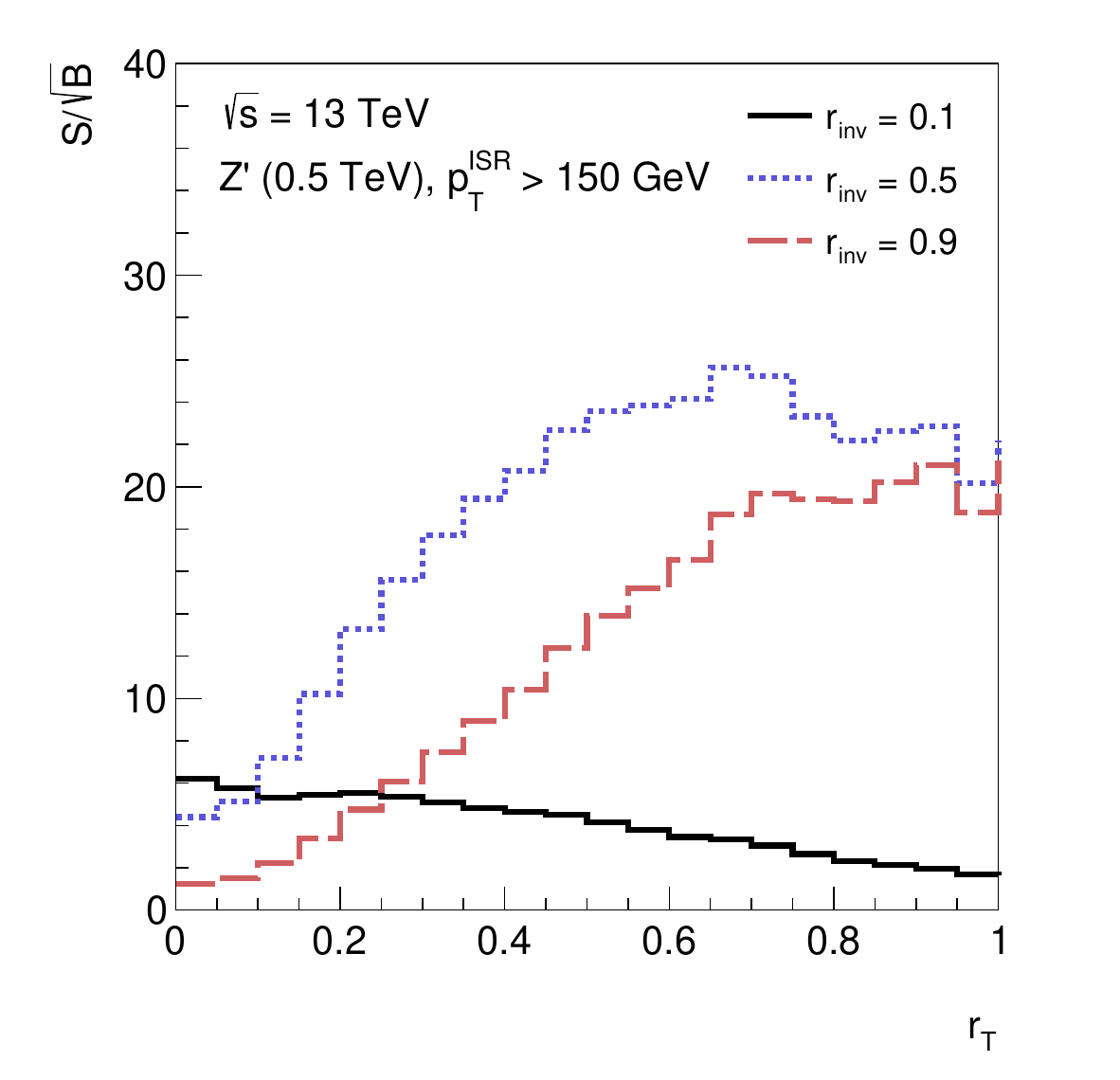}
 \includegraphics[width=0.45\columnwidth]{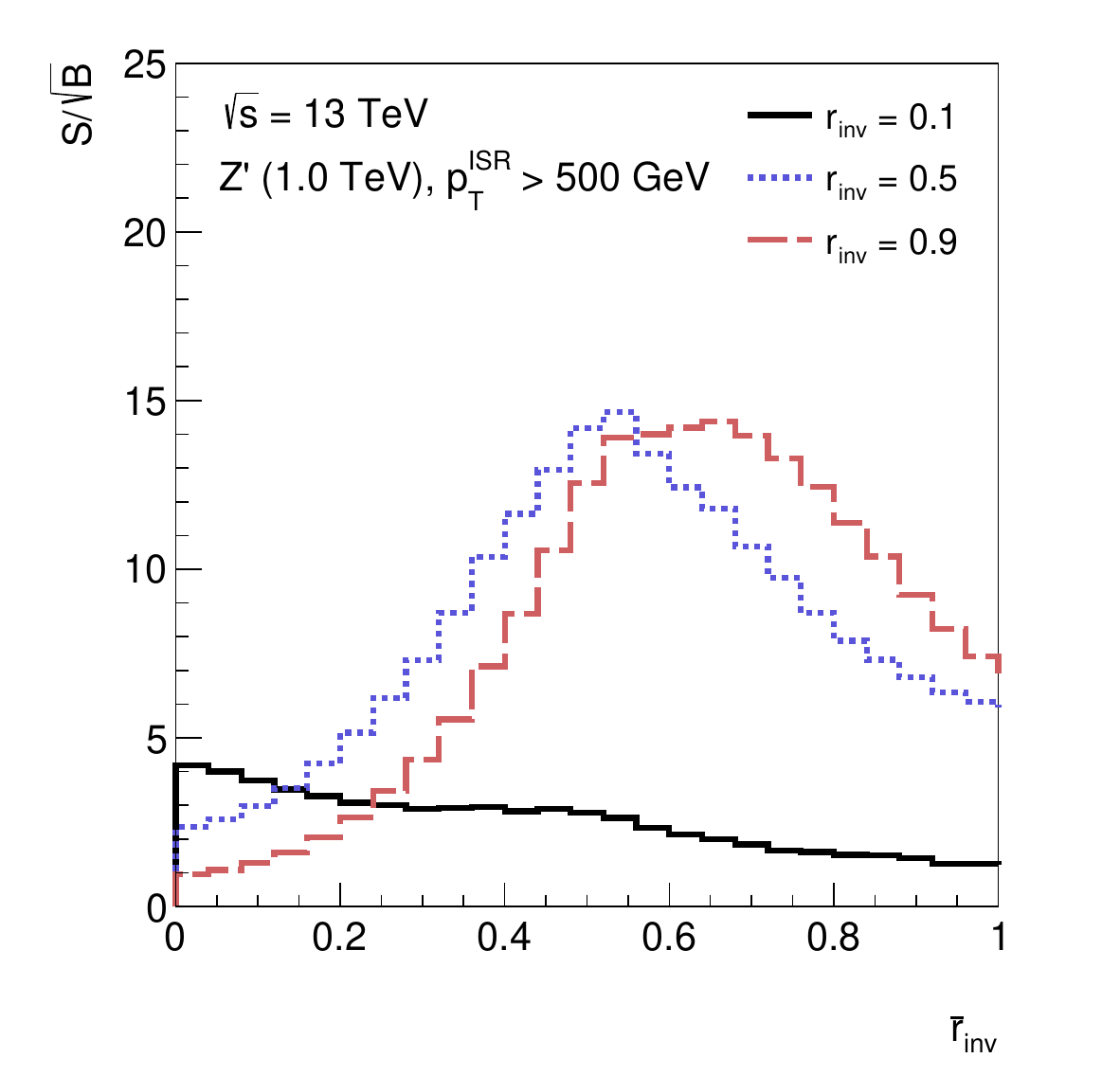}
\caption{Significance as a function of \rT (left) and \rinvave (right) for the photon ISR channel and jet ISR channel, respectively. In the photon (jet) ISR channel, the 0.5 (1.0) TeV \ZPrime sample is studied with three \rinv values.}
\label{fig:optimize}
\end{center}
\end{figure}

The photon and jet ISR channels are optimized separately using \rT and
\rinvave, respectively, as presented in Figure~\ref{fig:optimize}. We apply a
cut at 0.4 on \rT (\rinvave) in the photon (jet) ISR channels to estimate the
final sensitivity. In principle, tighter selections on \rT (\rinvave) can be
applied, but the reduced background efficiency leads to unacceptable statistical uncertainty in the generated background sample.
Table~\ref{tab:selections} summarizes all the selections applied to
achieve the final results.  The minimum \maos selections for each channel are
necessary to ensure a smoothly falling background spectrum.  For lower \maos
values, the multijet background has a peak, arising from Sudakov
suppression~\cite{Dasgupta:2013ihk} and any other sculpting from kinematic
selections.

\begin{table}[htbp]
  \begin{center}
  \caption{The complete set of signal region selections applied to both the photon ISR and jet ISR channels.}
\makebox[0pt]{
\renewcommand{\arraystretch}{1.2}
\begin{tabular}{|c|c|}
\hline
Photon ISR & Jet ISR \\
\hline
${\geq}1$ photon with $\Pt > 150\GeV$ & ${\geq}1$ Jet with $\Pt > 500\GeV$ \\
\hline
${\geq}2$ jets with $\Pt > 25\GeV$, $|\eta| < 4.5$ & ${\geq}3$ jets with $\Pt > 25\GeV$, $|\eta| < 4.5$ \\
\hline
Jet-photon overlap removal & - \\
\hline
Leading two jets as signal jets & Identify two signal jets via jet matching \\
\hline
\multicolumn{2}{|c|}{$|y*| < 0.8$} \\
\hline
$\rT > 0.4$ & $\rinvave > 0.4$ \\
\hline
$\maos > 200\GeV$ & $\maos > 400\GeV$ \\
\hline
\end{tabular}
}
  \label{tab:selections}
 \end{center}
\end{table}

\section{Sensitivity Estimation}
\label{sec:results}

\begin{figure}[ht]
\begin{center}
 \includegraphics[width=0.45\columnwidth]{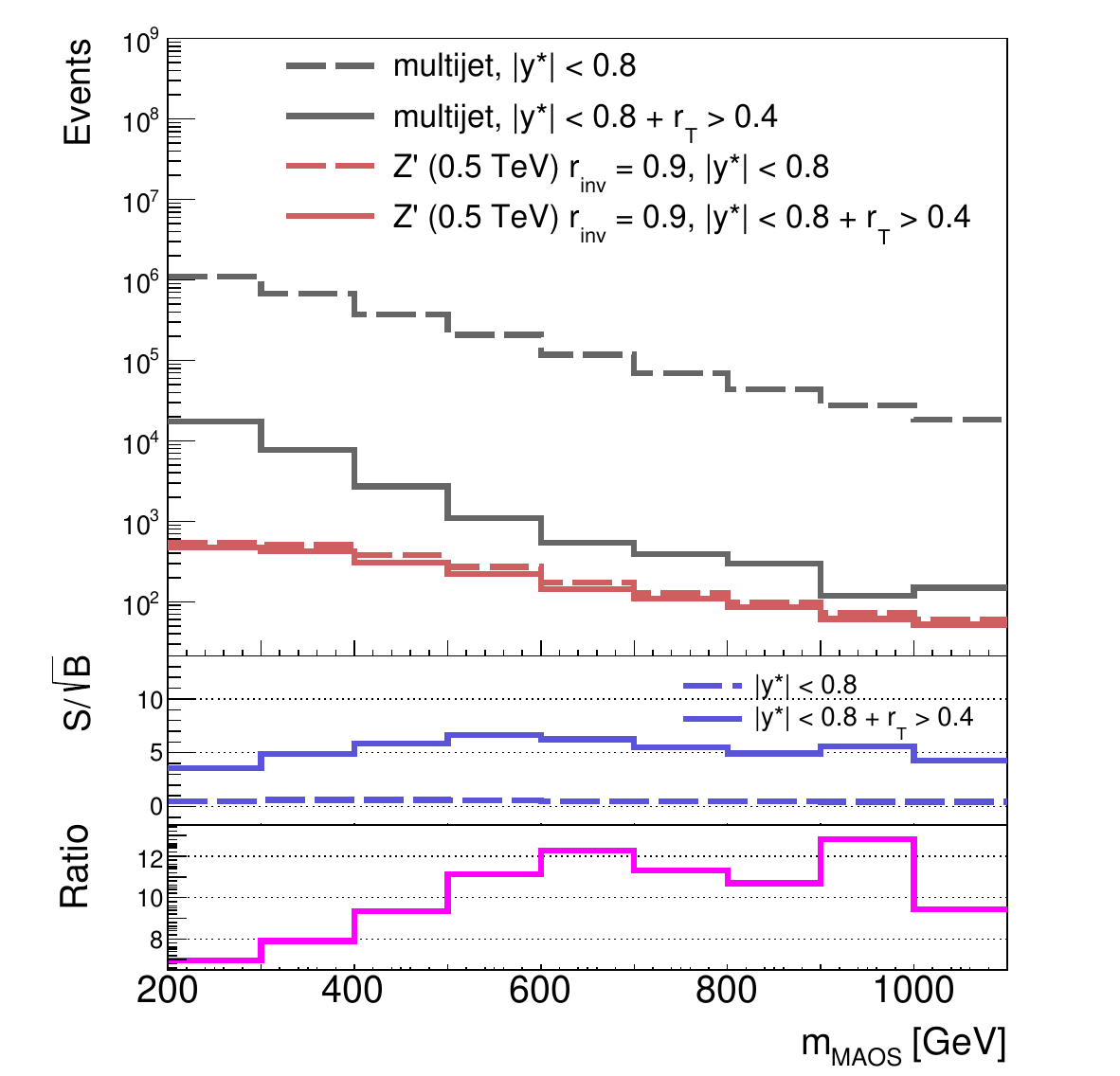}
 \includegraphics[width=0.45\columnwidth]{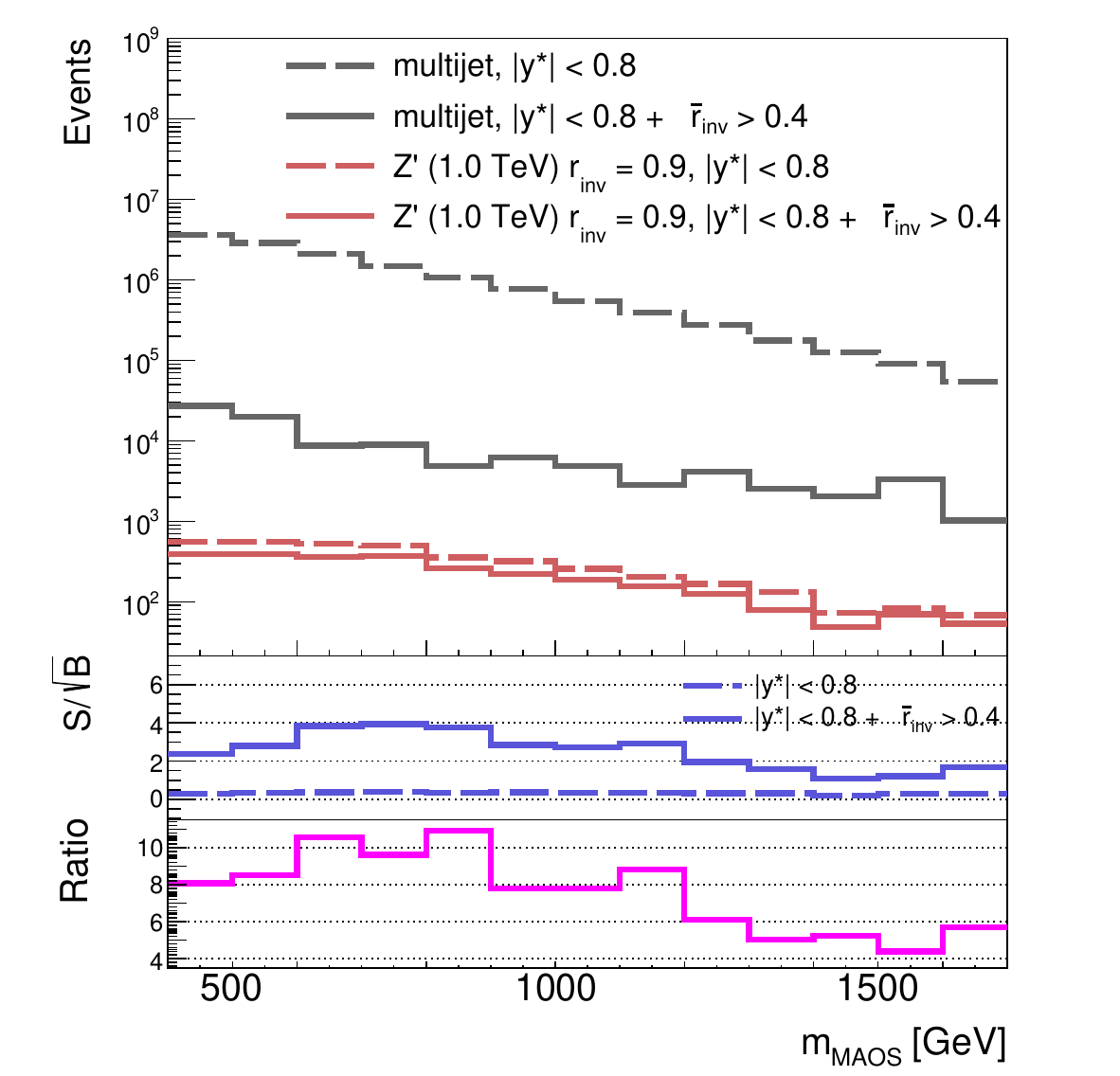}
\caption{Signal and background distributions in the photon ISR (left) and jet ISR (right) signal regions. The significance is compared between the signal region selections with and without the \rT (left) or the \rinvave (right) requirements. The bottom panel presents the ratio between the significance curves in the middle panel.}
\label{fig:results_summary}
\end{center}
\end{figure}

The traditional background modeling method is to perform a fit to data, as in the
CMS analysis~\cite{cmssvj}. Here we present the ideal sensitivities given
perfect background modeling. Since we have identified \maos as a better
observable, the sensitivities are presented as functions of \maos.
We apply the selections in Table~\ref{tab:selections}, optimized separately for each channel as discussed in Section~\ref{sec:SRs}.
As seen in Figure~\ref{fig:results_summary}, thanks to the \rinvave (\rT) selection, the
significance is increased by up to a factor of ten across the bulk of the \maos
distribution for 1.0 (0.5) TeV \ZPrime with \rinv = 0.9 in the jet (photon) ISR
channel. 

\begin{figure}[ht]
\begin{center}
 \includegraphics[width=0.32\columnwidth]{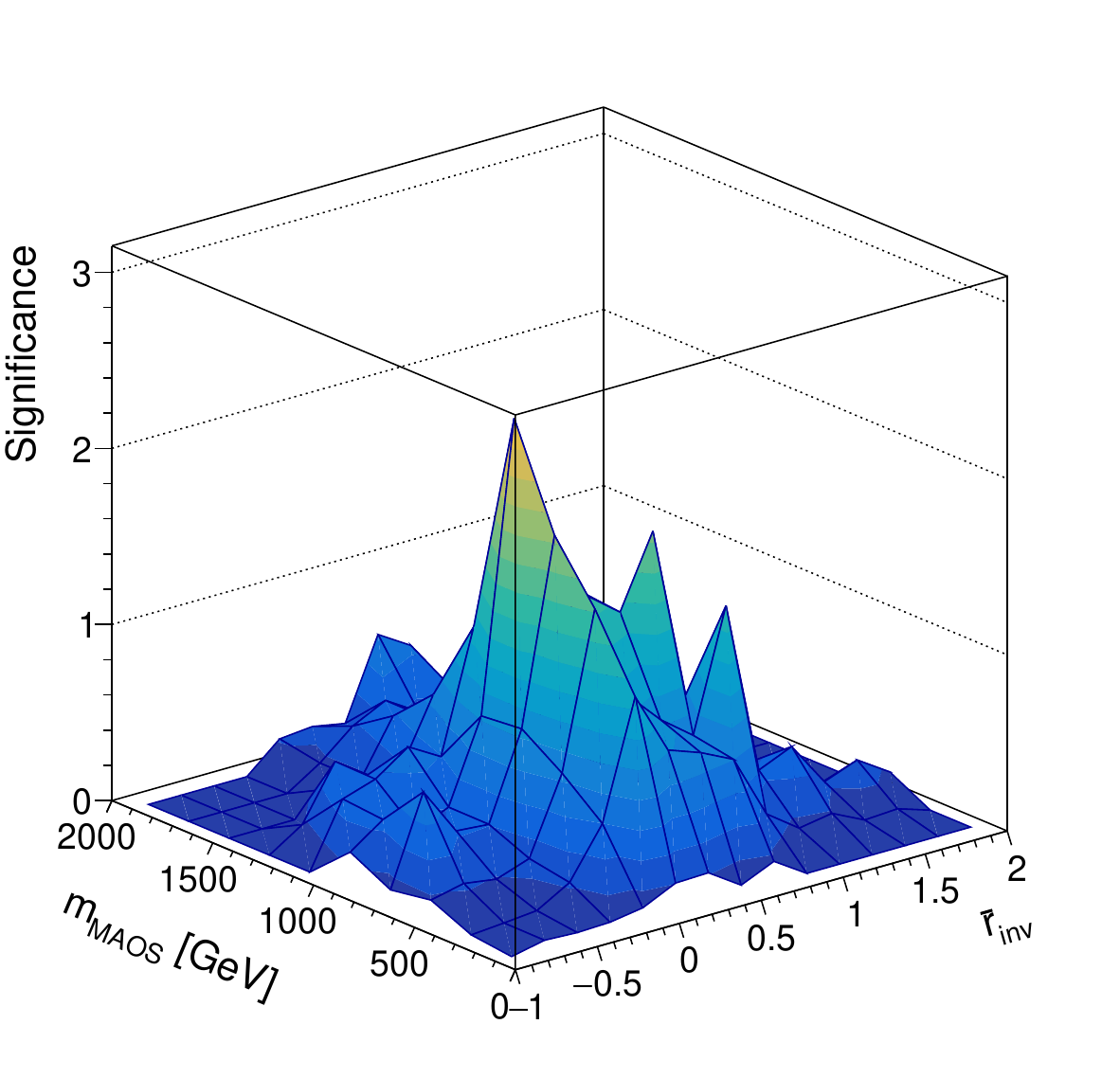}
 \includegraphics[width=0.32\columnwidth]{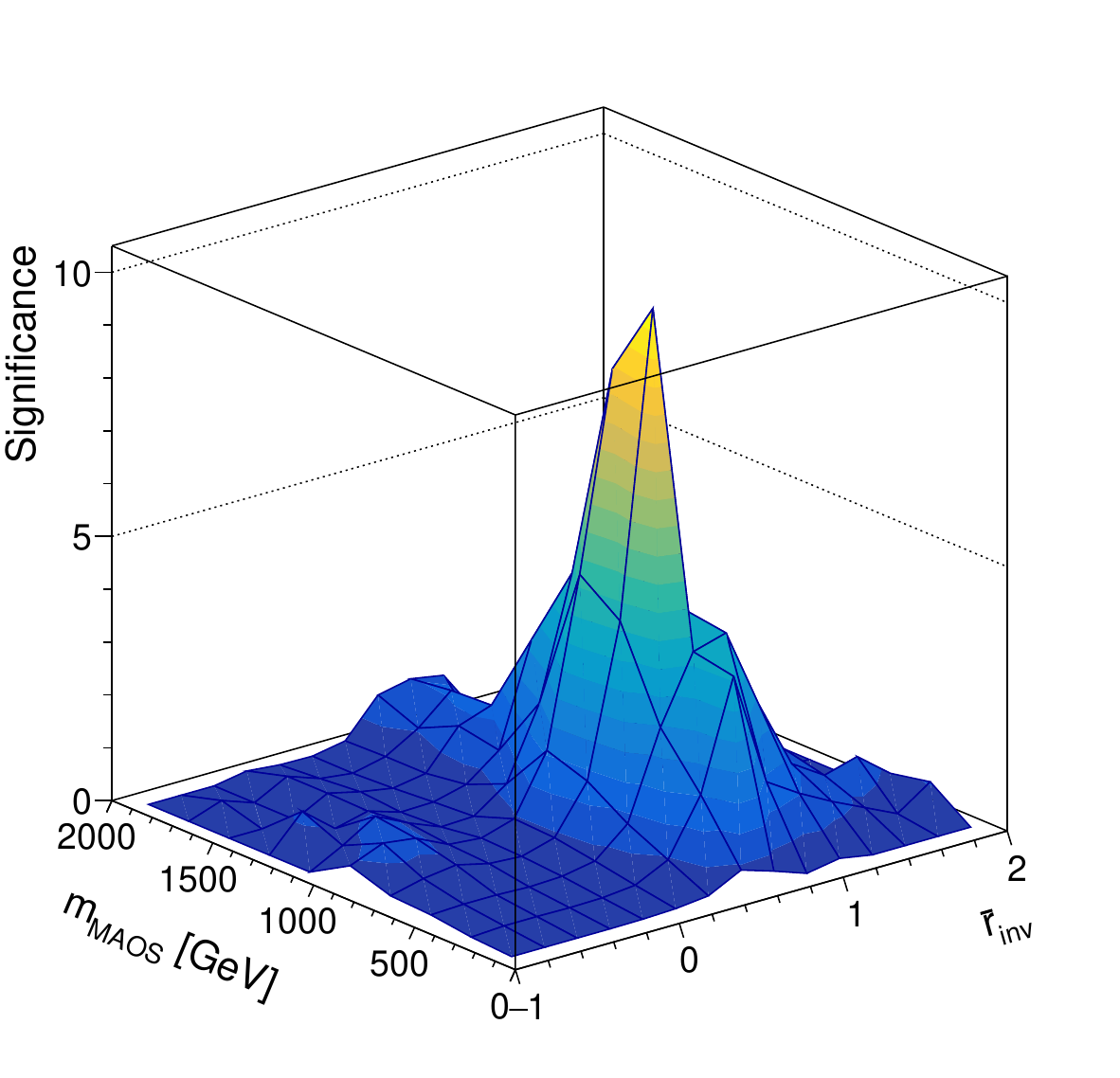}
 \includegraphics[width=0.32\columnwidth]{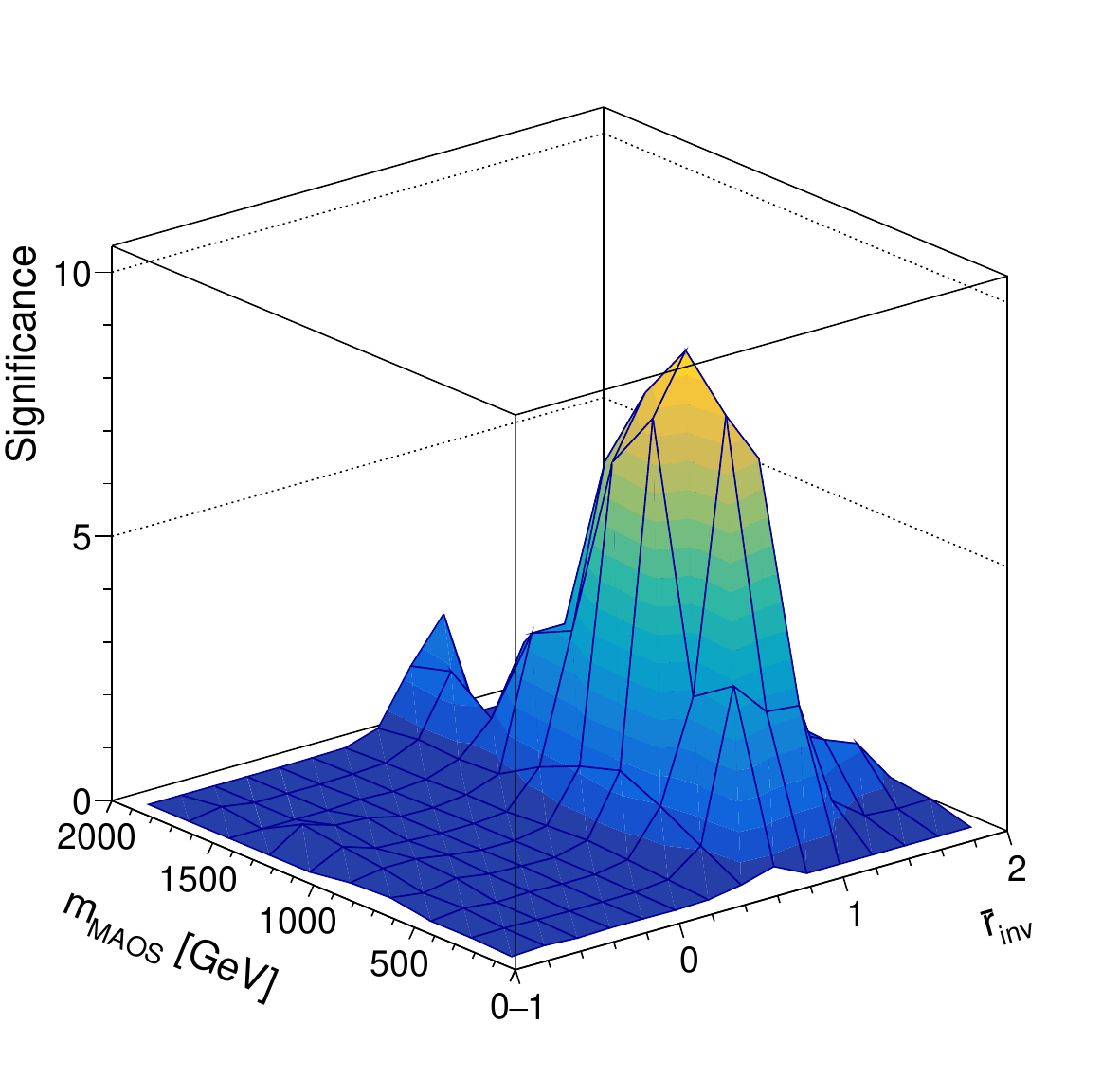}
\caption{The significance distribution in the \maos-\rinvave plane in the jet ISR channel for the 1.0 TeV \ZPrime with \rinv = 0.1 (left), 0.5 (middle) and 0.9 (right).}
\label{fig:twod_sig}
\end{center}
\end{figure}

The fact that \rinvave is a reasonable estimator of \rinv, forming a localized
feature, motivates an exploration of using its full distribution rather than
just as a selection. Figure~\ref{fig:twod_sig} shows the significance in the
\maos-\rinvave 2D plane, for the 1 TeV \ZPrime with various \rinv parameters,
where a local peak is formed in each case. Scanning this 2D plane is a viable
search strategy. However, the background modeling is more challenging.
Figures~\ref{fig:rinv1} and~\ref{fig:rinv2} show that the \rinvave distribution is smooth in
both multijet and signal, which makes it possible to model this variable via a
parameterization. Another advantage of this approach is that it provides a
unified strategy for various \rinv and mass combinations without choosing a particular
\rinvave selection. 

\begin{figure}[ht]
\begin{center}
 \includegraphics[width=0.6\columnwidth]{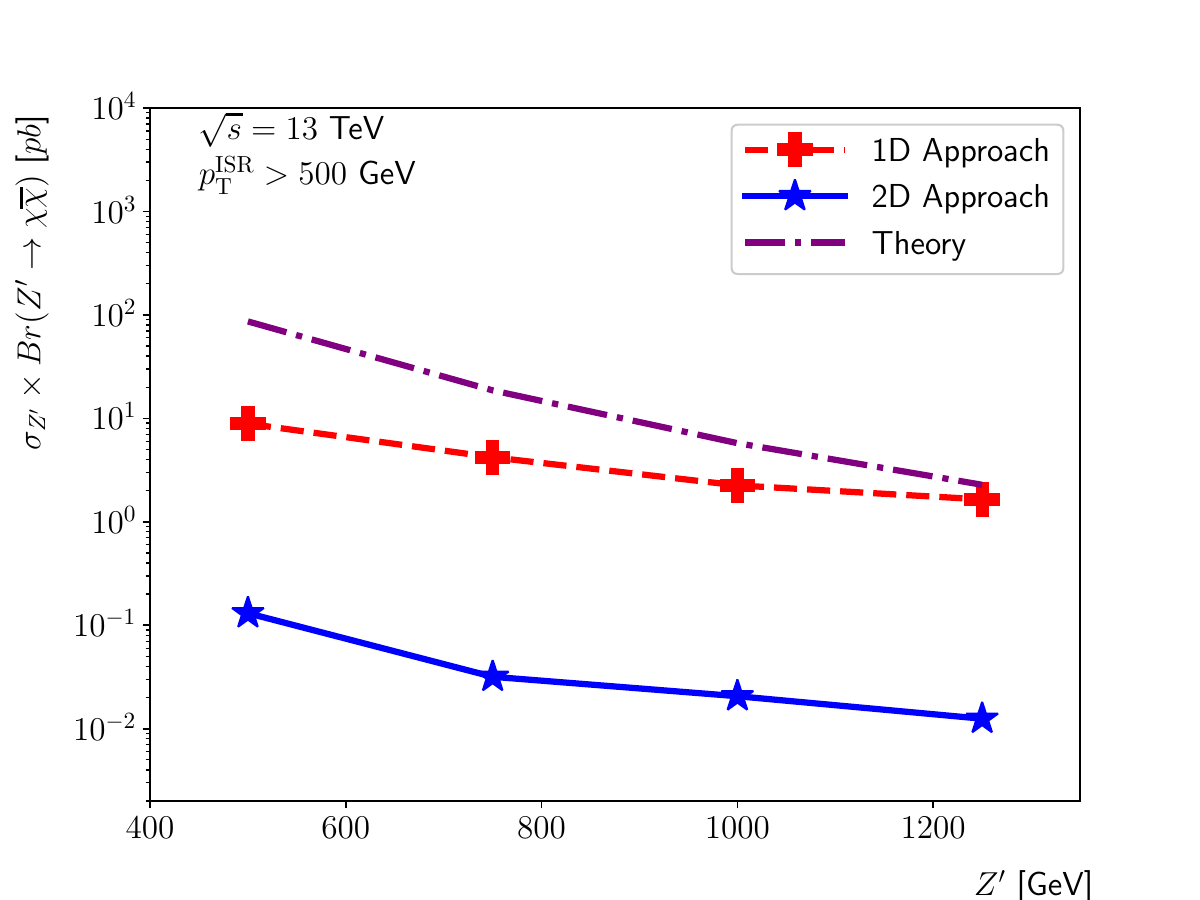}
\caption{Expected limits at 95\% CL obtained in the jet ISR channel, as functions of the \ZPrime mass. The 1D approach (dashed line) applies the $\rinvave > 0.4$ cut before scanning the \maos spectrum, and the 2D approach (solid line) directly considers the \maos-\rinvave 2D plane. The predicted $\sigma_{\ZPrime} \mathcal{B}$ from \MGMCatNLOV{3.5.1}~\cite{madgraph} is also displayed (dashed dotted line).}
\label{fig:limits}
\end{center}
\end{figure}

The sensitivity can be further quantified by computing the limits on
$\sigma_{\ZPrime} \mathcal{B}({\ZPrime\to\chi\overline{\chi}})$ at the 95\%
confidence level (CL) using the asymptotic approximation to the
$\text{CL}_{\text{s}}$ method~\cite{Cowan:2010js}.  The expected limits from
both the 1D and 2D approaches are evaluated for the jet ISR channel, as a
function of the \ZPrime mass. Two additional signal mass points, 750 GeV and
1250 GeV, are added to show the trend. The exclusion limits are calculated in
\PYHF{0.7.6}~\cite{pyhf, pyhf_joss}, where the background estimate is taken
directly from the simulated multijet samples. In the 1D approach, the \maos
distribution after applying the $\rinvave > 0.4$ selection is considered, while
the 2D approach utilizes the entire \maos-\rinvave 2D
plane\footnote{\PYHF{0.7.6} currently does not support an optimal 2D treatment,
where the correlations between neighboring bins are all taken into account.
The 2D input is flattened into a 1D dataset, i.e. a list of values in the
\rinvave-\maos bins.}. Figure~\ref{fig:limits} compares the expected limits
between these two strategies, which clearly motivates further development of
the 2D approach.

The absolute experimental accuracy of this study is subject to limitations in detector and reconstruction modeling in Delphes.
Jets misidentified as photons will complicate the event selection in the ISR photon channel,
potentially leading to additional backgrounds or reduced matching efficiency.
Mismeasured jets will induce spurious \ptvecmiss, increasing the multijet background.
However, the relative performance of the different strategies for the different signal models is accurately reflected here.

\section{Outlook}
\label{sec:summary}

This work considers the case where two semi-visible jets (SVJs) are produced
along with initial state radiation (ISR), proving that this production channel
is a promising avenue for semi-visible jet searches. Besides allowing us to
overcome the trigger threshold limitation, the presence of an energetic ISR
object brings other advantages. We derive several observables to reconstruct
properties of the semi-visible jet process that show great potential. Assuming
that each invisible decay component is aligned with the corresponding visible
jet, an analytic \ptvecmiss decomposition facilitates an estimation of the
invisible fraction for each jet, which can be averaged to find the event-level
invisible fraction \rinvave. The separation power of this observable is
enhanced as the ISR \Pt increases.  The reconstructed mediator mass \maos is
found using a different, numerical \ptvecmiss decomposition based on \MTtwo and
promoting the two invisible components to four-vectors with the \MTtwo-assisted
on shell (MAOS) scheme~\cite{maos}.  The signal \maos resolution suffers almost
no degradation as \rinv increases, when the ISR \Pt is significant. In
addition, the matching ambiguity in the jet ISR channel is mitigated thanks to
the alignment of \ETmiss with the signal jets, and we propose a method that
gives good matching efficiency with minimal background distortion. 

The analysis sensitivity is studied in detail for a 0.5 (1.0) TeV \ZPrime with
\rinv = 0.9 in the photon (jet) ISR channel. If only \maos is used as the
search variable, the sensitivity can be greatly increased by cutting on the
ratio of \ETmiss and \mT (\rinvave). Further, the \rinvave of the signal is
similarly localized, so a search in the \maos-\rinvave 2D plane merits
investigation. A preliminary examination of the expected limits shows that the
2D approach is an order of magnitude more sensitive. This conclusion applies to
mediator masses below 1 TeV and other \rinv values as well. Observables
constructed using \ETmiss are not invariant with respect to the ISR object \pT,
so a thorough survey is needed to optimize the search strategy. This analysis
is complementary to the $s$-channel SVJ search performed by CMS~\cite{cmssvj}
as it covers mediator masses below 1.5 TeV. The proposed 2D approach
is expected to achieve better sensitivity than a traditional monojet
search~\cite{cmsmonojet,atlasmonojet}, thanks to the dedicated new observables.

It is possible to apply a unified search strategy to explore a wide range of
the \rinv model parameter using the techniques developed in this work.  Even
when \rinv approaches unity, a mass resonance can be reconstructed.  As
recently pointed out in Ref.~\cite{jenpaper}, jet substructure variables suffer
from large modeling uncertainties in the context of dark QCD. This study
explores only simple physics objects, which makes it applicable to various
showering models. The advantages of jet substructure variables and machine
learning in SVJ searches were studied by previous
publications~\cite{cmssvj,svjgraph,svjauto,svjsub,svjml, forcedtoaddbyauthors}, and the findings in
this work could be combined with those methods.  In particular, a single
\ptvecmiss decomposition that provides optimal reconstruction for both the
mediator mass and invisible fraction could potentially be achieved using
interpretable machine learning~\cite{svjpedro}.  We point out that the
event-level topological features are similar between SVJ and other processes
such as Higgs decays with both visible and invisible final state particles.  We
look forward to seeing the ISR production channels probed in SVJ searches by
the LHC experiments.

\section*{{Acknowledgments}} We thank Benedikt Maier and Brendan Regnery for
initial discussions on this topic. B.X. Liu is supported by Shenzhen Campus of
the Sun Yat-sen University under project 74140-12240013. B.X. Liu appreciates
the support from Guangdong Provincial Key Laboratory of Gamma-Gamma Collider
and Its Comprehensive Applications, and the support from Guangdong Provincial
Key Laboratory of Advanced Particle Detection Technology. K. Pedro is supported
by Fermi Research Alliance, LLC under Contract No. DE-AC02-07CH11359 with the
U.S. Department of Energy, Office of Science, Office of High Energy Physics.

\appendix 
\section{Impacts of the Jet Clustering Radius}\label{app:radius}

As mentioned in Section~\ref{sec:intro}, many dark hadrons are created during
dark shower, a fraction of which subsequently decay to SM quarks.
The width of the jet increases with the dark hadron mass,
as heavier dark hadrons impart more momentum to their decay products,
leading to wider opening angles.
This effect was reported in Ref.~\cite{cmssvj}, where $R=0.8$ jets
were used to achieve better performance. The impacts of the jet clustering radius
are investigated for the ISR topology in this work. Five different
clustering radii are considered from $R = 0.4$ to $R = 0.8$ with a
step of 0.1. We label these different jet types as AK4 to AK8, since they all use anti-\kt clustering.

\begin{figure}[ht]
\begin{center}
 \includegraphics[width=0.32\columnwidth]{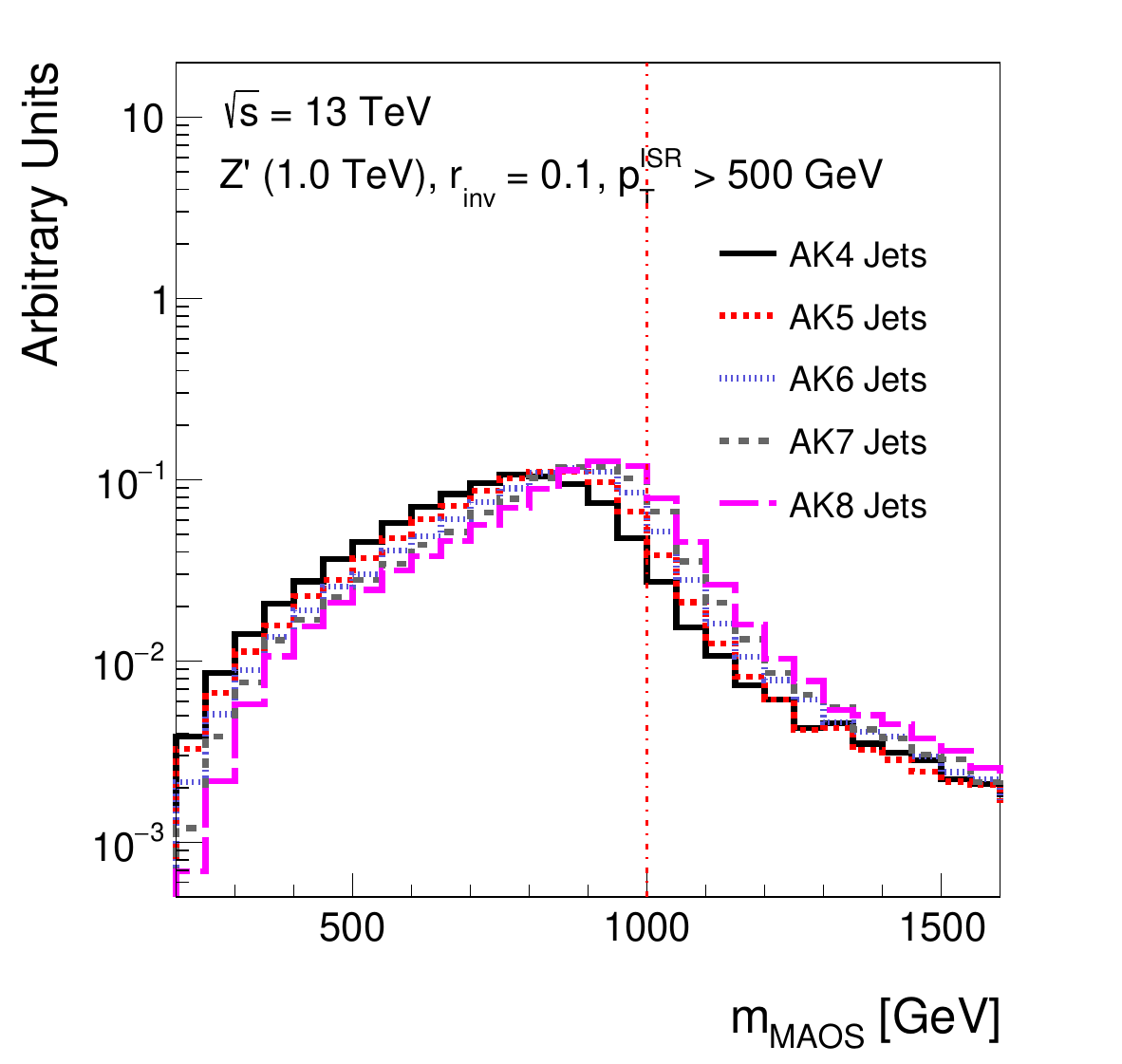}
 \includegraphics[width=0.32\columnwidth]{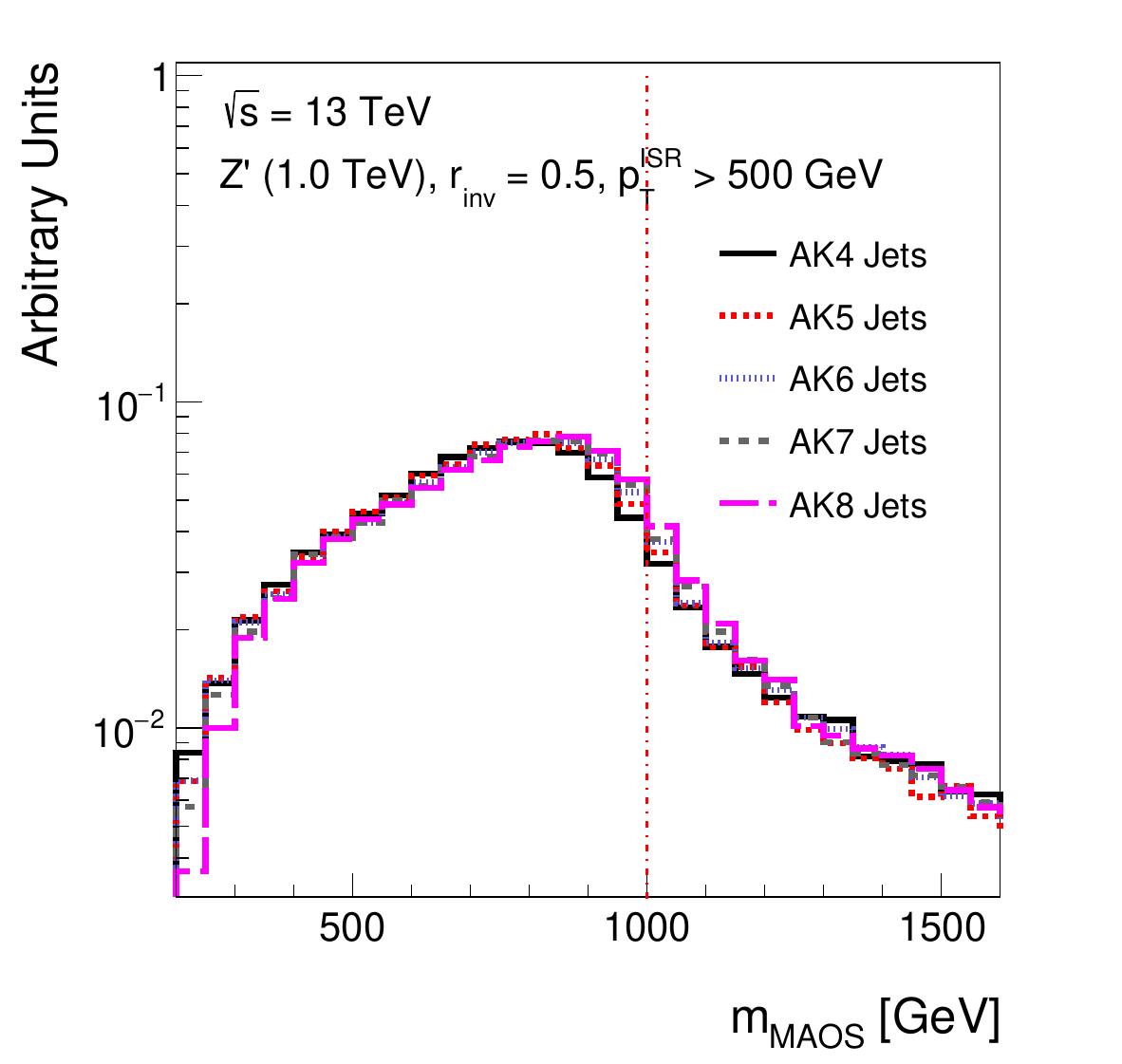}
 \includegraphics[width=0.32\columnwidth]{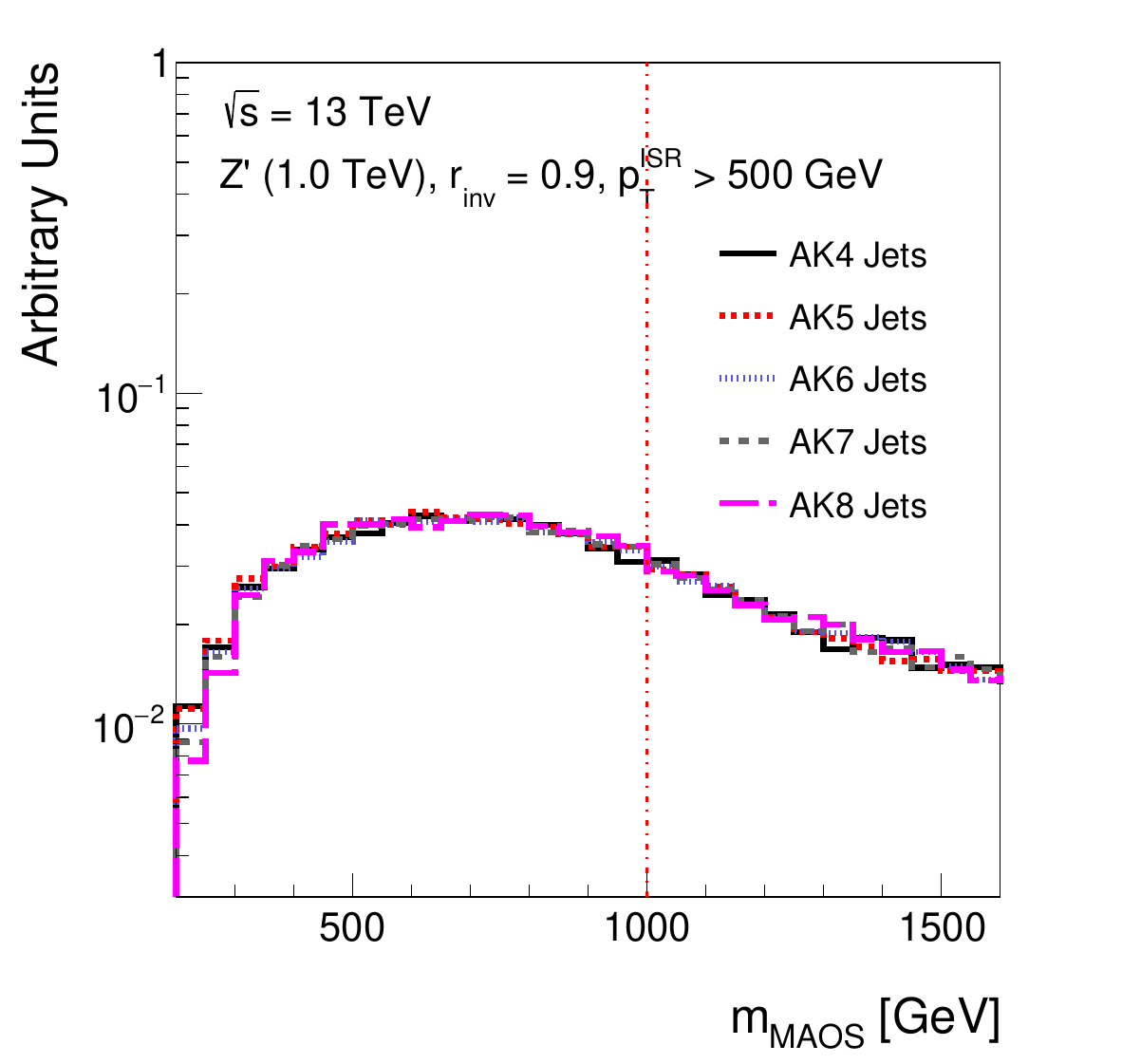}
\caption{Distributions of \maos for signals with $\mZPrime = 1.0\TeV$, $\PtISR = 500\GeV$, and $\rinv = 0.1$ (left), $\rinv = 0.5$ (center), and $\rinv = 0.9$ (right), using various jet clustering radii. The red dashed vertical line shows the actual \ZPrime mass.}
\label{fig:ak_maos}
\end{center}
\end{figure}

\begin{figure}[ht]
\begin{center}
 \includegraphics[width=0.32\columnwidth]{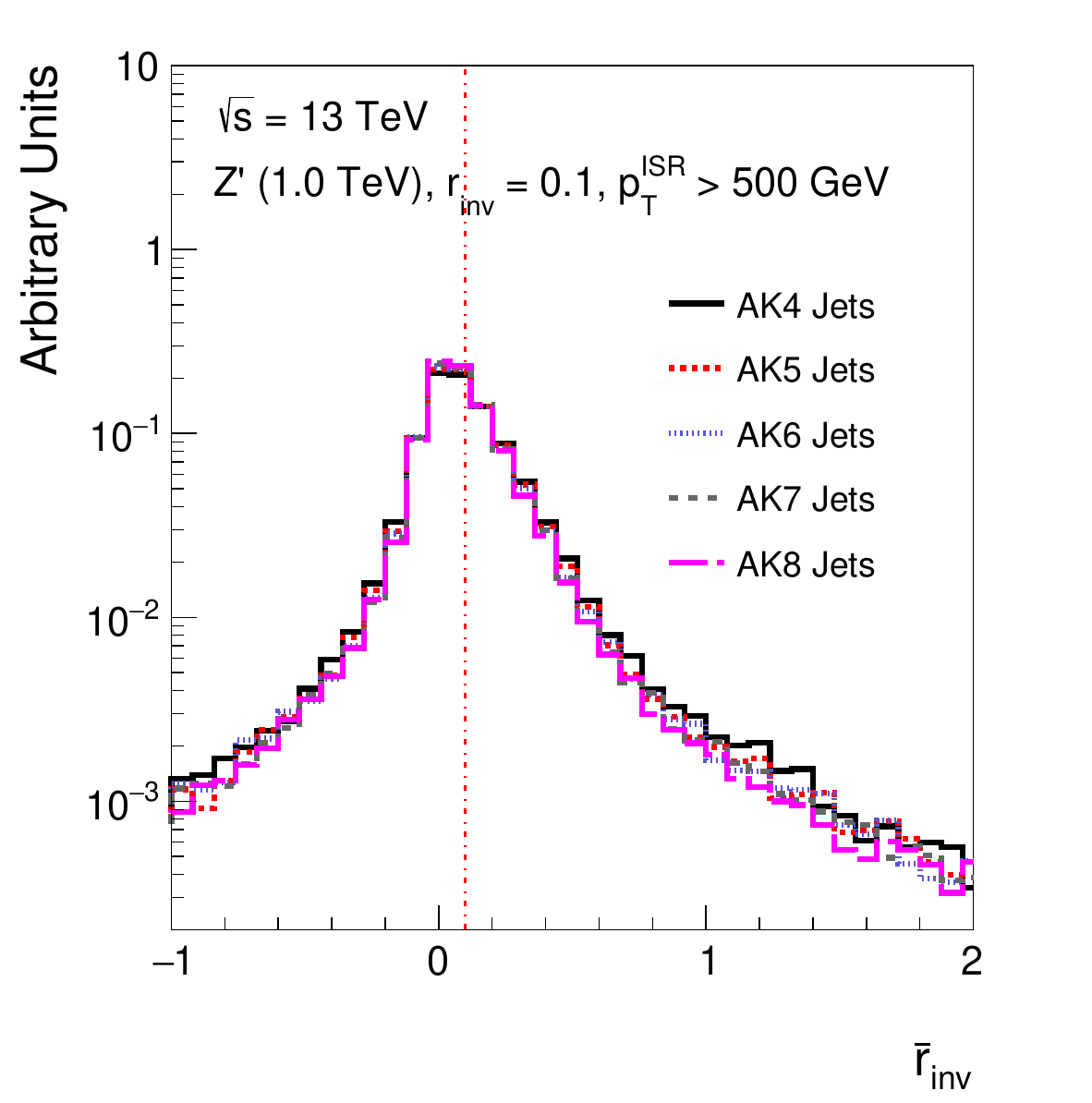}
 \includegraphics[width=0.32\columnwidth]{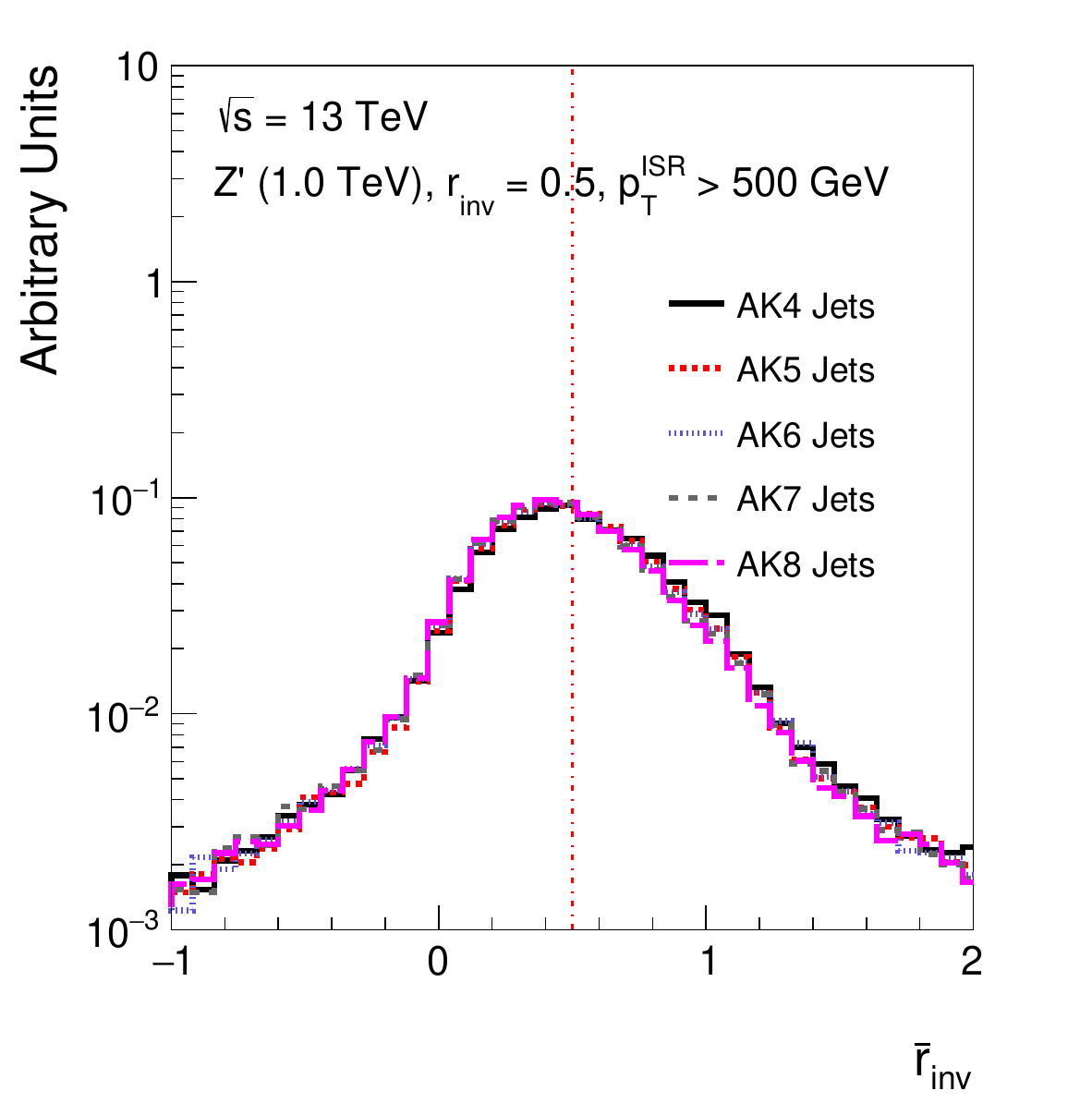}
 \includegraphics[width=0.32\columnwidth]{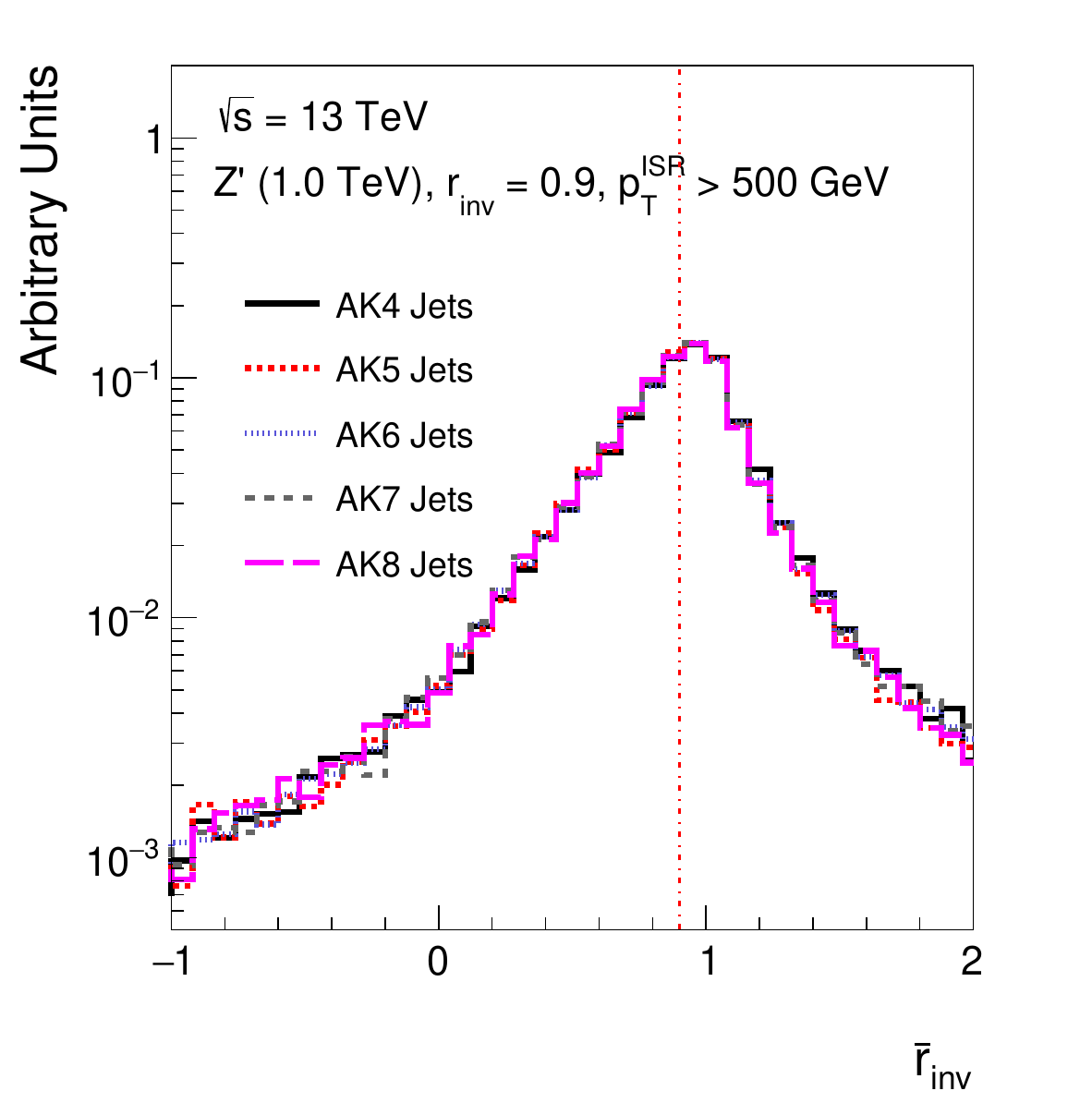}
\caption{Distributions of \rinv estimator for signals with $\mZPrime = 1.0\TeV$, $\PtISR = 500\GeV$, and $\rinv = 0.1$ (left), $\rinv = 0.5$ (center), and $\rinv = 0.9$ (right), using various jet clustering radii. The red dashed vertical line shows the theoretical \rinv value.}
\label{fig:ak_rinv}
\end{center}
\end{figure}

As shown in Fig.~\ref{fig:ak_maos}, the impact of
the jet clustering radius on \maos is obvious at low \rinv. However, as \rinv increases,
\ETmiss becomes more important and the impact of the clustering
radius decreases. The \maos distribution is not significantly affected when $\rinv=0.9$.
The new \rinvave observable proposed in this work is not
affected by the jet clustering radius, as seen in Fig.~\ref{fig:ak_rinv}, for
the same reasons. Therefore, the results in this paper apply to various
jet clustering choices.      

\clearpage
\addcontentsline{toc}{section}{References}
\bibliography{main.bib}

\providecommand{\href}[2]{#2}\begingroup\raggedright\begin{thebibliography}{10}

\bibitem{dm1}
V.C.~Rubin, N.~Thonnard and W.K.~Ford, Jr., \emph{{Rotational properties of 21 SC galaxies with a large range of luminosities and radii, from NGC 4605 /R = 4kpc/ to UGC 2885 /R = 122 kpc/}}, \href{https://doi.org/10.1086/158003}{\emph{Astrophys. J.} {\bfseries 238} (1980) 471}.

\bibitem{dm2}
M.~Persic, P.~Salucci and F.~Stel, \emph{The universal rotation curve of spiral galaxies: 1. the dark matter connection}, \href{https://doi.org/10.1093/mnras/278.1.27}{\emph{Mon. Not. Roy. Astron. Soc.} {\bfseries 281} (1996) 27} [\href{https://arxiv.org/abs/astro-ph/9506004}{{\ttfamily astro-ph/9506004}}].

\bibitem{dm3}
D.~Clowe, M.~Bradac, A.H.~Gonzalez, M.~Markevitch, S.W.~Randall, C.~Jones et~al., \emph{{A direct empirical proof of the existence of dark matter}}, \href{https://doi.org/10.1086/508162}{\emph{Astrophys. J. Lett.} {\bfseries 648} (2006) L109} [\href{https://arxiv.org/abs/astro-ph/0608407}{{\ttfamily astro-ph/0608407}}].

\bibitem{dm4}
{\scshape DES} collaboration, \emph{{Dark Energy Survey} year 1 results: Curved-sky weak lensing mass map}, \href{https://doi.org/10.1093/mnras/stx3363}{\emph{Mon. Not. Roy. Astron. Soc.} {\bfseries 475} (2018) 3165} [\href{https://arxiv.org/abs/1708.01535}{{\ttfamily 1708.01535}}].

\bibitem{dm5}
{\scshape Planck} collaboration, \emph{{Planck 2018 results. VI. Cosmological parameters}}, \href{https://doi.org/10.1051/0004-6361/201833910}{\emph{Astron. Astrophys.} {\bfseries 641} (2020) A6} [\href{https://arxiv.org/abs/1807.06209}{{\ttfamily 1807.06209}}].

\bibitem{wimps}
G.~Jungman, M.~Kamionkowski and K.~Griest, \emph{{Supersymmetric dark matter}}, \href{https://doi.org/10.1016/0370-1573(95)00058-5}{\emph{Phys. Rept.} {\bfseries 267} (1996) 195} [\href{https://arxiv.org/abs/hep-ph/9506380}{{\ttfamily hep-ph/9506380}}].

\bibitem{cmsmonojet}
{\scshape CMS} collaboration, \emph{{Search for new particles in events with energetic jets and large missing transverse momentum in proton-proton collisions at $ \sqrt{s} $ = 13 TeV}}, \href{https://doi.org/10.1007/JHEP11(2021)153}{\emph{JHEP} {\bfseries 11} (2021) 153} [\href{https://arxiv.org/abs/2107.13021}{{\ttfamily 2107.13021}}].

\bibitem{atlasmonojet}
{\scshape ATLAS} collaboration, \emph{{Search for new phenomena in events with an energetic jet and missing transverse momentum in $pp$ collisions at $\sqrt {s}$ =13 TeV with the ATLAS detector}}, \href{https://doi.org/10.1103/PhysRevD.103.112006}{\emph{Phys. Rev. D} {\bfseries 103} (2021) 112006} [\href{https://arxiv.org/abs/2102.10874}{{\ttfamily 2102.10874}}].

\bibitem{cmsmonotop}
{\scshape CMS} collaboration, \emph{{Search for dark matter produced in association with a single top quark or a top quark pair in proton-proton collisions at $ \sqrt{s}=13 $ TeV}}, \href{https://doi.org/10.1007/JHEP03(2019)141}{\emph{JHEP} {\bfseries 03} (2019) 141} [\href{https://arxiv.org/abs/1901.01553}{{\ttfamily 1901.01553}}].

\bibitem{atlasmonotop}
{\scshape ATLAS} collaboration, \emph{{Search for new phenomena in events with two opposite-charge leptons, jets and missing transverse momentum in pp collisions at $ \sqrt{\mathrm{s}} $ = 13 TeV with the ATLAS detector}}, \href{https://doi.org/10.1007/JHEP04(2021)165}{\emph{JHEP} {\bfseries 04} (2021) 165} [\href{https://arxiv.org/abs/2102.01444}{{\ttfamily 2102.01444}}].

\bibitem{atlasmonotoppair}
{\scshape ATLAS} collaboration, \emph{{Search for new phenomena with top quark pairs in final states with one lepton, jets, and missing transverse momentum in $pp$ collisions at $ \sqrt{s} $ = 13 TeV with the ATLAS detector}}, \href{https://doi.org/10.1007/JHEP04(2021)174}{\emph{JHEP} {\bfseries 04} (2021) 174} [\href{https://arxiv.org/abs/2012.03799}{{\ttfamily 2012.03799}}].

\bibitem{cmsmonoz}
{\scshape CMS} collaboration, \emph{{Search for dark matter produced in association with a leptonically decaying Z boson in proton-proton collisions at $\sqrt{s} =$ 13 TeV}}, \href{https://doi.org/10.1140/epjc/s10052-020-08739-5}{\emph{Eur. Phys. J. C} {\bfseries 81} (2021) 13} [\href{https://arxiv.org/abs/2008.04735}{{\ttfamily 2008.04735}}].

\bibitem{atlasmonov}
{\scshape ATLAS} collaboration, \emph{{Search for dark matter in events with a hadronically decaying vector boson and missing transverse momentum in $pp$ collisions at $\sqrt{s} = 13$ TeV with the ATLAS detector}}, \href{https://doi.org/10.1007/JHEP10(2018)180}{\emph{JHEP} {\bfseries 10} (2018) 180} [\href{https://arxiv.org/abs/1807.11471}{{\ttfamily 1807.11471}}].

\bibitem{cmsmonophoton}
{\scshape CMS} collaboration, \emph{{Search for new physics in final states with a single photon and missing transverse momentum in proton-proton collisions at $\sqrt{s} =$ 13 TeV}}, \href{https://doi.org/10.1007/JHEP02(2019)074}{\emph{JHEP} {\bfseries 02} (2019) 074} [\href{https://arxiv.org/abs/1810.00196}{{\ttfamily 1810.00196}}].

\bibitem{cmsmonohiggs}
{\scshape CMS} collaboration, \emph{{Search for dark matter produced in association with a Higgs boson decaying to a pair of bottom quarks in proton\textendash{}proton collisions at $\sqrt{s}=13\,\text {Te}\text {V} $}}, \href{https://doi.org/10.1140/epjc/s10052-019-6730-7}{\emph{Eur. Phys. J. C} {\bfseries 79} (2019) 280} [\href{https://arxiv.org/abs/1811.06562}{{\ttfamily 1811.06562}}].

\bibitem{atlasmonohiggstautau}
{\scshape ATLAS} collaboration, \emph{{Search for dark matter produced in association with a Higgs boson decaying to tau leptons at $ \sqrt{s} $ = 13 TeV with the ATLAS detector}}, \href{https://doi.org/10.1007/JHEP09(2023)189}{\emph{JHEP} {\bfseries 09} (2023) 189} [\href{https://arxiv.org/abs/2305.12938}{{\ttfamily 2305.12938}}].

\bibitem{atlasmonohiggsbb}
{\scshape ATLAS} collaboration, \emph{{Search for dark matter produced in association with a Standard Model Higgs boson decaying into b-quarks using the full Run 2 dataset from the ATLAS detector}}, \href{https://doi.org/10.1007/JHEP11(2021)209}{\emph{JHEP} {\bfseries 11} (2021) 209} [\href{https://arxiv.org/abs/2108.13391}{{\ttfamily 2108.13391}}].

\bibitem{darkqcd}
Y.~Bai and P.~Schwaller, \emph{{Scale of dark QCD}}, \href{https://doi.org/10.1103/PhysRevD.89.063522}{\emph{Phys. Rev. D} {\bfseries 89} (2014) 063522} [\href{https://arxiv.org/abs/1306.4676}{{\ttfamily 1306.4676}}].

\bibitem{hiddenvalleydm}
H.~Beauchesne, E.~Bertuzzo and G.~Grilli Di~Cortona, \emph{Dark matter in hidden valley models with stable and unstable light dark mesons}, \href{https://doi.org/10.1007/JHEP04(2019)118}{\emph{JHEP} {\bfseries 04} (2019) 118} [\href{https://arxiv.org/abs/1809.10152}{{\ttfamily 1809.10152}}].

\bibitem{svjtheory}
T.~Cohen, M.~Lisanti and H.K.~Lou, \emph{Semivisible jets: Dark matter undercover at the {LHC}}, \href{https://doi.org/10.1103/PhysRevLett.115.171804}{\emph{Phys. Rev. Lett.} {\bfseries 115} (2015) 171804} [\href{https://arxiv.org/abs/1503.00009}{{\ttfamily 1503.00009}}].

\bibitem{svjlhc}
T.~Cohen, M.~Lisanti, H.K.~Lou and S.~Mishra-Sharma, \emph{{LHC} searches for dark sector showers}, \href{https://doi.org/10.1007/JHEP11(2017)196}{\emph{JHEP} {\bfseries 11} (2017) 196} [\href{https://arxiv.org/abs/1707.05326}{{\ttfamily 1707.05326}}].

\bibitem{cmssvj}
{\scshape CMS} collaboration, \emph{{Search for resonant production of strongly coupled dark matter in proton-proton collisions at 13 TeV}}, \href{https://doi.org/10.1007/JHEP06(2022)156}{\emph{JHEP} {\bfseries 06} (2022) 156} [\href{https://arxiv.org/abs/2112.11125}{{\ttfamily 2112.11125}}].

\bibitem{atlassvj}
{\scshape ATLAS} collaboration, \emph{{Search for non-resonant production of semi-visible jets using Run 2 data in ATLAS}}, \href{https://doi.org/10.1016/j.physletb.2023.138324}{\emph{Phys. Lett. B} {\bfseries 848} (2024) 138324} [\href{https://arxiv.org/abs/2305.18037}{{\ttfamily 2305.18037}}].

\bibitem{snowmasssvj}
G.~Albouy et~al., \emph{{Theory, phenomenology, and experimental avenues for dark showers: a Snowmass 2021 report}}, \href{https://doi.org/10.1140/epjc/s10052-022-11048-8}{\emph{Eur. Phys. J. C} {\bfseries 82} (2022) 1132} [\href{https://arxiv.org/abs/2203.09503}{{\ttfamily 2203.09503}}].

\bibitem{atlasdarkjets}
{\scshape ATLAS} collaboration, \emph{{Search for resonant production of dark quarks in the dijet final state with the ATLAS detector}}, \href{https://doi.org/10.1007/JHEP02(2024)128}{\emph{JHEP} {\bfseries 02} (2024) 128} [\href{https://arxiv.org/abs/2311.03944}{{\ttfamily 2311.03944}}].

\bibitem{atlasdijet}
{\scshape ATLAS} collaboration, \emph{{Search for new resonances in mass distributions of jet pairs using 139 fb$^{-1}$ of $pp$ collisions at $\sqrt{s}=13$ TeV with the ATLAS detector}}, \href{https://doi.org/10.1007/JHEP03(2020)145}{\emph{JHEP} {\bfseries 03} (2020) 145} [\href{https://arxiv.org/abs/1910.08447}{{\ttfamily 1910.08447}}].

\bibitem{cmsdarksector}
{\scshape CMS} collaboration, \emph{{Dark sector searches with the CMS experiment}},  \href{https://arxiv.org/abs/2405.13778}{{\ttfamily 2405.13778}}.

\bibitem{cmsdijetisr}
{\scshape CMS} collaboration, \emph{{Search for dijet resonances using events with three jets in proton-proton collisions at s=13TeV}}, \href{https://doi.org/10.1016/j.physletb.2020.135448}{\emph{Phys. Lett. B} {\bfseries 805} (2020) 135448} [\href{https://arxiv.org/abs/1911.03761}{{\ttfamily 1911.03761}}].

\bibitem{atlasdijetisr}
{\scshape ATLAS} collaboration, \emph{{Search for low-mass resonances decaying into two jets and produced in association with a photon or a jet at $\sqrt{s}=13$ TeV with the ATLAS detector}},  \href{https://arxiv.org/abs/2403.08547}{{\ttfamily 2403.08547}}.

\bibitem{madgraph}
J.~Alwall, R.~Frederix, S.~Frixione, V.~Hirschi, F.~Maltoni, O.~Mattelaer et~al., \emph{{The automated computation of tree-level and next-to-leading order differential cross sections, and their matching to parton shower simulations}}, \href{https://doi.org/10.1007/JHEP07(2014)079}{\emph{JHEP} {\bfseries 07} (2014) 079} [\href{https://arxiv.org/abs/1405.0301}{{\ttfamily 1405.0301}}].

\bibitem{pythia}
T.~Sj\"ostrand, S.~Ask, J.R.~Christiansen, R.~Corke, N.~Desai, P.~Ilten et~al., \emph{{An introduction to PYTHIA 8.2}}, \href{https://doi.org/10.1016/j.cpc.2015.01.024}{\emph{Comput. Phys. Commun.} {\bfseries 191} (2015) 159} [\href{https://arxiv.org/abs/1410.3012}{{\ttfamily 1410.3012}}].

\bibitem{delphes}
{\scshape DELPHES 3} collaboration, \emph{{DELPHES 3, A modular framework for fast simulation of a generic collider experiment}}, \href{https://doi.org/10.1007/JHEP02(2014)057}{\emph{JHEP} {\bfseries 02} (2014) 057} [\href{https://arxiv.org/abs/1307.6346}{{\ttfamily 1307.6346}}].

\bibitem{Lonnblad:2011xx}
L.~Lonnblad and S.~Prestel, \emph{{Matching tree-level matrix elements with interleaved showers}}, \href{https://doi.org/10.1007/JHEP03(2012)019}{\emph{JHEP} {\bfseries 03} (2012) 019} [\href{https://arxiv.org/abs/1109.4829}{{\ttfamily 1109.4829}}].

\bibitem{Albert:2017onk}
A.~Albert et~al., \emph{Recommendations of the {LHC} dark matter working group: Comparing {LHC} searches for dark matter mediators in visible and invisible decay channels and calculations of the thermal relic density}, \href{https://doi.org/10.1016/j.dark.2019.100377}{\emph{Phys. Dark Univ.} {\bfseries 26} (2019) 100377} [\href{https://arxiv.org/abs/1703.05703}{{\ttfamily 1703.05703}}].

\bibitem{Lester:1999tx}
C.G.~Lester and D.J.~Summers, \emph{{Measuring masses of semiinvisibly decaying particles pair produced at hadron colliders}}, \href{https://doi.org/10.1016/S0370-2693(99)00945-4}{\emph{Phys. Lett. B} {\bfseries 463} (1999) 99} [\href{https://arxiv.org/abs/hep-ph/9906349}{{\ttfamily hep-ph/9906349}}].

\bibitem{Lester:2014yga}
C.G.~Lester and B.~Nachman, \emph{{Bisection-based asymmetric M$_{T2}$ computation: a higher precision calculator than existing symmetric methods}}, \href{https://doi.org/10.1007/JHEP03(2015)100}{\emph{JHEP} {\bfseries 03} (2015) 100} [\href{https://arxiv.org/abs/1411.4312}{{\ttfamily 1411.4312}}].

\bibitem{svjdecomp}
H.~Beauchesne and G.~Grilli~di Cortona, \emph{{Event-level variables for semivisible jets using anomalous jet tagging}},  in \emph{{Snowmass 2021}}, 11, 2021 [\href{https://arxiv.org/abs/2111.12156}{{\ttfamily 2111.12156}}].

\bibitem{maos}
K.~Choi, J.S.~Lee and C.B.~Park, \emph{Measuring the {Higgs} boson mass with transverse mass variables}, \href{https://doi.org/10.1103/PhysRevD.82.113017}{\emph{Phys. Rev. D} {\bfseries 82} (2010) 113017}.

\bibitem{svjpedro}
K.~Pedro and P.~Shyamsundar, \emph{{Optimal mass variables for semivisible jets}}, \href{https://doi.org/10.21468/SciPostPhysCore.6.4.067}{\emph{SciPost Phys. Core} {\bfseries 6} (2023) 067} [\href{https://arxiv.org/abs/2303.16253}{{\ttfamily 2303.16253}}].

\bibitem{htautau_colinear}
R.~Ellis, I.~Hinchliffe, M.~Soldate and J.~{Van Der Bij}, \emph{{Higgs decay to $\tau^{+}\tau^{-}$--A possible signature of intermediate mass Higgs bosons at high energy hadron colliders}}, \href{https://doi.org/https://doi.org/10.1016/0550-3213(88)90019-3}{\emph{Nucl. Phys. B} {\bfseries 297} (1988) 221}.

\bibitem{Cacciari:2008gp}
M.~Cacciari, G.P.~Salam and G.~Soyez, \emph{{The anti-\kt jet clustering algorithm}}, \href{https://doi.org/10.1088/1126-6708/2008/04/063}{\emph{JHEP} {\bfseries 04} (2008) 063} [\href{https://arxiv.org/abs/0802.1189}{{\ttfamily 0802.1189}}].

\bibitem{Cacciari:2011ma}
M.~Cacciari, G.P.~Salam and G.~Soyez, \emph{{FastJet user manual}}, \href{https://doi.org/10.1140/epjc/s10052-012-1896-2}{\emph{Eur. Phys. J. C} {\bfseries 72} (2012) 1896} [\href{https://arxiv.org/abs/1111.6097}{{\ttfamily 1111.6097}}].

\bibitem{Dasgupta:2013ihk}
M.~Dasgupta, A.~Fregoso, S.~Marzani and G.P.~Salam, \emph{{Towards an understanding of jet substructure}}, \href{https://doi.org/10.1007/JHEP09(2013)029}{\emph{JHEP} {\bfseries 09} (2013) 029} [\href{https://arxiv.org/abs/1307.0007}{{\ttfamily 1307.0007}}].

\bibitem{Cowan:2010js}
G.~Cowan, K.~Cranmer, E.~Gross and O.~Vitells, \emph{{Asymptotic formulae for likelihood-based tests of new physics}}, \href{https://doi.org/10.1140/epjc/s10052-011-1554-0}{\emph{Eur. Phys. J. C} {\bfseries 71} (2011) 1554} [\href{https://arxiv.org/abs/1007.1727}{{\ttfamily 1007.1727}}].

\bibitem{pyhf}
L.~Heinrich, M.~Feickert and G.~Stark, ``{pyhf: v0.7.6}.''
\newblock 10.5281/zenodo.1169739.

\bibitem{pyhf_joss}
L.~Heinrich, M.~Feickert, G.~Stark and K.~Cranmer, \emph{pyhf: pure-python implementation of histfactory statistical models}, \href{https://doi.org/10.21105/joss.02823}{\emph{J. Open Source Softw.} {\bfseries 6} (2021) 2823}.

\bibitem{jenpaper}
T.~Cohen, J.~Roloff and C.~Scherb, \emph{{Dark sector showers in the Lund jet plane}}, \href{https://doi.org/10.1103/PhysRevD.108.L031501}{\emph{Phys. Rev. D} {\bfseries 108} (2023) L031501} [\href{https://arxiv.org/abs/2301.07732}{{\ttfamily 2301.07732}}].

\bibitem{svjgraph}
E.~Bernreuther, T.~Finke, F.~Kahlhoefer, M.~Kr\"amer and A.~M\"uck, \emph{{Casting a graph net to catch dark showers}}, \href{https://doi.org/10.21468/SciPostPhys.10.2.046}{\emph{SciPost Phys.} {\bfseries 10} (2021) 046} [\href{https://arxiv.org/abs/2006.08639}{{\ttfamily 2006.08639}}].

\bibitem{svjauto}
F.~Canelli, A.~de~Cosa, L.L.~Pottier, J.~Niedziela, K.~Pedro and M.~Pierini, \emph{{Autoencoders for semivisible jet detection}}, \href{https://doi.org/10.1007/JHEP02(2022)074}{\emph{JHEP} {\bfseries 02} (2022) 074} [\href{https://arxiv.org/abs/2112.02864}{{\ttfamily 2112.02864}}].

\bibitem{svjsub}
D.~Kar and S.~Sinha, \emph{{Exploring jet substructure in semi-visible jets}}, \href{https://doi.org/10.21468/SciPostPhys.10.4.084}{\emph{SciPost Phys.} {\bfseries 10} (2021) 084}.

\bibitem{svjml}
T.~Faucett, S.-C.~Hsu and D.~Whiteson, \emph{{Learning to identify semi-visible jets}}, \href{https://doi.org/10.1007/JHEP12(2022)132}{\emph{JHEP} {\bfseries 12} (2022) 132} [\href{https://arxiv.org/abs/2208.10062}{{\ttfamily 2208.10062}}].

\bibitem{forcedtoaddbyauthors}
M.~Park and M.~Zhang, \emph{{Tagging a jet from a dark sector with Jet-substructures at colliders}}, \href{https://doi.org/10.1103/PhysRevD.100.115009}{\emph{Phys. Rev. D} {\bfseries 100} (2019) 115009} [\href{https://arxiv.org/abs/1712.09279}{{\ttfamily 1712.09279}}].

\end{thebibliography}\endgroup

\end{document}